\documentclass[a4paper,11pt]{article}

\usepackage{./auxi/jheppub}
\usepackage[utf8]{inputenc}
\usepackage[ngerman, english]{babel}
\usepackage[T1]{fontenc}
\usepackage[titletoc]{appendix}
\usepackage{titletoc}
\usepackage{graphicx}
\usepackage{subcaption} 
\usepackage{xcolor}
\usepackage{amsmath,amssymb,amsthm}
\usepackage{bm}
\usepackage{float}
\usepackage[format=plain,labelfont={bf,it},textfont=it]{caption}
\usepackage{multirow}
\usepackage{slashed} 
\usepackage{tabularx}
\usepackage{bbold}
\usepackage[shortlabels]{enumitem}
\usepackage{tikz}
\usetikzlibrary{positioning,arrows,calc}
\usetikzlibrary{arrows.meta}
\usetikzlibrary{decorations.pathmorphing}
\usetikzlibrary{decorations.markings}
\usetikzlibrary{shapes.geometric}
\usetikzlibrary{patterns}
\tikzset{
  snake it/.style={
    decorate, 
    decoration=snake,
    segment length=3
  }
}
\usepackage{xparse}
\usepackage{pdfpages} 
\usepackage{titlesec} 
\usepackage{fancyhdr} 
\usepackage{emptypage} 
\usepackage{dsfont} 

\usepackage{booktabs}
\usepackage{tabularx}
\definecolor{DarkBlueGrey}{RGB}{76,94,107}
\definecolor{MediumBlueGrey}{RGB}{110,135,153}
\definecolor{LightBlueGrey}{RGB}{134,163,184}
\definecolor{VeryLightBlueGrey}{RGB}{242,249,255}
\definecolor{WCOrange}{RGB}{242,146,29}
\definecolor{VeryLightOrange}{RGB}{255,245,233}
\definecolor{SCRed}{RGB}{179,48,48}
\definecolor{VeryLightRed}{RGB}{255,239,239}
\definecolor{VertexColor}{RGB}{242,146,29}
\definecolor{GluonColor}{RGB}{255,172,172}
\definecolor{SEColor}{RGB}{134,163,184}
\definecolor{EuclideanColor}{RGB}{0, 255, 255}

\definecolor{BGBox}{RGB}{255,254,230}
\definecolor{PlaneColor}{RGB}{230,230,230}
\definecolor{BlobColor}{RGB}{190,180,230}

\newcommand{\vev}[1]{\langle\, #1 \, \rangle}

\def\veps{\varepsilon}

\def\Cm{{\mathcal{C}}}

\def\Fm{{\mathcal{F}}}

\def\Im{{\mathcal{I}}}

\def\Nm{{\mathcal{N}}}
\def\Om{{\mathcal{O}}}

\def\Zm{{\mathcal{Z}}}

\usepackage{bbm}

\def\eps{\epsilon}
\def\veps{\varepsilon}

\newcommand\zb{{\bar{z}}}

\newif\ifstartcompletesineup
\newif\ifendcompletesineup
\pgfkeys{
    /pgf/decoration/.cd,
    start up/.is if=startcompletesineup,
    start up=true,
    start up/.default=true,
    start down/.style={/pgf/decoration/start up=false},
    end up/.is if=endcompletesineup,
    end up=true,
    end up/.default=true,
    end down/.style={/pgf/decoration/end up=false}
}
\pgfdeclaredecoration{complete sines}{initial}
{
    \state{initial}[
        width=+0pt,
        next state=upsine,
        persistent precomputation={
            \ifstartcompletesineup
                \pgfkeys{/pgf/decoration automaton/next state=upsine}
                \ifendcompletesineup
                    \pgfmathsetmacro\matchinglength{
                        0.5*\pgfdecoratedinputsegmentlength / (ceil(0.5* \pgfdecoratedinputsegmentlength / \pgfdecorationsegmentlength) )
                    }
                \else
                    \pgfmathsetmacro\matchinglength{
                        0.5 * \pgfdecoratedinputsegmentlength / (ceil(0.5 * \pgfdecoratedinputsegmentlength / \pgfdecorationsegmentlength ) - 0.499)
                    }
                \fi
            \else
                \pgfkeys{/pgf/decoration automaton/next state=downsine}
                \ifendcompletesineup
                    \pgfmathsetmacro\matchinglength{
                        0.5* \pgfdecoratedinputsegmentlength / (ceil(0.5 * \pgfdecoratedinputsegmentlength / \pgfdecorationsegmentlength ) - 0.4999)
                    }
                \else
                    \pgfmathsetmacro\matchinglength{
                        0.5 * \pgfdecoratedinputsegmentlength / (ceil(0.5 * \pgfdecoratedinputsegmentlength / \pgfdecorationsegmentlength ) )
                    }
                \fi
            \fi
            \setlength{\pgfdecorationsegmentlength}{\matchinglength pt}
        }] {}
    \state{downsine}[width=\pgfdecorationsegmentlength,next state=upsine]{
        \pgfpathsine{\pgfpoint{0.5\pgfdecorationsegmentlength}{0.5\pgfdecorationsegmentamplitude}}
        \pgfpathcosine{\pgfpoint{0.5\pgfdecorationsegmentlength}{-0.5\pgfdecorationsegmentamplitude}}
    }
    \state{upsine}[width=\pgfdecorationsegmentlength,next state=downsine]{
        \pgfpathsine{\pgfpoint{0.5\pgfdecorationsegmentlength}{-0.5\pgfdecorationsegmentamplitude}}
        \pgfpathcosine{\pgfpoint{0.5\pgfdecorationsegmentlength}{0.5\pgfdecorationsegmentamplitude}}
}
    \state{final}{}
}

\tikzset{
corner/.style={line width=1pt,dashed,draw=black,dash pattern=on 6pt off 4pt},
scalar/.style={line width=1pt,draw=black},
gluon/.style={line width=1pt,decorate, draw=GluonColor,
    decoration={complete sines,aspect=0,amplitude=1.25mm,segment length=1.5mm,start up,end up}},
gluontwo/.style={line width=1pt,decorate, draw=GluonColor,
    decoration={complete sines,aspect=0,amplitude=.7mm,segment length=1mm,start up,end up}},
ghost/.style={line width=1pt,loosely dotted,draw=black},
wilson/.style={line width=2pt,draw=black},
 }
\NewDocumentCommand\semiloop{O{black}mmmO{}O{above}}
{%
\draw[#1] let \p1 = ($(#3)-(#2)$) in (#3) arc (#4:({#4+180}):({0.5*veclen(\x1,\y1)})node[midway, #6] {#5};)
}

\newcommand\x{.75}

\newcommand{\GMIProofFig}{\begin{tikzpicture}[scale = 1, baseline={([yshift=-.5ex]current bounding box.center)}]
\draw[scalar] (5.8,.7) -- (5.8,1);
\draw[scalar] (5.8,.7) -- (6.1,.7);
\node[] at (6,.9) {$\zb$};
\draw[scalar,dashed,gray] (1,-1) -- (1,1);
\node[gray] at (1,1.25) {\footnotesize $-2\beta$};
\draw[scalar,dashed,gray] (2,-1) -- (2,1);
\node[gray] at (2,1.25) {\footnotesize $-\beta$};
\draw[scalar,dashed,gray] (4,-1) -- (4,1);
\node[gray] at (4,1.25) {\footnotesize $\beta$};
\draw[scalar,dashed,gray] (5,-1) -- (5,1);
\node[gray] at (5,1.25) {\footnotesize $2\beta$};
\draw[scalar,-Latex] (3,0) -- (6.2,0);
\draw[scalar,-Latex] (3,-1) -- (3,1);
\draw[wilson,EuclideanColor] (3,0) -- (4,0);
\draw[-, decorate, decoration={zigzag, segment length=6, amplitude=3},line width=1,SCRed] (0,0) -- (3,0);
\draw[fill=WCOrange,WCOrange] (3,0) circle (.075);
\draw[fill=WCOrange,WCOrange] (4,0) circle (.075);
\draw[wilson,-Latex] (7,0) -- (8,0);
\draw[scalar] (5.8+9,.7) -- (5.8+9,1);
\draw[scalar] (5.8+9,.7) -- (6.1+9,.7);
\node[] at (6+9,.9) {$w$};
\draw[scalar,dashed,gray] (.7+3+9,-1) -- (.7+3+9,1);
\node[gray] at (.7+3+9,1.25) {\footnotesize $r$};
\draw[scalar,dashed,gray] (-.7+3+9,-1) -- (-.7+3+9,1);
\node[gray] at (-.7+3+9,1.25) {\footnotesize $-r$};
\draw[scalar,dashed,gray] (2+3+9,-1) -- (2+3+9,1);
\node[gray] at (2+3+9,1.25) {\footnotesize $1/r$};
\draw[scalar,dashed,gray] (-2+3+9,-1) -- (-2+3+9,1);
\node[gray] at (-2+3+9,1.25) {\footnotesize $-1/r$};
\draw[scalar] (3+9,0) -- (6+9,0);
\draw[scalar,-Latex] (6+9,0) -- (6.2+9,0);
\draw[scalar,-Latex] (3+9,-1) -- (3+9,1);
\draw[wilson,EuclideanColor] (3+9,0) circle (.9);
\draw[-, decorate, decoration={zigzag, segment length=6, amplitude=3},line width=1,SCRed] (0+9,0) -- (3+9,0);
\draw[scalar] (0,-1.75-1.75/2) -- (3,-1.75-1.75/2);
\draw[scalar,-Latex] (6,-1.75-1.75/2) -- (6.2,-1.75-1.75/2);
\draw[scalar,-Latex] (3,-3.5) -- (3,-1.5);
\draw[scalar,dashed,gray] (1,-1-1.75-1.75/2) -- (1,1-1.75-1.75/2);
\draw[scalar,dashed,gray] (2,-1-1.75-1.75/2) -- (2,1-1.75-1.75/2);
\draw[scalar,dashed,gray] (4,-1-1.75-1.75/2) -- (4,1-1.75-1.75/2);
\draw[scalar,dashed,gray] (5,-1-1.75-1.75/2) -- (5,1-1.75-1.75/2);
\draw[wilson,EuclideanColor] (3,-1.75-1.75/2) -- (4,-1.75-1.75/2);
\draw[-, decorate, decoration={zigzag, segment length=6, amplitude=3},line width=1,SCRed] (4,-1.75-1.75/2) -- (6,-1.75-1.75/2);
\draw[fill=WCOrange,WCOrange] (3,-1.75-1.75/2) circle (.075);
\draw[fill=WCOrange,WCOrange] (4,-1.75-1.75/2) circle (.075);
\draw[wilson,-Latex] (7,0-1.75-1.75/2) -- (8,0-1.75-1.75/2);
\draw[scalar,dashed,gray] (.7+3+9,-1-1.75-1.75/2) -- (.7+3+9,1-1.75-1.75/2);
\draw[scalar,dashed,gray] (-.7+3+9,-1-1.75-1.75/2) -- (-.7+3+9,1-1.75-1.75/2);
\draw[scalar,dashed,gray] (2+3+9,-1-1.75-1.75/2) -- (2+3+9,1-1.75-1.75/2);
\draw[scalar,dashed,gray] (-2+3+9,-1-1.75-1.75/2) -- (-2+3+9,1-1.75-1.75/2);
\draw[scalar] (0+9,0-1.75-1.75/2) -- (3+9,0-1.75-1.75/2);
\draw[scalar,-Latex] (3.7+9,0-1.75-1.75/2) -- (6.2+9,0-1.75-1.75/2);
\draw[scalar,-Latex] (3+9,-1-1.75-1.75/2) -- (3+9,1-1.75-1.75/2);
\draw[wilson,EuclideanColor] (3+9,0-1.75-1.75/2) circle (.9);
\draw[-, decorate, decoration={zigzag, segment length=6, amplitude=3},line width=1,SCRed] (3+9,-1.75-1.75/2) -- (3.7+9,-1.75-1.75/2);
\end{tikzpicture}}

\newcommand{\AnalyticStructureThermal}{\begin{tikzpicture}[scale = 1, baseline={([yshift=-.5ex]current bounding box.center)}]
\draw[scalar,-Latex] (0,0) -- (6.2,0);
\draw[scalar,-Latex] (3,-2) -- (3,2);
\node[] at (5.7,1.7) {$w$};
\draw[scalar] (5.4,1.9) -- (5.4,1.45);
\draw[scalar] (5.4,1.45) -- (6,1.45);
\draw[wilson,EuclideanColor] (3,0) circle (1.5);
\draw[fill=WCOrange,WCOrange] (3,0) circle (.075);
\draw[-, decorate, decoration={zigzag, segment length=6, amplitude=3},line width=1,SCRed] (3,0) -- (4,0);
\draw[fill=WCOrange,WCOrange] (4,0) circle (.075);
\node[WCOrange] at (4,-.5) {$r$};
\draw[-, decorate, decoration={zigzag, segment length=6, amplitude=3},line width=1,SCRed] (5,0) -- (6,0);
\draw[fill=WCOrange,WCOrange] (5,0) circle (.075);
\node[WCOrange] at (5,-.5) {$1/r$};
\end{tikzpicture}}
\newcommand{\AnalyticStructureGeneral}{\begin{tikzpicture}[scale = 1, baseline={([yshift=-.5ex]current bounding box.center)}]
\draw[scalar,-Latex] (0,0) -- (6.2,0);
\draw[scalar,-Latex] (3,-2) -- (3,2);
\node[] at (5.7,1.7) {$w$};
\draw[scalar] (5.4,1.9) -- (5.4,1.45);
\draw[scalar] (5.4,1.45) -- (6,1.45);
\draw[wilson,EuclideanColor] (3,0) circle (1.5);
\draw[-, decorate, decoration={zigzag, segment length=6, amplitude=3},line width=1,SCRed] (2,0) -- (4,0);
\draw[-, decorate, decoration={zigzag, segment length=6, amplitude=3},line width=1,SCRed] (0,0) -- (1,0);
\draw[-, decorate, decoration={zigzag, segment length=6, amplitude=3},line width=1,SCRed] (5,0) -- (6,0);
\node[WCOrange] at (1,-.4) {$-1/r$};
\node[WCOrange] at (2,-.4) {$-r$};
\node[WCOrange] at (4,-.4) {$r$};
\node[WCOrange] at (5,-.4) {$1/r$};
\draw[fill=WCOrange,WCOrange] (1,0) circle (.075);
\draw[fill=WCOrange,WCOrange] (2,0) circle (.075);
\draw[fill=WCOrange,WCOrange] (3,0) circle (.075);
\draw[fill=WCOrange,WCOrange] (4,0) circle (.075);
\draw[fill=WCOrange,WCOrange] (5,0) circle (.075);
\end{tikzpicture}}

\begin{document} 

\thispagestyle{empty}

\vspace*{-.6in}
\begin{flushright}
    DESY-26-094\\
    YITP-SB-2026-13
\end{flushright}

\vspace{1cm}
{\large
\begin{center}
    {\Large \bf Analytic thermal bootstrap in momentum space: \\ \large From thermal OPE to QNMs
    }\\
\end{center}}

\vspace{0.5cm}

\begin{center}
    {Julien Barrat,$^{a,}\footnote{julien.barrat@desy.de}$ Deniz N. Bozkurt,$^{a,}\footnote{deniz.bozkurt@desy.de}$ Enrico Marchetto,$^{a,}\footnote{enrico.marchetto@desy.de}$ \\ Alessio Miscioscia,$^{b,}\footnote{alessio.miscioscia@stonybrook.edu}$ and Elli Pomoni$^{a,}\footnote{elli.pomoni@desy.de}$}\\[0.5cm] 
    { \small
    $^{a}$ Deutsches Elektronen-Synchrotron DESY, Notkestr. 85, 22607 Hamburg, Germany\\
    \small $^{b}$
    C. N. Yang Institute for Theoretical Physics, Stony Brook University, Stony Brook, NY 11794, USA
    }
\vspace{1cm}

   \bf Abstract
\end{center}

\begin{abstract}
\noindent We initiate a bootstrap program that relates ultraviolet  data, encoded in the thermal OPE, to infrared observables, namely, the low-frequency behavior and quasinormal modes.
Starting from KMS-symmetric completions of individual thermal OPE blocks, which play the role of thermal Polyakov blocks, we construct their
Fourier transform,  yielding an asymptotic expansion of retarded thermal correlators valid at any spatial momentum. 
We use these results to derive inversion formulae and connect thermal OPE data to the analytic structure of retarded correlators in the complex frequency plane. 
Under the assumption of meromorphicity, the inversion formulae express OPE coefficients in terms of the quasinormal-mode frequencies, leading to nontrivial sum rules, constraints on the quasinormal spectrum, and its asymptotics at large spatial momentum. 
We illustrate these results in free theories, two-dimensional CFTs, the large-$N$ limit and $\veps$--expansion of the $\mathrm{O}(N)$ model, and the $R$-current correlator of strongly coupled $\Nm = 4$ SYM at zero spatial momentum.
As a byproduct, we derive universal asymptotic formulae for thermal OPE coefficients of heavy operators, resolving their dependence on spin and extending previous results at zero spatial separation.
We test these formulae in the three-dimensional Ising CFT, finding good agreement between the resulting truncated correlators and Monte Carlo data.
\end{abstract}

\newpage

\setcounter{tocdepth}{3}

\noindent

\tableofcontents

\setcounter{page}{1}

\newpage
\section{Introduction and summary}

Thermal effects in quantum field theories (QFTs) are of fundamental importance from several perspectives.
Experimentally, physical systems are typically probed at finite temperature, making thermal observables directly relevant for comparison with laboratory measurements (see~\cite{Sachdev:2011fcc,pelissetto2002critical} and references therein).
In the context of holography, finite-temperature states of conformal field theories (CFTs) are naturally related to black-hole geometries in the dual Anti-de Sitter (AdS) spacetime~\cite{Maldacena:1997re,Witten:1998zw}.
More formally, studying a QFT at finite temperature amounts to placing the theory on the simplest non-trivial Euclidean thermal geometry, namely $S_\beta^1 \times \mathbb R^{d-1}$ where the thermal circle has circumference $\beta = 1/T$, with $T$ the temperature.
In the context of CFTs, this latter viewpoint has recently proved fruitful, leading to new insights into the structure of thermal observables and the properties of conformal data~\cite{Benjamin:2023qsc,Benjamin:2024kdg,Buric:2024kxo,Diatlyk:2024qpr,Anand:2025mfh,Buric:2025uqt,Komargodski:2026ain,Buric:2026pes,Parmentier:2026aqh}.

Recent progress on finite-temperature CFTs has produced both new analytic tools~\cite{Iliesiu:2018fao,Alday:2020eua,Marchetto:2023xap,Marchetto:2023fcw,Barrat:2025nvu,Niarchos:2025cdg,Barrat:2024aoa,Ghosh:2026xnp} and new results for strongly coupled theories, ranging from the O(N) models \cite{Iliesiu:2018zlz,David:2023uya,David:2024naf,Barrat:2025wbi} to holographic correlators \cite{Dodelson:2022yvn,Parisini:2023nbd,Dodelson:2023vrw,Ceplak:2024bja,Buric:2025anb,Buric:2025fye,Dodelson:2025jff,Dodelson:2025rng,Barrat:2025twb, Arnaudo:2026der,Buric:2026pes,Araya:2026shz}. 
Thermal correlation functions are highly constrained objects due to the interplay between their analytic structure, the local operator product expansion (OPE), and the Kubo--Martin--Schwinger (KMS) condition \cite{Kubo:1957mj,Martin:1959jp,El-Showk:2011yvt}. This translates into powerful constraints on the spectrum and the OPE coefficients, and suggests a bootstrap strategy: starting from a minimal set of input data -- either zero-temperature conformal data or a small amount of holographic information -- one attempts to reconstruct the full thermal two-point function.

The recent success of this program has so far been largely confined to the OPE regime, namely to regions of the parameter space where the short-distance expansion converges and dominates.
Nevertheless, many physical properties of finite-temperature correlators are encoded in their large-distance behavior, where the OPE is not expected to converge. In principle, an exact resummation of the OPE would give access to this information; in practice, however, this is often difficult to implement. This is due both to the approximations typically involved in bootstrap analyses -- which often rely on asymptotic control at large spin, large scaling dimension, or asymptotic thermal/conformal data as input -- and to the fact that many bootstrap results are numerical in nature. 

\bigskip

In this paper we take a first step towards understanding the relation between short-distance data and large-distance observables in CFTs at finite temperature and we initiate a bootstrap program that relates ultraviolet  data, encoded in the thermal OPE, to infrared observables, namely, the low-frequency behavior and quasinormal modes.
To do so, we start from the momentum-space OPE. In momentum space, the OPE is usually understood as an asymptotic expansion of the correlator at large Euclidean momenta, i.e.,  \(\omega,k \to \infty\) \cite{Manenti:2019wxs}. 
This limitation comes from the fact that in position space the OPE has a finite radius of convergence, and the Fourier transform of individual OPE blocks only carries short-distance information, which translates into large-momentum information. Nevertheless, previous work suggests that the OPE contains information beyond its radius of convergence, in particular at fixed spatial momentum, and even at \(k=0\) \cite{Caron-Huot:2009ypo,Sachdev:2011fcc}. Thus, one wonders if a globally defined, manifestly KMS-invariant decomposition of thermal correlators can be extracted from the OPE, leading to a globally defined decomposition in momentum space.

We answer this question in this work. We focus on thermal scalar two-point functions:
\begin{equation} \label{eq: twoptintro}
    g_{E}(\tau, x)=\langle \phi(\tau,x) \phi(0,0) \rangle_\beta \ ,
\end{equation}
and our starting point will be the KMS-symmetric inversion of individual OPE blocks appearing in the expansion of \eqref{eq: twoptintro}~\cite{Barrat:2025nvu}.
We show that this object admits a natural representation in terms of thermal images of the original block. As already emphasized in~\cite{Barrat:2025nvu}, this KMS-symmetric inversion automatically satisfies several bootstrap axioms of CFTs at finite temperature and plays a role analogous to that of the Polyakov blocks in the conformal bootstrap~\cite{Polyakov:1974gs} (or the crossing-symmetric inversion of single conformal blocks provided in \cite{Mazac:2018qmi}). Based on this analogy, in this paper we will refer to these objects as \emph{thermal Polyakov blocks}. This decomposition provides a controlled way to reorganize the short-distance expansion into building blocks that are compatible with the global structure of the thermal correlator.

We then study these thermal Polyakov blocks in detail. In particular, we construct their analytic continuation to Lorentzian signature and compute their Fourier transform explicitly. This leads to a representation of the retarded correlator of the form: 
\begin{equation}\label{eq:OPEmomentum}
g_R(\omega,k)
=
\sum_{\mathcal O}
\frac{\tilde a_{\mathcal O}}{\beta^{\Delta_{\mathcal O}}}
\left(k^2-\omega^2\right)^{\Delta_\phi-\Delta_{\mathcal O}/2-d/2}
C_J^{(\nu)}
\left(
\frac{i\omega}{\sqrt{k^2-\omega^2}}
\right)
+
\tilde{g}_{\rm arcs}(\omega,k) \, ,
\end{equation}
where \(\nu=(d-2)/2\) and $\Delta_{\phi}$ and $\Delta_{\mathcal{O}}$ are the conformal dimensions of the external operator $\phi$ and the operator $\mathcal{O}$ produced by the OPE, respectively. The coefficient \(\tilde a_{\mathcal O}\) contains the relevant zero-temperature OPE coefficient together with the thermal one-point function \(b_{\mathcal O}\) (see Equation~\eqref{eq: momopeco} for the precise definition). The term \(\tilde{g}_{\rm arcs}\) denotes an additional contribution that is exponentially suppressed at large frequency and encodes the information not captured by the Fourier transform of the thermal Polyakov blocks. 

An important result of \cite{Barrat:2025nvu} is that \(g_{\rm arcs}\) is not an independent dynamical input. Rather, it is uniquely fixed, up to a single constant associated with the clustering limit $x \gg \beta$ of the position-space correlator \( g_R(\tau,x)\), by the bootstrap axioms for the thermal correlators together with the dynamical data \(\tilde a_{\mathcal O}\).\footnote{A simple example of how the arc contribution is fixed is provided by the holographic correlators at \(x=0\) discussed in \cite{Barrat:2025twb}. In that case the arcs are strictly related to the so-called bouncing singularity~\cite{Fidkowski:2003nf,Parisini:2023nbd,Ceplak:2024bja}. In this work, we provide an explicit example at $x \neq 0$ in Section \ref{subsec:ApplicationLargeNONModel}.} Let us emphasize that the expansion in Equation~\eqref{eq:OPEmomentum} is derived for any value of $k$, but the sum over operators can (and in most cases is expected to) be an asymptotic series.  

As a by-product of this analysis, we also use the thermal Polyakov blocks for the individual OPE blocks to derive universal asymptotic formulae for thermal OPE coefficients. At $x=0$, our result reduces to the \textit{Tauberian approximation} proposed in \cite{Marchetto:2023xap}. However, away from this limit, the formula refines the asymptotics by resolving the spin dependence of the exchanged operators. In this form, it naturally matches the large-spin perturbation theory developed in \cite{Iliesiu:2018fao}.
We then apply these results to the 3$d$ $\mathrm O(N)$ model at large $N$ and the 3$d$ Ising CFT.
For the latter, using the numerical predictions of~\cite{Barrat:2025wbi}, we approximate the full Euclidean finite-temperature two-point functions of \(\sigma\) and \(\epsilon\), extending the analysis beyond the \(x=0\) limit. Since we work with a truncated OPE, this approximation cannot be expected to describe the correlators at arbitrarily large spatial separations. Nevertheless, we show that within the OPE regime the resulting correlators are in good agreement with Monte Carlo simulations.

Having obtained an explicit momentum-space decomposition valid at arbitrary spatial momentum \(k\), we can formulate an inversion problem for the retarded correlator, extracting the OPE coefficients directly from \(g_R(\omega,k)\). The derivation is analogous to other Euclidean\footnote{Here ``Euclidean'' does not refer to the spacetime signature, but rather to the fact that the inversion is performed directly on the correlator, independently of its analytic structure in complex frequency space.} inversion formulae \cite{Caron-Huot:2017vep,Simmons-Duffin:2017nub,Iliesiu:2018fao} and reads:
\begin{equation}\label{eq:abstractinversion}
    \tilde{a}_{\mathcal O}
=
\frac{1}{
h_J^{(\nu)}
}
\operatorname*{Res}_{\alpha=2s-d}
\int_{\Lambda}^{\infty}
\frac{\text{d}q}{q}\,
q^{-\alpha}
\int_{-1}^{1} \text{d}\eta\,
(1-\eta^2)^{\nu-\frac12}
C_J^{(\nu)}(\eta)\ 
g_R(q,\eta) \ ,
\end{equation}
where $q = \sqrt{k^2-\omega^2}$,  $\eta = i \omega/q$ and $s=\Delta_\phi-\Delta_\Om/2$. Overall coefficients and other details are given in Section \ref{sec:non-zeromom}.
Notably, we show explicitly that the contribution of $\tilde{g}_{\rm arcs}(\omega,k)$ in Equation~\eqref{eq:OPEmomentum} to the OPE coefficients vanishes identically. 

The real power of inversion formulae, however, usually comes from exploiting the analytic structure of the correlator. For thermal correlation functions this structure is, in general, still poorly understood, except in special cases such as holographic theories, where the analytic structure is controlled by the quasinormal modes (QNMs) of the dual black hole geometry; see for example \cite{Berti:2009kk,Faulkner:2009wj,Dodelson:2023vrw,Sachdev:2011fcc} and references therein. As a first step, we therefore consider the simplest possible analytic ansatz and assume that the retarded correlator \(g_R(\omega,k)\) is meromorphic in complex frequency space. This is the expected structure in holographic models dual to classical gravity, where the poles of the retarded correlator are the quasinormal modes. We show that the inversion formula constrains the high-energy asymptotic behavior of both the quasinormal mode spectrum and their residues. 

To explore the consequences of this assumption, we specialize to the case \(k=0\). In this limit the inversion formula takes a particularly simple form, relating the OPE coefficients \(\tilde a_\Delta = \sum_J \tilde a_{\mathcal O} \, C_J^{(\nu)}(1)\) to the QNM frequencies \(\omega_n\) and their residues \(r_n\): 
\begin{equation}
\tilde a_\Delta
= e^{ -i \pi  (s-d/2)} \operatorname*{Res}_{u=2s-d+1}
\left[
\frac{\pi e^{- i \pi u}}{\sin(\pi u)}
\,\mathcal Z(u)
\right],
\qquad
\mathcal Z(u)
=
\sum_n r_n^{\vphantom{u}} \omega_n^{-u} \, . \label{eq: invabstra}
\end{equation}
This is a direct relation between UV and IR observables, i.e., between the thermal OPE and the QNM spectrum.
To appreciate this form of the inversion more deeply, we focus on the spectral function \(\mathcal Z(u)\) of the QNM spectrum, which has several remarkable properties. First, it obeys a set of vanishing conditions:
\begin{equation}
\mathcal Z(m)=0,
\qquad m\in\mathbb Z^{<0}, \nonumber
\end{equation}
except when an operator of integer dimension contributes at the corresponding value, in which case \(\mathcal Z(m)\) is fixed by the associated OPE coefficient. This striking formula provides an infinite set of constraints on the QNM spectrum. Possible poles at negative integer values are related to conformal anomalies which we will explain in Section \ref{sec:UVIR}. Second, \(\mathcal Z(u)\) develops simple poles at the locations:
\begin{equation}
u=2\Delta_\phi-\Delta \ ,\nonumber
\end{equation}
with residues fixed by the OPE coefficients. Third, the values \(\mathcal Z(m)\) with $m \in \mathbb Z^{>0}$ are related to the coefficients of the small-frequency expansion of the retarded correlator.

After briefly discussing the case in which only finitely many QNMs contribute, we turn to the much richer situation of an infinite QNM spectrum. In this case, the pole conditions of \(\mathcal Z(u)\), together with mild assumptions on the large-\(n\) behavior of \(\omega_n\) and \(r_n\), determine the asymptotic distribution of the QNM frequencies and residues. The vanishing conditions at negative integers then provide additional constraints on the low-lying part of the spectrum. In this way, the inversion formula naturally gives rise to a bootstrap problem for QNMs, conceptually similar to the numerical bootstrap problem studied in \cite{Barrat:2024fwq} (and assumptions made in these steps are deeply connected to Tauberian theorems \cite{Mukhametzhanov:2018zja,Qiao:2017xif,Marchetto:2023xap}). The $3d$ large $N$ $\mathrm{O}(N)$ model and the \(k=0\) limit for the \(R\)-current correlator in \(\mathcal N=4\) SYM serve as benchmarks, given the exact knowledge of the retarded correlators.

We then turn to the full inversion formula at finite spatial momentum, namely the formula in Equation~\eqref{eq:abstractinversion}. In this case, the quasinormal frequencies and their residues become momentum-dependent functions, \(\omega_n(k)\) and \(r_n(k)\). The structure of the inversion formula suggests that the dominant contribution is controlled by the large-\(k\) regime. We therefore analyse this limit in detail and discuss how it is encoded in the momentum-space inversion formula. A more systematic attempt to solve this problem is left for future work. Besides the rich holographic case, we test the inversion formula in several other examples, where the retarded correlator is explicitly known and can be used as input: $4d$ free theory,
the $3d$ large $N$ \(\mathrm{O} (N)\) model, the $\mathrm{O} (N)$ model in the $\veps$-expansion and $2d$ CFTs.

Finally, we explore the possibility of using the inversion formula to invert deep IR retarded correlators to UV OPE coefficients.
First, we perform this analysis in 2d CFTs. In higher dimensions, although physical diffusive behaviors should not be expected for scalar thermal correlation functions, we consider inverting toy models of diffusion with the goal of understanding how much information about the UV the inversion formula can capture, and whether meaningful patterns emerge.
We show that, even if the inversion formula does not capture the correct OPE coefficients, the results retain a meaningful physical structure and resemble those of consistent theories.
A more detailed analysis is left for future studies.

\paragraph{Organization of the paper and future directions.}
The paper is organized as follows. 
In Section~\ref{secLThermalPolyakovBlocksAndAsymptoticOPEData}, we derive the thermal Polyakov block for the contribution of a single operator. As a first application, we analyse the thermal two-point functions of the energy and magnetization operators in the $3d$ Ising model, comparing our predictions with Monte Carlo simulations.
In Section~\ref{sec:OPEInMomentumSpace}, we Fourier transform the thermal Polyakov blocks and decompose the correlator in momentum space in terms of its OPE data. The thermal Polyakov blocks allow us to extend the regime of validity of the OPE in momentum space and we discuss various implications of this decomposition.
In Section~\ref{sec:UVIR}, we focus on the zero-momentum ($k=0$) inversion formula. We derive the general relation, discuss its physical implications, and illustrate its use with explicit examples.
In Section~\ref{sec:non-zeromom}, we derive the inversion formula at arbitrary spatial momentum, obtaining a relation that reconstructs the full retarded correlator as a function of both frequency and momentum. We then explore its main consequences, including the constraints on the large-momentum asymptotics of poles and residues. We present several explicit applications, including four-dimensional free scalar theory and the critical $\mathrm{O}(N)$ model both at large $N$ and in the $\veps$-expansion, as well as $2d$ CFTs. Finally, we initiate the study of correlators in the deep IR regime using the inversion formula by considering toy models in four dimensions.

\bigskip

Several interesting questions remain open and deserve further investigation:
\begin{itemize}
    \item[$\star$]
    Can the KMS-symmetric inversion of individual thermal conformal blocks and the momentum-space inversion formulae developed in this work be exploited in holographic CFTs to better understand the generic analytic structure of thermal correlators, both in position and momentum space? Our attempt to answer this question will be presented in a separate paper \cite{FutureHolo}. It would also be interesting to adapt the method beyond conformality, for example, in the context of the SYK model \cite{Sachdev:1992fk,Kitaev:2015SYK1,Kitaev:2015SYK2,Dodelson:2024atp,Dodelson:2026gak,Buric:2026qsp}.
    
    \item[$\star$] In this work we comment on a non-meromorphic extension of the inversion formula \eqref{eq: invabstra}, which includes the case of the critical $\mathrm O(N)$ model beyond the large $N$ limit. With non-planar corrections, branch cuts are expected to coexist with isolated quasinormal modes. It would be interesting to explore this direction in the future.

    \item[$\star$]
    The present analysis should be generalized to CFTs in the presence of line defects, along the lines of ~\cite{Barrat:2024aoa,Giombi:2026kdz}. In particular, it would be worth understanding whether analogous inversion formulae exist for defect thermal correlators and what constraints they impose on the corresponding defect CFT data.

    \item[$\star$]
    To what extent are the hydrodynamic properties of a thermal quantum field theory constrained by the analytic structure of its retarded Green’s functions in the complex frequency plane? In particular, can the inversion formulae developed in this paper be combined with general principles -- including causality, analyticity, and the KMS condition -- to derive new constraints on transport coefficients or on the spectrum of hydrodynamic and non-hydrodynamic modes? Some of these questions are currently being investigated and will appear in a forthcoming work based on the analysis of thermal current–current and stress-tensor two-point functions \cite{futureJJTT}.
\end{itemize}

\noindent \textbf{Note added.}{ \emph{During the final stages of this work, we became aware of the independent work~\cite{NewProjectCristo}, which overlaps with some of the results presented in this paper, in particular in Section~\ref{sec:UVIR}. We coordinated with the authors for a simultaneous submission to arXiv.}}

\section{Thermal Polyakov blocks and asymptotic OPE data}
\label{secLThermalPolyakovBlocksAndAsymptoticOPEData}

This section develops a bootstrap formalism based on the OPE, periodicity in position space, and the analytic structure of the two-point functions.
In the first part, we summarize the dispersion relation techniques developed in~\cite{Barrat:2025nvu} and extend them to non-zero spatial coordinates.
This framework relies on the analytic structure of the correlator and can be used to derive an OPE expansion in KMS-symmetric blocks, which we refer to as thermal Polyakov blocks.
We show in the second part that it can be used to extract novel expressions for asymptotic OPE coefficients of heavy operators, with resolution in scaling dimensions and spin.
We then present applications to the large-$N$ $\mathrm{O}(N)$ model and the three-dimensional critical Ising.

\begin{table}[t]
    \centering
    \footnotesize
    \setlength{\tabcolsep}{2pt}
    \renewcommand{\arraystretch}{1.20}

    \caption[Dictionary of thermal OPE coefficients]{
        Dictionary of thermal OPE coefficients. Here $\Delta$ and $J$
        denote the dimension and spin of $\mathcal O$, respectively.
    }
    \label{tab:notation}

    \begin{tabular}{@{}c@{\hspace{1.6em}}c@{\hspace{1.6em}}c@{}}
        \toprule
        \multicolumn{1}{c}{\textit{Kinematic parameters}} &
        \multicolumn{1}{c}{\textit{Position-space coefficients}} &
        \multicolumn{1}{c}{\textit{Momentum-space coefficients}} \\
        \cmidrule(lr){1-1}
        \cmidrule(lr){2-2}
        \cmidrule(lr){3-3}
        \addlinespace[0.25em]

        \begin{tabular}[t]{@{}r@{\;}l@{}}
            $s$ & $= \Delta_\phi-\frac{\Delta}{2}$ \\[6pt]
            $\nu$ & $= \frac{d}{2}-1$ \\[6pt]
            $\mathcal N_{\mathcal O}^{(\nu)}$
            & $= (-1)^{J/2}4^{\nu-s}
            \left(s+\frac{J}{2}\right)_{\nu+1-2s}$
        \end{tabular}
        &
        \begin{tabular}[t]{@{}r@{\;}l@{}}
            $a_{\mathcal O}$
            & $= \dfrac{J!}{2^J(\nu)_J}
            \dfrac{f_{\phi\phi\mathcal O}b_{\mathcal O}}
            {c_{\mathcal O}}$ \\[6pt]
            $a_{\mathcal O}^{2d}$
            & $= \dfrac{2}{1+\delta_{J,0}}\,
            \dfrac{1}{2^J}\,
            \dfrac{f_{\phi\phi\mathcal O}b_{\mathcal O}}
            {c_{\mathcal O}}$ \\[6pt]
            $a_\Delta$
            & $= \sum_{J}^{\Delta\,\mathrm{fixed}}
            a_{\mathcal O}C_J^{(\nu)}(1)$
        \end{tabular}
        &
        \begin{tabular}[t]{@{}r@{\;}l@{}}
            $\tilde a_{\mathcal O}$
            & $= \Gamma(\nu)\,
            \mathcal N_{\mathcal O}^{(\nu)}a_{\mathcal O}$ \\[6pt]
            $\tilde a_{\mathcal O}^{2d}$
            & $= 4\pi\,
            \mathcal N_{\mathcal O}^{(0)}a_{\mathcal O}^{2d}$ \\[6pt]
            $\tilde a_\Delta$
            & $= \sum_{J}^{\Delta\,\mathrm{fixed}}
            \tilde a_{\mathcal O}C_J^{(\nu)}(1)$
        \end{tabular}
        \\
        \bottomrule
    \end{tabular}
\end{table}

\subsection{Dispersion relation and OPE}
\label{subsec:DispersionRelationAndOPE}

Our starting point is a review of Euclidean two-point correlators and of the generalized method of images proposed in~\cite{Barrat:2025nvu}.

\paragraph{Euclidean two-point function and bootstrap axioms.}
Our main observable of interest is the Euclidean correlator at finite temperature of two identical operators $\phi$, for which we define the shorthand notation:
\begin{equation}
    g(z,\zb)
    =
    \vev{\phi (\tau,x) \phi(0,0)}_\beta\,.
    \label{eq:g_Definition}
\end{equation}
On the left-hand side we have used the dimensionless variables:
\begin{equation}
    z = \frac{\tau + i x}{\beta}\,, \qquad \zb = \frac{\tau - i x}{\beta}\,,
    \label{eq:zzb_Definition}
\end{equation}
while $x=|\vec{x}|$ refers to the norm of the vector in $\mathbb{R}^{d-1}$.

\begin{figure}
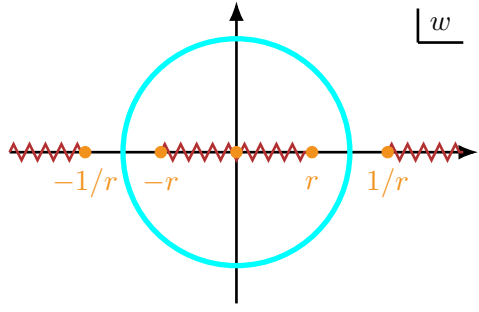

    \centering
    \AnalyticStructureGeneral
    \caption{Analytic structure of the correlator $g(z,\zb)$ in the $w$-plane.
    The Euclidean correlator is located at $|w|=1$, here represented by a blue circle.}
    \label{fig:AnalyticStructureGeneral}
\end{figure}

The Euclidean correlator satisfies a number of conditions, which we refer to as \textit{thermal bootstrap axioms} in this work.
We list their essential ingredients below:
\begin{itemize}
    \item[$\star$] \textbf{OPE consistency}.
    Any Euclidean correlator admits an OPE that takes the form:
    \begin{equation}
        g(z,\zb)
        =
        \frac{1}{\beta^{2\Delta_\phi}}
        \sum_\Om a_\Om\, f_\Om (z,\zb)\,.
        \label{eq:OPE}
    \end{equation}
    Here the (dimensionless) functions $f_\Om (z,\zb)$ are known as thermal blocks, associated with the exchanged operator $\Om \in \phi \times \phi$.
    They are fixed by the residual symmetry to take the form:
    \begin{equation}
        f_\Om (z,\zb)
        =
        \frac{1}{(z\zb)^s}\, C_{J_\Om}^{(\nu)} \left( \frac{z + \zb}{2\sqrt{z\zb}} \right)\,, \qquad s = \Delta_\phi - \frac{\Delta_\Om}{2}\,,
\label{eq:ThermalBlocks}
    \end{equation}
    where $\nu = (d-2)/2$ and $C_{J_\Om}^{(\nu)}$ are the Gegenbauer polynomials.
    The quantum numbers of the exchanged operator are its scaling dimension $\Delta_\Om$ and its spin $J_\Om$, while $\Delta_\phi$ refers to the scaling dimension of the external operator.
    The OPE coefficients $a_\Om$ are defined as:
    \begin{equation}
        a_\Om
        =
        \frac{J_\Om!}{2^{J_\Om} (\nu)_{J_\Om}} \frac{f_{\phi\phi\Om} b_\Om}{c_\Om}\,,
        \label{eq:aO_Definition}
    \end{equation}
    where $f_{\phi \phi \Om}$ is the three-point function OPE coefficient at zero temperature, $c_\Om$ is the normalization of the operator $\Om$, and $b_\Om$ is the thermal one-point function coefficient:
    \begin{equation}
        \vev{\Om^{\mu_1 \ldots \mu_{J_\Om}}}_\beta
        =
        \frac{b_\Om}{\beta^{\Delta_\Om}} (e^{\mu_1} \ldots e^{\mu_{J_\Om}} - \text{traces})\,.
        \label{eq:bO_Definition}
    \end{equation}
    Note that the zero-temperature limit simply keeps the identity contribution to the block expansion:
    \begin{equation}
        \lim\limits_{\beta \to \infty} g(z,\zb)
        =
        \frac{1}{(\tau^2 + x^2)^{\Delta_\phi}}\,.
        \label{eq:ZeroTLimit}
    \end{equation}
    \item[$\star$] \textbf{KMS condition}.
    The Euclidean correlator satisfies the KMS condition, namely:
    \begin{equation}
        g(z,\zb)
        =
        g(1 - z,1 - \zb)\,.
        \label{eq:KMS}
    \end{equation}
    This implies that the OPE~\eqref{eq:OPE} admits an expansion in the regime $z,\zb \sim 1$:
    \begin{equation}
        g(z,\zb)
        =
        \frac{1}{\beta^{2\Delta_\phi}}
        \sum_\Om a_\Om\, f_\Om (1 - z,1 - \zb)\,.
        \label{eq:OPE_tChannel}
    \end{equation}
    We call this expansion the \textit{$t$-channel}, while we refer to the regime $z,\zb \sim 0$ as \textit{$s$-channel}, in analogy to zero-temperature four-point functions.
    Note that the thermal blocks are not KMS-symmetric individually:
    \begin{equation}
        f_\Om (z,\zb)
        \neq
        f_\Om (1 - z,1 - \zb)\,.
        \label{eq:BlocksAreNotKMSSymmetric}
    \end{equation}
    \item[$\star$] \textbf{Analytic structure}.
    The analytic structure of the Euclidean correlator in the complex $w$-plane was studied in~\cite{Iliesiu:2018fao,Barrat:2025nvu} and takes the form presented in Figure~\ref{fig:AnalyticStructureGeneral}.
    Here we use the variables:
    \begin{equation}
        r^2 = z \zb\,, \qquad
        w^2 = \frac{z}{\zb} \,.
        \label{eq:rw_Definition}
    \end{equation}
    \item[$\star$] \textbf{Regge boundedness}.
    Even though it has not been rigorously proven, a version of Regge boundedness is expected to hold for thermal correlation functions~\cite{Iliesiu:2018fao}.
    Concretely, the assumption that we make throughout this paper is that the function $g(z,\zb)$ does not grow faster than $w^{J_\star}$ at large $w$ with $J_\star$ a model-dependent constant.
    In the presence of marginal deformations, the value of $J_\star$ can vary over the conformal manifold, namely, it may depend on the marginal couplings. Furthermore, Regge boundedness is expected to be a non-perturbative property of thermal correlation functions, although it does not need to hold at any finite order in perturbation theory.
    
    \item[$\star$] \textbf{Clustering condition}.
    Thermal correlators are expected to cluster at large spatial distance, i.e.,
    \begin{equation}
        \lim\limits_{x \to \infty} g(z,\zb)
        =
        \vev{\phi}_\beta^2\,.
        \label{eq:Clustering_LargeSpatialDistance}
    \end{equation}
    Similarly, we can take the large-separation limit along the real time direction:
    \begin{equation}
        \lim\limits_{\tau \to i\infty} g(z,\zb)
        =
        \vev{\phi}_\beta^2\,.
        \label{eq:Clustering_LargeRealTime}
    \end{equation}
    Clustering is also expected to be a non-perturbative property of thermal correlation functions, but it does not need to hold at any finite order in perturbation theory as, for example, in the $\veps$-expansion of the $\mathrm O(N)$ model~\cite{Barrat:2025nvu}.
\end{itemize}

\paragraph{Generalized method of images.}
In~\cite{Barrat:2025nvu}, it was argued that the correlator can be expressed as a manifestly KMS-symmetric integral of its discontinuity.
This is useful, as the discontinuity of the correlator is often a simpler object than the full correlator.
In Appendix~\ref{app:DerivationOfTheGeneralizedMethodOfImages} we provide a proof of the generalized method of images, while here we summarize the main idea and its consequences.

The starting point is to rewrite the Euclidean correlator as a Cauchy integral:
\begin{equation}
    g(z,\zb)
    =
    \oint_{\Cm} \frac{\mathrm{d} w'}{2 \pi i} \frac{g(rw',r/w')}{w' -w}\,.
    \label{eq:Cauchy_Step1}
\end{equation}
Here $\Cm$ is a contour around the point $w$ (at fixed $r$) such that $g(z,\zb)$ is holomorphic in the associated region.
We now restrict ourselves to the strip $-\beta/2 \leq \text{Re} (\tau) < \beta/2$ and deform the contours around the cut located at $\text{Re} (\tau) = 0$.
Using periodicity, we can rewrite the full Euclidean correlator as a sum of images, i.e.,
\begin{equation}
    g(z,\zb)
    =
    \sum_{m=-\infty}^\infty
    \oint_{\mathcal{C}_0} \frac{\mathrm{d} w'}{2 \pi i} \frac{g(r_m w',r_m/w')}{w' - w_m}\,,
    \label{eq:Cauchy_SumOfImages}
\end{equation}
where we used the shorthand variables:
\begin{equation}
    r_m^2 = (z+m)(\zb+m)\,, \qquad
    w_m^2 = \frac{z+m}{\zb+m}\,.
    \label{eq:rmwm_Definition}
\end{equation}
We now express the contour integral as a real integral by introducing the discontinuity of the correlator.
In this process we drop the arcs, allowing us to write the generalized method of images:\footnote{Note that in~\cite{Barrat:2025nvu} $g_{\text{dr}} (z,\zb)$ was used to refer to the dispersion relation without the generalized method of images (see also~\cite{Alday:2020eua}).
Since in this paper we show that the generalized method of images arises naturally by considering the relation between the analytic structure and the KMS invariance, we extend the definition to include the sum of images (see Appendix \ref{app:DerivationOfTheGeneralizedMethodOfImages} for details).}
\begin{align}
    g_\text{dr}(z,\zb)
    &=
    \frac{1}{2}
    \sum_{m=-\infty}^\infty
    \int_{0}^{r_m} \frac{\text{d} w'}{2 \pi i}
    \frac{w_m^2 (1-w'^4)}{w' (w'-w_m)(w'+w_m) (1-w_m^2 w'^2)}
    \operatorname{Disc} \left[ g(r_m w',r_m/w') \right]\,,
    \label{eq:GMI}
\end{align}
with
\begin{equation}
    \operatorname{Disc} g(z,\zb) = g(z+ i 0,\zb) - g(z-i 0,\zb).
    \label{eq:Disc_Definition}
\end{equation}
The full correlator is then given by
\begin{equation}
    g(z,\zb)
    =
    g_\text{dr}(z,\zb) + g_\text{arcs} (z,\zb)\,.
    \label{eq:GMI_FullCorrelator}
\end{equation}
Since $g_\text{dr}(z,\zb)$ is KMS-symmetric on its own, note that $g_\text{arcs} (z,\zb)$ should also be KMS-symmetric.

\paragraph{Manifestly periodic OPE in position space.}
We now go one step further and insert the $t$-channel OPE~\eqref{eq:OPE_tChannel} in~\eqref{eq:GMI}.
In order to make progress, we commute the sum of operators with the integral and the sum of images.
This interchange is not generally justified and leads to a redefinition of the function $g_\text{arcs}(z,\zb)$.
The generalized method of images can then be expressed as:
\begin{equation}
    g_\text{dr}(z,\zb)
    =
    \frac{1}{\beta^{2\Delta_\phi}}
    \sum_\Om a_\Om\, F_\Om(z,\zb)\,,
    \label{eq:GMI_PolyakovBlocks}
\end{equation}
where the functions $F_\Om(z,\zb)$ are thermal Polyakov blocks, i.e., KMS-symmetric blocks, associated with the exchanged operators $\Om$.
They are defined as:
\begin{align}
    F_\Om(z,\zb)
    &=
    \frac{1}{2}
    \sum_{m=-\infty}^\infty
    \int_{0}^{r_m} \frac{\mathrm{d} w'}{2 \pi i}
    \frac{w_m^2 (1-w'^4)}{w' (w'-w_m)(w'+w_m) (1-w_m^2 w'^2)} \notag \\
    &\phantom{=\ } \times
    \operatorname{Disc} \left[ f_\Om (1 - r_m w',1 - r_m/w') \right]\,.
    \label{eq:GMI_PolyakovBlocks_Step1}
\end{align}
In fact, we argue in Appendix~\ref{app:DerivationOfTheGeneralizedMethodOfImages} that this expression can be simplified to:
\begin{equation}
    F_\Om (z,\zb)
    =
    \sum_{m=-\infty}^\infty
    f_\Om \left( r_m w_m, \frac{r_m}{w_m} \right)\,,
    \label{eq:GMI_PolyakovBlocks_Definition}
\end{equation}
i.e., the thermal Polyakov blocks $F_\Om(z,\zb)$ consist of a \textit{sum of images of the thermal blocks $f_\Om(z,\zb)$}.

It is important to notice that the (infinite) sum over operators in $g_\text{dr}(z,\zb)$ is not guaranteed to preserve the analytic structure of the correlator.
This occurs, for instance, in holographic correlators~\cite{Barrat:2025twb}.
In this case the arc contribution must compensate for this effect and restore the analytic structure of the full correlator $g(z,\zb)$.

In the following we study some of the essential properties of the thermal Polyakov blocks.

\paragraph{Analytic continuation of thermal Polyakov blocks.}
The sum of images in~\eqref{eq:GMI_PolyakovBlocks_Definition} converges only for certain values of $s$.
In order to make use of this expansion we need to regularize the blocks and take their analytic continuation to our regimes of interest.

For this purpose, it is useful to consider the Epstein--Hurwitz zeta function:
\begin{equation}
    \zeta_{\mathrm{EH}}(s,x,\tau)
    =
    \sum_{m = -\infty}^\infty
    \frac{1}{\left(x^2+(\tau+m\beta)^2\right)^s}\,.
    \label{eq:EHZeta_Definition}
\end{equation}
This corresponds to the thermal two-point function of generalized free fields (GFF) theory for $s = \Delta_\phi$.
This function appears naturally in the thermal Polyakov blocks and can be used to illustrate (and solve) the problem of convergence.
Indeed, Equation~\eqref{eq:EHZeta_Definition} converges only for $\operatorname{Re}(s) > 1/2$.
Beyond this regime, its analytic continuation is given by:
\begin{equation}
    \zeta_{\mathrm{EH}}(s,x,\tau)
    =
    \frac{\sqrt{\pi}\,\Gamma\!\left(s-\frac{1}{2}\right)}
    {\Gamma(s)}
    \,\frac{x^{1-2s}}{\beta}
    +
    \frac{4\pi^s}{\Gamma(s)}
    \left(\beta x\right)^{1/2-s}
    \sum_{n=1}^{\infty}
    n^{\,s-\frac12}
    K_{\,s-\frac12}\!\left(\frac{2\pi n x}{\beta}\right)
    \cos \left(\frac{2\pi n \tau}{\beta} \right)\,.
    \label{eq:EHZeta_AnalyticContinuation}
\end{equation}
This definition provides a meromorphic continuation of $\zeta_{\mathrm{EH}}(s,x,\tau)$ in the domain $\operatorname{Re}(s) < 1/2$.
From there we observe that the analytic continuation of individual blocks is given by
\begin{equation}
    F_\Om (z,\zb)
    =
    \beta^{2s}
    \sum_{k=0}^{\lfloor J_\Om/2 \rfloor}
    \sum_{r=0}^{\frac{J_\Om-2k}{2}}
    (-1)^{k+r}
    \frac{
    2^{J_\Om-2k}\Gamma(J_\Om-k+\nu)
    }{
    \Gamma(\nu)\,k!\,(J_\Om-2k)!
    }
    \binom{\frac{J_\Om-2k}{2}}{r}
    x^{2r}
    \,
    \zeta_{\mathrm{EH}}
    \!\left(
    s+r,
    x,\tau
    \right)\,,
    \label{eq:GMI_maindef}
\end{equation}
where $s$ here is defined in~\eqref{eq:ThermalBlocks}. 

\paragraph{Reduction to $x=0$.}
As a consistency check, we now consider Equation~\eqref{eq:GMI_maindef} in the limit $x \to 0$, which should reduce to Equation~(2.27) in~\cite{Barrat:2025twb}.
One can first easily verify that:
\begin{equation}
    \zeta_{\mathrm{EH}}(s,0,\tau)
    =
    \frac{1}{\beta^{2s}}
    \sum_{n=1}^{\infty}
    \frac{
    2\,\pi^{\,2s-\frac12}
    \,n^{\,2s-1}
    \,\Gamma\!\left(\frac12-s\right)
    }{
    \Gamma(s)
    }
    \cos \left( \frac{2\pi n\tau}{\beta} \right)\,,
    \label{eq:EHZeta_Zerox}
\end{equation}
which is precisely the Fourier expansion of the function
\begin{equation}
    F_\Om (z,z)
    =
    C_{J_\Om}^{(\nu)} (1)
    \left[
    \zeta_H \left( 2s,\frac{\tau}{\beta} \right)
    +
    \zeta_H \left( 2s,1-\frac{\tau}{\beta} \right)
    \right]\,,
    \label{eq:GMI_ReductionToZerox}
\end{equation}
after analytic continuation.
Typically we absorb the constant $C_{J_\Om}^{(\nu)} (1)$ in a new definition of the OPE coefficient (see~\eqref{eq:aDelta_Definition}).
Moreover, we see that the Gegenbauer polynomial reduces to a constant, $C_{J_\Om}^{(\nu)}(1)$, which exactly matches the result of the $x=0$ dispersion relation \cite{Barrat:2025nvu}.

\subsection{Asymptotic OPE coefficients for heavy operators}
\label{subsec:AsymptoticOPECoefficientsForHeavyOperators}

\begin{figure}
    \centering
    \includegraphics[width=0.75\linewidth]{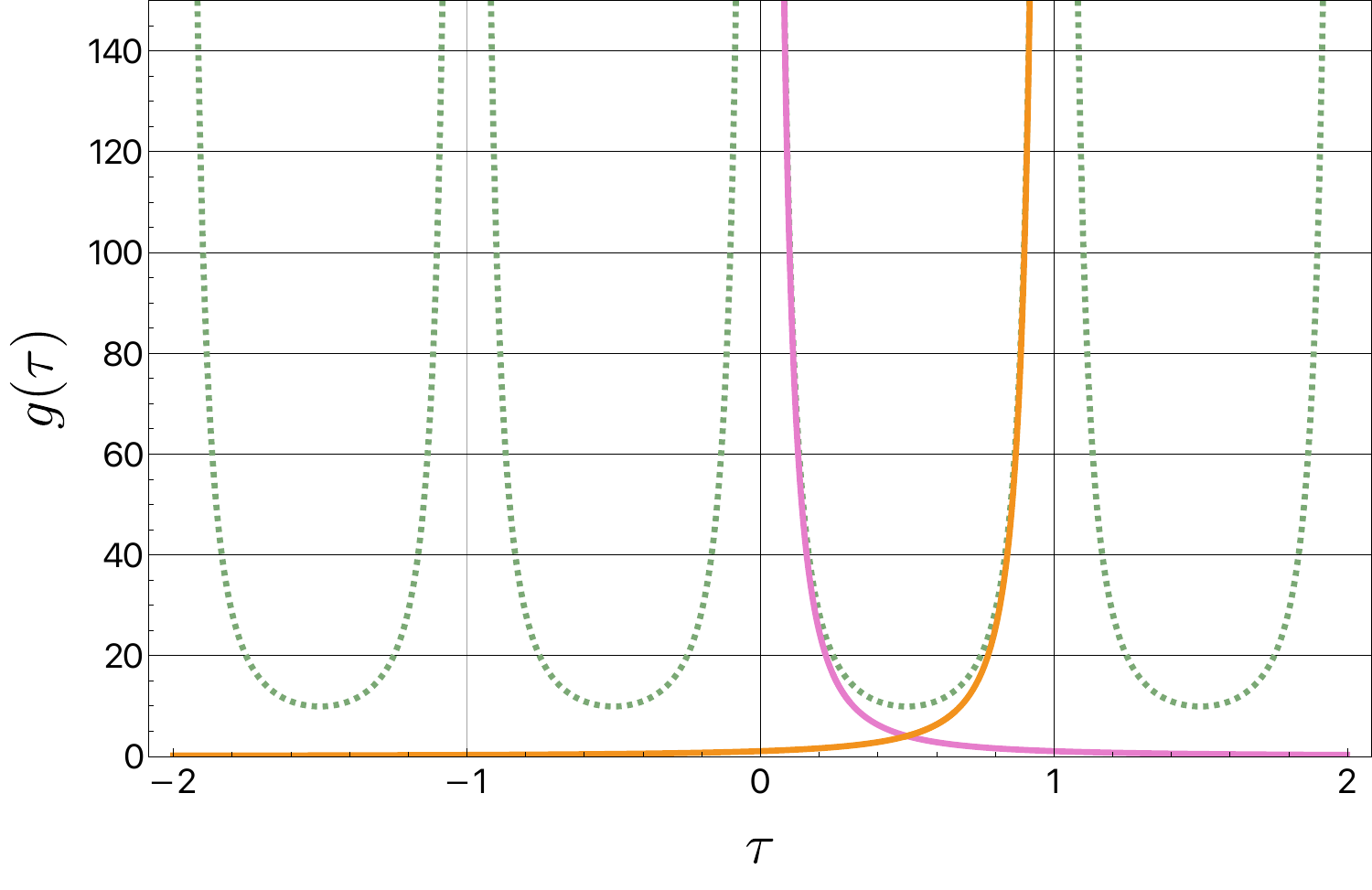}
    \caption{This plot illustrates the contribution of heavy operators for a given thermal Polyakov block $F_\Om (z,\zb)$, here the one associated with the identity operator $\Om = \mathds{1}$.
    We set $x=0$ for clarity.
    The pink curve shows the behavior of the thermal block in the $s$-channel, i.e., around $z,\zb \sim 0$ or $\tau \sim 0$.
    The green curve displays the thermal Polyakov block defined in~\eqref{eq:GMI_PolyakovBlocks_Definition}.
    We see that an infinite number of heavy operators is needed in order to generate a $t$-channel expansion, i.e., around $z,\zb \sim 1$ or $\tau \sim \beta$ (the orange curve), that agrees with the $s$-channel one.}
    \label{fig:ExampleHeavyOps}
\end{figure}

As a byproduct of the generalized method of images~\eqref{eq:GMI_PolyakovBlocks_Definition}, we are able to derive asymptotics for heavy operators as functions of the OPE coefficients associated with light operators.
This can be understood as follows: each periodic block $F_\Om (z,\zb)$ takes the ``light'' contribution $a_\Om$ and generates ``heavy'' double-twist operators to ensure that the $t$-channel has the same OPE as the $s$-channel.
A plot comparing the thermal block $f_\mathds{1}(\tau,\tau)$ and the thermal Polyakov block $F_\mathds{1}(\tau,\tau)$ is provided in Figure~\ref{fig:ExampleHeavyOps}.
This extends previous results in this direction obtained in~\cite{Iliesiu:2018fao,Marchetto:2023xap,Barrat:2025nvu} and already applied successfully to concrete setups in~\cite{Iliesiu:2018zlz,Barrat:2025wbi,Barrat:2025twb}.

\paragraph{Asymptotic OPE coefficients.}
The key step is to realize that a thermal Polyakov block~\eqref{eq:GMI_PolyakovBlocks_Definition} admits another natural analytic continuation, which can be expressed using the $z, \zb$ variables as:
\begin{equation}
    F_\Om (z,\zb)
    =
    f_\Om (z,\zb)
    +
    \sum_{k=0}^{J_\Om} \frac{(\nu)_k (\nu)_{J_\Om - k}}{k! (J_\Om - k)!} S_{J_\Om/2 -s-k,-J_\Om/2-s+k} (z,\zb)\,,
    \label{eq:RewriteGMIAsBlockPlusDT}
\end{equation}
where $s=\Delta_\phi-\Delta_{\mathcal{O}}/2$ and $f_\Om (z,\zb)$ is the thermal OPE block given in \eqref{eq:ThermalBlocks}, while we also defined:
\begin{equation}
    S_{s_1, s_2} (z,\zb)
    =
    \sum_{n_1 = 0}^\infty \sum_{n_2 = 0}^\infty
    \binom{s_1}{n_1} \binom{s_2}{n_2} (1+ (-1)^{n_1 + n_2}) \zeta_{n_1 + n_2 - s_1 - s_2} z^{n_1} \zb^{n_2}\,.
    \label{eq:S_Definition}
\end{equation}
Here $J_\Om$ is the spin of the exchanged operator $\Om$ in the OPE.
In words, Equation~\eqref{eq:RewriteGMIAsBlockPlusDT} shows that each periodic block associated with an operator $\Om$ can be rewritten as \textit{the thermal block for $\Om$ itself plus a fully determined expansion in integer powers of $z, \zb$}.
The latter contributes to the OPE in the form of classical double-twist operators, i.e., operators of dimensions:
\begin{equation}
    \Delta_{[\phi\phi]_{n,J}} = 2\Delta_\phi + 2n + J\,.
    \label{eq:DoubleTwist_Dimensions}
\end{equation}
The periodic block of a given operator $\Om$ therefore generates infinite families of double-twist operators which can be arbitrarily heavy.
In order to quantify these contributions, we need to compare~\eqref{eq:RewriteGMIAsBlockPlusDT} to an expansion in blocks of the form:
\begin{equation}
    F_\Om (z,\zb)
    =
    f_\Om (z,\zb)
    +
    \sum_{n,J} \hat{a}^{(\Om)}_{[\phi\phi]_{n,J}} f^{\phantom{(\Om)}}_{[\phi\phi]_{n,J}} (z,\zb)\,.
    \label{eq:GMIBlock_DecompositionIntoBlockPlusDT}
\end{equation}
One can expand~\eqref{eq:RewriteGMIAsBlockPlusDT} and~\eqref{eq:GMIBlock_DecompositionIntoBlockPlusDT} around $z,\zb \sim 0$ to obtain an infinite system of equations for the asymptotic coefficients $\hat{a}^{(\Om)}_{[\phi\phi]_{n,J}}$.
The system can be solved recursively and results in a recurrence formula for the OPE coefficients of the heavy operators associated with a light operator $\Om$:
\begin{align}
    \hat{a}^{(\Om)}_{[\phi\phi]_{n,J}}
    &=
    \frac{1}{d_{J/2} (J)} \biggl(
    \sum_{k=0}^{J_\Om} \frac{(\nu)_k (\nu)_{J_\Om - k}}{k! (J_\Om - k)!}
    c_{J+n,n} (J_\Om/2-s-k,-J_\Om/2-s+k) \notag \\
    &\phantom{=\ }
    -
    \sum_{m=0}^{n-1} \hat{a}^{(\Om)}_{[\phi\phi]_{m,2n+J-2m}}\, d_{J/2} (2n+J-2m)
    \biggr)\,,
    \label{eq:Tauberian_RecursionRelation}
\end{align}
with the starting value:
\begin{equation}
    \hat{a}^{(\Om)}_{[\phi\phi]_{0,J}}
    =
    \frac{1}{d_{J/2} (J)}
    \sum_{k=0}^{J_\Om} \frac{(\nu)_k (\nu)_{J_\Om - k}}{k! (J_\Om - k)!}
    c_{J,0} (J_\Om/2-s-k,-J_\Om/2-s+k)\,.
    \label{eq:Tauberian_StartingValue}
\end{equation}
We have also defined the following shorthand functions to streamline the formula~\eqref{eq:Tauberian_RecursionRelation}:
\begin{align}
    c_{n_1, n_2} (s_1, s_2) &=
    \left( 1 + (-1)^{n_1 + n_2} \right)
    \binom{s_1}{n_1} \binom{s_2}{n_2}
    \zeta_{n_1 + n_2 - s_1 - s_2}\,, \label{eq:c_Definition} \\
    d_m (J) &=
    \frac{(\nu)_{J/2+m} (\nu)_{J/2-m}}{(1)_{J/2+m} (1)_{J/2-m}}\,.
    \label{eq:d_Definition}
\end{align}
In spite of its intricate appearance, the recurrence relation~\eqref{eq:Tauberian_RecursionRelation} can be solved very efficiently in practical models, as we illustrate with the $3d$ Ising model in Section~\ref{subsec:Application3dIsingModel}.

For the sake of clarity, let us summarize the notation of the formula above: $s=\Delta_\phi-\Delta_\Om/2$ contains the dependence on the scaling dimension of the ``light'' operator $\Om$, with spin $J_\Om$, while $n$ and $J$ label the quantum numbers of the ``heavy'' double-twist operator $[\phi\phi]_{n,J}$.
The hat on the OPE coefficient in~\eqref{eq:Tauberian_RecursionRelation} is used to emphasize that these operators should not be considered the actual operators of the theory.
Generically, strongly-coupled models indeed do not contain exact classical double-twist operators in their spectrum.\footnote{Notable examples in which exactly classical double-twist operators appear are generalized free fields and holographic models.
In these cases, the asymptotic coefficients $\hat{a}^{(\Om)}_{[\phi\phi]_{0,J}}$ coincide with the real OPE coefficients $a^{(\Om)}_{[\phi\phi]_{0,J}}$ (up to potential contribution from the arcs).
See~\cite{Barrat:2025nvu,Barrat:2025twb} for additional details.}
This formula is nevertheless useful because it provides a good estimate of the correlator given a handful of light operators, up to the contribution of the arcs which has to be determined through other means.
In many cases the arcs do not contribute, as we illustrate for the example of the $3d$ Ising model in Section~\ref{subsec:Application3dIsingModel}.

\paragraph{Comparison with previous work.}
Note that~\eqref{eq:Tauberian_RecursionRelation} can be understood as the periodic version of Equation~(6.18) in~\cite{Iliesiu:2018fao}, here given using our conventions:
\begin{align}
    \hat{a}^{(\Om)}_{[\phi\phi]_{n,J}}
    &=
    a_\Om
    \frac{(1+(-1)^{J}) \Gamma(J+1)}{J \, \Gamma(J+\nu)}
    \sum_{r=0}^n \sum_{k=0}^{J_\Om}
    (-1)^n
    \binom{J_\Om/2 - s - k}{n - r} \notag \\
    &\phantom{=\ } \times
    \frac{(J + 2r) (J)_r (\nu+1-r)_r \Gamma(J_\Om -k+\nu) \Gamma(k + \nu) \Gamma(n + J + r + s - J_\Om/2 - k)}{r! (J+\nu+1)_r \Gamma(J_\Om -k+1) \Gamma(k + 1) \Gamma(s + J_\Om/2 - k)\Gamma(n + J +r+1)}\,.
    \label{eq:AsymptoticCoefficients_DSD}
\end{align}
The main difference is the appearance of zeta functions in~\eqref{eq:Tauberian_RecursionRelation}.
In fact the two formulas agree very well in the large-dimension/spin limit, in which the zeta functions approach $1$ rapidly.
A comparison between the methods in the case of the $3d$ Ising model is provided in Table~\ref{tab:LIFvsGMI}.

\paragraph{Reduction to $x=0$.}
Finally, one may be interested in considering the limit $x=0$.
In this case the recurrence relation~\eqref{eq:Tauberian_RecursionRelation} can be solved and we recover the result for double-twist operators which was already derived in~\cite{Barrat:2025nvu}, namely:
\begin{equation}
    \hat{a}_{n}
    =
    a_\Delta
    \frac{2 \zeta_{2\Delta_\phi - \Delta + 2n} \Gamma(\Delta-2\Delta_\phi + 1)}{(2n)!\, \Gamma(\Delta-2\Delta_\phi+1-2n)}\,.
    \label{eq:Tauberian_Zerox}
\end{equation}
Here we used the definition of the degenerate OPE coefficients
\begin{equation}
    a_\Delta
    =
    \sum_{\Delta_\Om = \Delta} C_{J_\Om}^{(\nu)} (1)\, a_\Om\,.
    \label{eq:aDelta_Definition}
\end{equation}
When the spatial dependence is suppressed, the correlator loses sensitivity to the spin quantum numbers, and the OPE data therefore enter only through the weighted sum denoted by $a_\Delta$. In the following section, we consider momentum-space correlators. Upon setting the spatial momentum to zero, $k=0$, we obtain an analogous weighted sum over OPE coefficients. We emphasize, however, that this quantity is distinct from the weighted sum defined here.

\subsection{Application: $\mathrm{O}(N)$ model at large $N$ }
\label{subsec:ApplicationLargeNONModel}

It is instructive to test the decomposition into thermal Polyakov blocks in the large $N$ $\mathrm{O}(N)$ model.
Its useful features make this theory an interesting playground for thermal physics \cite{Iliesiu:2018fao, Marchetto:2023fcw, Barrat:2024fwq, David:2024N1, David:2024N2, David:2025N3, Mauro:2026zus}. In particular, we focus on the thermal Polyakov decomposition of the two-point function:
\begin{equation} \label{eq: twoptlargeN}
    g(\tau,x)=\langle \phi(\tau,x) \phi(0,0) \rangle_\beta \ ,
\end{equation}
where $\phi$ is a fundamental scalar of the theory; we will implicitly assume that the two scalars are identical.
We also set $\beta = 1$ without loss of generality.

The OPE spectrum is well known: in the large $N$ limit we are studying, only two operator trajectories contribute and they can be identified with double-twist operators and with powers of the Hubbard-Stratonovich auxiliary field:
\begin{align*}
    [\phi \phi]_{\ell} & : \quad \Delta=2\Delta_\phi+\ell \ , \quad   J=\ell \ , \\
    [\sigma^m] & : \quad \Delta=2m  \ , \quad  J=0 \ . 
\end{align*}
It is important to notice that only the $[\sigma^m]$ trajectory contributes to the discontinuity of the correlator in the sense explained in Equation~\eqref{eq:GMI}. Hence, we can specialize the thermal Polyakov blocks~\eqref{eq:GMI_maindef} to these operators, obtaining a striking simplification:
\begin{equation}
    F_{[\sigma^m]} (\tau,x)
    =
    \zeta_{\mathrm{EH}}
    \!\left(
    \frac{1}{2}-m,x,\tau
    \right)\,. \nonumber
\end{equation}
Crucially, we notice that the Epstein-Hurwitz zeta function has poles precisely at the values $s=\frac12-m$, as is evident from its analytic continuation~\eqref{eq:EHZeta_AnalyticContinuation}.
We can introduce a regulator by assuming that the external dimension $\Delta_\phi$ is slightly different from its correct value:
\begin{equation}
    \Delta_\phi=\frac12 \to \frac{1}{2}+\varepsilon \ , \quad \varepsilon \ll 1 \ .\nonumber
\end{equation}
The thermal two-point function then simply reduces to a decomposition in terms of Epstein-Hurwitz zeta functions, completed with the arc contribution introduced in Equation~\eqref{eq:GMI_FullCorrelator}:
\begin{equation}
    g(\tau,x)=\sum_{m=0}^{\infty} a_{[\sigma^m]} \, \zeta_{\mathrm{EH}}
    \!\left(
    \frac{1}{2}-m+\varepsilon,x,\tau
    \right)+ g_{\text{arcs}}(\tau,x) \ .\nonumber
\end{equation}
The thermal OPE coefficients are known:
\begin{equation}
    a_{[\sigma^m]}=\frac{m_{\text{th}}^{2m}}{\Gamma(2m+1)} \ , \label{eq:ON-ope}
\end{equation}
where $m_{\text{th}}=\log \varphi^2$ is the thermal mass and $\varphi$ is the golden ratio, which is explicitly given in~\eqref{eq:ON-mth}. 
\paragraph{Resumming the thermal Polyakov block decomposition.} To resum the Polyakov block decomposition, we will make use of the analytic continuation of the Epstein-Hurwitz zeta function.
Specializing it to the $[\sigma^m]$ trajectory, this reads:
\begin{multline}
    \zeta_{\mathrm{EH}}\left(\frac12-m + \varepsilon,x,\tau\right)
    =
    \frac{\sqrt{\pi}\,\Gamma\!\left(-m+\varepsilon\right)}
    {\Gamma\left(\frac12-m + \varepsilon\right)}
    \,|x|^{2m-2 \varepsilon}
    \\+
    \frac{4\pi^{\frac12-m + \varepsilon}}{\Gamma\left(\frac12-m + \varepsilon\right)}
    |x|^{m-\varepsilon}
    \sum_{n=1}^{\infty}
    n^{\varepsilon-m}
    K_{\varepsilon-m}\!\left(2\pi n x\right)
    \cos \left(2\pi n \tau \right)\,. \nonumber
\end{multline}
Only the first term develops a pole in $\varepsilon$: expanding around $\varepsilon=0$ returns a regularized result for the analytic continuation of the Epstein-Hurwitz zeta function:
\begin{multline}
    \zeta_{\mathrm{EH}}\left(\frac12-m + \varepsilon,x,\tau\right)
    =
    \frac{\sqrt{\pi} (-1)^m}
    {m! \Gamma\left(\frac12-m\right)}
    \,|x|^{2m} \left[\frac{1}{\varepsilon}+\log\left(\frac{4}{x^2}\right)+2H_m-2H_{2m}  \right]
    \\+
    \frac{4\pi^{\frac12-m}}{\Gamma\left(\frac12-m\right)}
    |x|^{m}
    \sum_{n=1}^{\infty}
    n^{-m}
    K_{m}\!\left(2\pi n x\right)
    \cos \left(2\pi n \tau \right)\,. \nonumber
\end{multline}
We can now perform the resummation, splitting the first term, encoding the divergence, from the second term, which is regular. Let us introduce the notation:

\begin{align}
    \mathcal{B}_{1}^{(\varepsilon)}(m,x)&=\frac{\sqrt{\pi} (-1)^m}
    {m! \Gamma\left(\frac12-m\right)}
    \,x^{2m} \left[\frac{1}{\varepsilon}+\log\left(\frac{4}{x^2}\right)+2H_m-2H_{2m}  \right] \ ,  \nonumber\\
    \mathcal{B}_{2}^{\vphantom{(\varepsilon)}}(m,\tau,x)&=\frac{4\pi^{\frac12-m}}{\Gamma\left(\frac12-m\right)}
    x^{m}
    \sum_{n=1}^{\infty}
    n^{-m}
    K_{m}\!\left(2\pi n x\right)
    \cos \left(2\pi n \tau \right) \ , \nonumber
\end{align}
such that the correlator takes the following form:
\begin{align}
g(\tau,x)&=\sum_{m=0}^{\infty} a_{[\sigma^m]} \,\mathcal{B}_{1}^{(\varepsilon)}(m,x)+\sum_{m=0}^{\infty} a_{[\sigma^m]} \,\mathcal{B}_{2}^{\vphantom{(\varepsilon)}}(m,\tau,x) + g_{\text{arcs}}(\tau,x)\;. \nonumber
\end{align}
The first sum reads:
    \begin{multline}
       \sum_{m=0}^{\infty} a_{[\sigma^m]} \,\mathcal{B}_{1}^{(\varepsilon)}(m,x)= 
    \left[\frac{1}{\varepsilon}+\log\left(\frac{4}{x^2}\right) \right]I_{0}(m_{\text{th}} x)+2\sum_{m=0}^{\infty} \frac{H_m-H_{2m}}
    { (m!)^2 }
    \left( \frac{m_{\text{th}}^2 x^2}{4}\right)^{m}  \ , \nonumber
    \end{multline}
    where the modified Bessel function of the first kind $I_{0}(z)$ appears. This expression can be significantly simplified by using the series expansion of the modified Bessel function of the second kind $K_{0}(z)$:
    \begin{equation}
        K_{0}(z)=-\left(\log \frac{z}{2} +\gamma_{E} \right)I_{0}(z)+\sum_{m=0}^{\infty} \frac{H_m}
    { (m!)^2 }
    \left( \frac{z^2}{4}\right)^{m} \ . \nonumber
    \end{equation}
    Plugging this into the previous result and defining the function:
    \begin{equation} \label{eq: arcdef}
        F(z)=\sum_{m=0}^{\infty} \frac{H_{2m}}
    { (m!)^2 }
    \left( \frac{z^2}{4}\right)^{m} \ ,
    \end{equation}
    the expression simplifies to
    \begin{equation} \label{eq: firstsum}
        \sum_{m=0}^{\infty} a_{[\sigma^m]} \,\mathcal{B}_{1}^{(\varepsilon)}(m,x)=2K_0(m_{\mathrm{th}} x)
        +\left(\frac{1}{\varepsilon}+2\gamma_E+2\log m_{\mathrm{th}}\right)\,I_0(m_{\mathrm{th}} x)
        -2F(m_{\mathrm{th}} x) \ .
    \end{equation}
    After interchanging the sums, the second sum reads
    \begin{align}
       \sum_{m=0}^{\infty} a_{[\sigma^m]} \,\mathcal{B}_{2}^{\vphantom{(\varepsilon)}}(m,\tau,x)
       &= \sum_{m=0}^{\infty}\sum_{n=1}^{\infty} \frac{4}{ m!}
    \left(-\frac{m_{\text{th}}^{2}x}{ 4 \pi n}\right)^{m}
    K_{m}\!\left(2\pi n x\right)
    \cos \left(2\pi n \tau \right) \notag \\
    &=\sum_{n=1}^{\infty} 4  K_{0}\left(x \sqrt{m_{\text{th}}^2+(2 \pi n)^2}\right) \cos(2 \pi n \tau) \ . \nonumber
    \end{align}
    The sum over $n$ can be evaluated through Poisson resummation, leading to the final result for the second sum:
    \begin{equation} \label{eq: secondsum}
        \sum_{m=0}^{\infty} a_{[\sigma^m]} \,\mathcal{B}_{2}^{\vphantom{(\varepsilon)}}(m,\tau,x)=\sum_{\ell \in \mathbb{Z}} \frac{e^{- m_{\text{th}}\sqrt{x^2+(\tau+\ell)^2}}}{\sqrt{x^2+(\tau+\ell)^2}}- 2K_{0}(m_{\text{th}}x) \ .
    \end{equation}
Combining the two expressions \eqref{eq: firstsum}, \eqref{eq: secondsum}, we fully resum the thermal Polyakov blocks decomposition of the thermal two-point function \eqref{eq: twoptlargeN}:
\begin{multline} 
    \sum_{m=0}^{\infty} a_{[\sigma^m]} \, \zeta_{\mathrm{EH}}
    \!\left(
    \frac{1}{2}-m+\varepsilon,x,\tau
    \right)=\sum_{\ell \in \mathbb{Z}} \frac{e^{- m_{\text{th}}\sqrt{x^2+(\tau+\ell)^2}}}{\sqrt{x^2+(\tau+\ell)^2}}\\+\left(\frac{1}{\varepsilon}+2\gamma_E+2\log m_{\mathrm{th}}\right)\,I_0(m_{\mathrm{th}} x)
-2F(m_{\mathrm{th}} x) \ . \nonumber
\end{multline}
Finally, we drop the divergence and the scheme-dependent term of the expression, proportional to $I_0(m_{\text{th}} x)$, and we obtain an intermediate expression for the full thermal two-point function:
\begin{equation}\label{eq: fullGMIlargeN}
    g(\tau,x)=\sum_{\ell \in \mathbb{Z}} \frac{e^{- m_{\text{th}}\sqrt{x^2+(\tau+\ell)^2}}}{\sqrt{x^2+(\tau+\ell)^2}}-2F(m_{\mathrm{th}} x)+g_{\text{arcs}}(\tau,x) \ .
\end{equation}
Although we recognize the first term on the right-hand side of Equation~\eqref{eq: fullGMIlargeN} as the correct answer, already known in the literature~\cite{Iliesiu:2018fao, Barrat:2025nvu}, let us consider it as an ansatz here: it can be shown that it satisfies the bootstrap axioms listed in Section \ref{subsec:DispersionRelationAndOPE}.
Moreover, the clustering condition rules out an additive-constant ambiguity.
We can conclude that the final correlator reads:
\begin{equation}
    g(\tau,x)=\sum_{\ell \in \mathbb{Z}} \frac{e^{- m_{\text{th}}\sqrt{x^2+(\tau+\ell)^2}}}{\sqrt{x^2+(\tau+\ell)^2}} \ . \nonumber
\end{equation}

\begin{figure}
    \centering
    \includegraphics[width=0.75\linewidth]{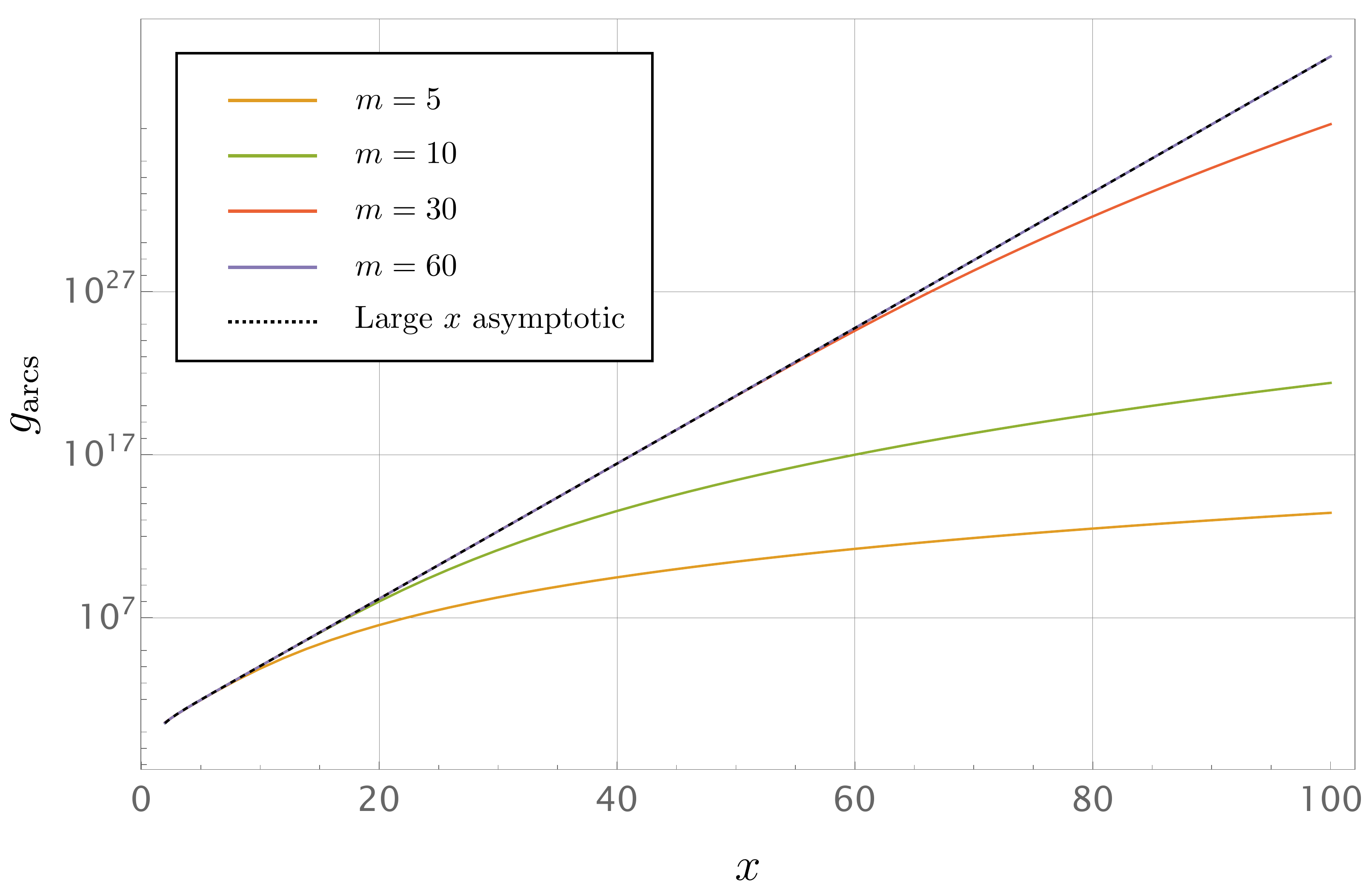}
    \caption{Convergence of $g_{\text{arcs}}(x)$, expressed as a Taylor series as in Equation~\eqref{eq: taylorarc}, to the asymptotic behavior \eqref{eq: asym arc} in the large-$x$ limit.}
    \label{fig:largeNarc}
\end{figure}

\paragraph{The arc function.}
Remarkably, the final result also fixes the arc function, which is possible because of the complete knowledge of the thermal OPE coefficients of the operators contributing to the discontinuity. The arc function simply reads:
\begin{equation}
    g_{\text{arcs}}(\tau,x)=g_{\text{arcs}}(x)=2 F(m_{\mathrm{th}} x) \ . \nonumber
\end{equation}
Following the definition \eqref{eq: arcdef}, we know its Taylor expansion around $x=0$:
\begin{equation} \label{eq: taylorarc}
    g_{\text{arcs}}(x)
    =
    2
    \sum_{m=1}^{\infty} \frac{H_{2m}}
    { (m!)^2 }
    \left( \frac{m^2_{\mathrm{th}} x^2}{4}\right)^{m} \ .
\end{equation}
Although this series does not allow for a closed form in terms of known functions, we can study its asymptotic behavior by considering a large-$m$ limit of the series coefficients.
Using the identity
\begin{equation}
    H_{2m}=\frac12\left( H_m+H_{m-\frac12} \right)+\log 2\,,  \nonumber
\end{equation}
we can write $H_{2m} \approx H_m+\log 2$.
In this case, the sum can be performed and we can extract the large-distance asymptotic behavior:
\begin{align*}
    2\sum_{m=1}^{\infty} \frac{H_{m}+\log 2}
    { (m!)^2 }
    \left( \frac{m^2_{\mathrm{th}} x^2}{4}\right)^{m}&=2K_{0}(m_{\mathrm{th}} x)+2(\log(m_{\mathrm{th}} x)+\gamma_{E})I_{0}(m_{\mathrm{th}} x)-2\log 2 \\
    &\approx 2\frac{\log(m_{\mathrm{th}} x)+\gamma_{E}}{\sqrt{2 \pi m_{\mathrm{th}} x}}e^{m_{\mathrm{th}} x}-2\log 2+\dots  \ , \quad x \gg 1  \ .
\end{align*}
The arc function behaves asymptotically as:
\begin{equation}  \label{eq: asym arc}
    g_{\text{arcs}}(x) \approx 2\frac{\log(m_{\mathrm{th}} x)+\gamma_{E}}{\sqrt{2 \pi m_{\mathrm{th}} x}}e^{m_{\mathrm{th}} x}-2\log 2 \ , \quad x \gg 1 \ ,
\end{equation}
and the asymptotic behavior can be visualized in Figure \ref{fig:largeNarc}.

In conclusion, the large-$N$ $\mathrm{O}(N)$ model provides an interesting example where the arc function $g_{\text{arcs}}(z,\bar{z})$, introduced in Equation~\eqref{eq:GMI_FullCorrelator}, can be determined exactly, supporting the claim that this contribution is uniquely fixed from bootstrap axioms combined with dynamical data.

\subsection{Application: $3d$ Ising model}
\label{subsec:Application3dIsingModel}

\begin{figure}
    \centering
    \includegraphics[width=0.75\linewidth]{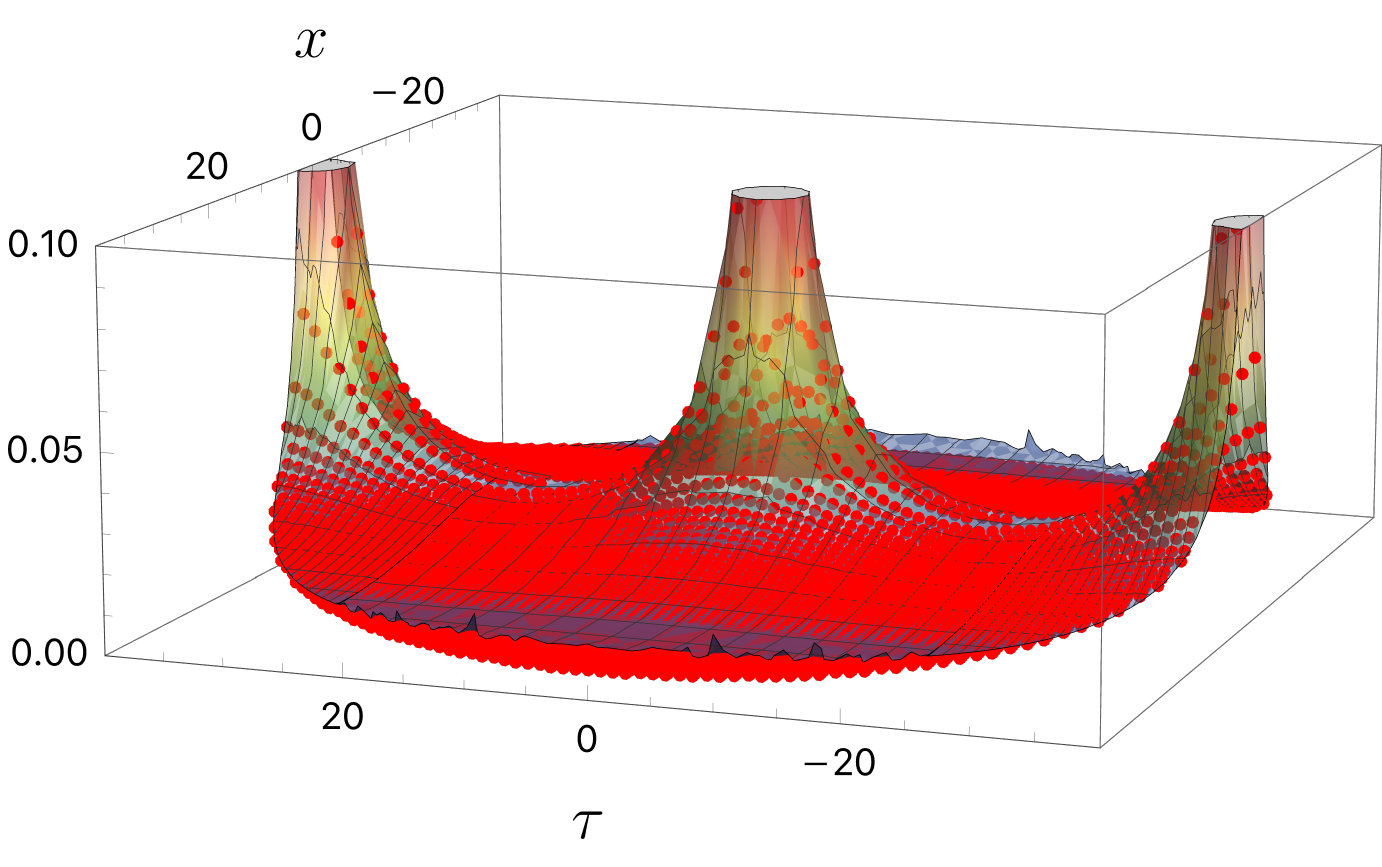}
    \caption{Three-dimensional plot of $g(\tau, x)=\vev{\sigma(\tau, x)\sigma(0,0)}_\beta$ in the OPE regime $\tau^2 + x^2 < \beta^2$ with $\beta = 40$.
    The continuous surface represents the truncated expansion in thermal Polyakov blocks, while the red dots represent the Monte Carlo results.
    We observe very good agreement between the results, further supported by the plot of the discrepancy in Figure~\ref{fig:DensityPlots_SigmaSigma}.}
    \label{fig:Plot3d_SigmaSigma}
\end{figure}

As an application, we consider the case of the $3d$ critical Ising model at finite temperature.
The idea here is the following: we insert the bootstrap results of~\cite{Barrat:2025wbi} for the one-point functions of the lightest operators $\mathds{1}$, $\eps$, $T$, $\eps'$ in the (truncated) expansion in thermal Polyakov blocks~\eqref{eq:GMI_PolyakovBlocks} and compare the resulting correlator with that obtained from Monte Carlo simulations.
Such a comparison was already performed in~\cite{Barrat:2025nvu} for the $\vev{\sigma\sigma}_\beta$ and $\vev{\eps\eps}_\beta$ correlators at $x=0$, where a good agreement was observed.
Here we investigate whether this agreement extends to non-zero spatial separation.

\paragraph{Correlator $\vev{\sigma\sigma}_\beta$.}
We first consider the two-point function $\vev{\sigma\sigma}_\beta$.
The expansion in thermal Polyakov blocks reads:
\begin{equation}
    g^{\vev{\sigma\sigma}}_\text{dr} (z,\zb)
    =
    \frac{1}{\beta^{2\Delta_\sigma}}
    \sum_{\Om=\mathds{1},\eps,T,\eps'}
    a^{\vev{\sigma\sigma}}_\Om\, F^{\vev{\sigma\sigma}}_\Om (z,\zb)\,.
    \label{eq:3dIsing_SS_GMI}
\end{equation}
Here we use the OPE data from the zero-temperature bootstrap~\cite{Simmons-Duffin:2016wlq} (for $\Delta_\Om$) and from the previous thermal-bootstrap calculation~\cite{Barrat:2025wbi} (for $a_\Om$).
The external operator has dimension $\Delta_\sigma = 0.5181489(10)$, while the low-lying exchanged operators have quantum numbers and OPE coefficients:
\begin{align}
    (\Delta_\mathds{1}, J_\mathds{1}) &= (0,0): && a^{\vev{\sigma\sigma}}_\mathds{1} = 1\,, \label{eq:3dIsing_SS_Input1} \\
    (\Delta_\eps, J_\eps) &= (1.412625(10),0): && a^{\vev{\sigma\sigma}}_\eps = 0.75(15)\,, \label{eq:3dIsing_SS_Input2} \\
    (\Delta_T, J_T) &= (3, 2): && a^{\vev{\sigma\sigma}}_T = 1.97(7)\,, \label{eq:3dIsing_SS_Input3} \\
    (\Delta_{\eps'}, J_{\eps'}) &= (3.82951(61),0): && a^{\vev{\sigma\sigma}}_{\eps'} = 0.19(6)\,. \label{eq:3dIsing_SS_Input4}
\end{align}
Meanwhile, the Monte Carlo (MC) simulation is run for $\beta = 40$. We use $841880$ lattice points in the region $\tau^2 + x^2 < \beta^2$.\footnote{We display only $100000$ points in Figures~\ref{fig:Plot3d_SigmaSigma} and~\ref{fig:DensityPlots_SigmaSigma} to improve the clarity. Details on the Monte Carlo simulations are provided in Appendix E of \cite{Barrat:2025nvu}. The code used to perform the simulations is publicly available on \href{https://github.com/alemiscio/Monte_Carlo_Ising_3d.git}{GitHub}, including the routines required to analyze the data and reproduce the results presented in this work.}

\begin{figure}
    \centering
    \begin{minipage}{.5\textwidth}
        \centering
        \includegraphics[width=0.95\linewidth]{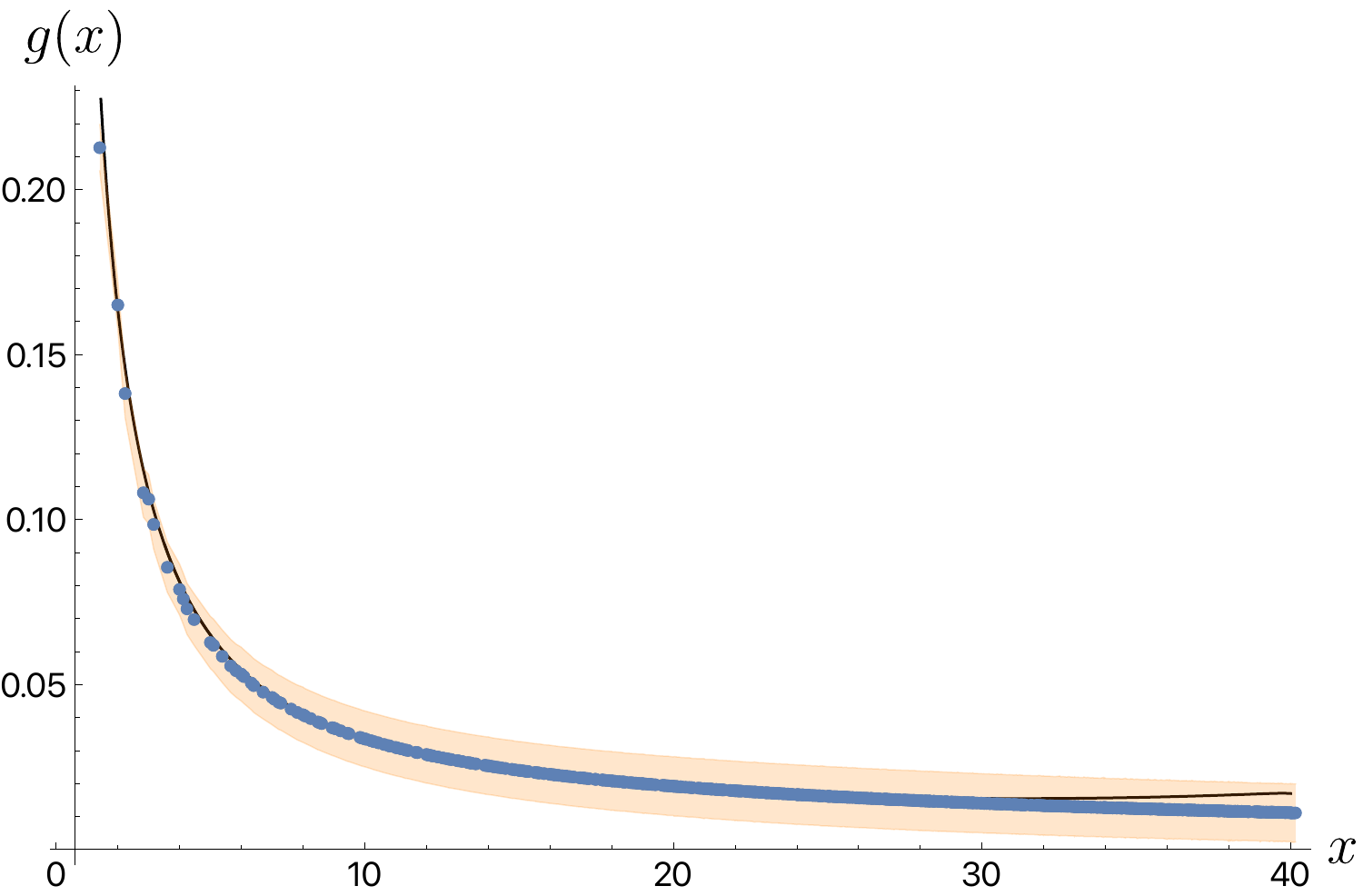}
    \end{minipage}%
    \begin{minipage}{0.5\textwidth}
        \centering
        \includegraphics[width=0.95\linewidth]{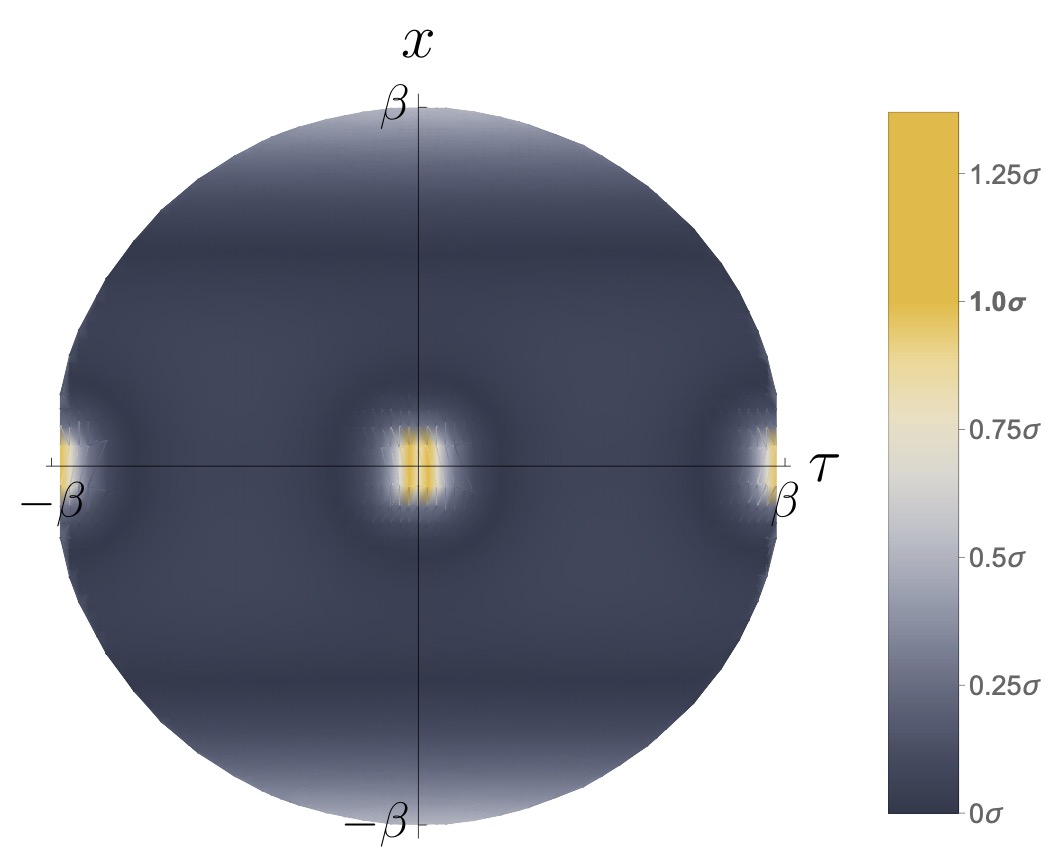}
    \end{minipage}
    \caption{
    \textbf{Left}: Plot of the Monte Carlo simulations (blue points with orange error bands) against the truncated expansion in thermal Polyakov blocks (black solid line) for the correlator $g(x)=\vev{\sigma(0,x)\sigma(0,0)}_\beta$ in the spatial direction $x$ for $\tau=0$.
    We observe a very good agreement in most of the OPE regime $x^2 < \beta^2$.
    As expected the two approaches start to disagree close to $x^2 \sim \beta^2$.
    \textbf{Right}: Discrepancy between the analytic prediction of the truncated expansion in thermal Polyakov blocks and the Monte Carlo data for the correlator $\vev{\sigma(\tau, x)\sigma(0,0)}_\beta$, measured in units of the Monte Carlo statistical uncertainty $\sigma$.
    The color at each point represents $|\delta g|/\sigma$.
    The analytic prediction lies well below $1\sigma$ for most of the OPE region ($\tau^2 + x^2 < \beta^2$), with the only exception occurring near the singularities, i.e., when $(\tau,x)=(m\beta,0)$.
    The large distance behavior $x > \beta$ is not expected to be well captured by the truncated expansion in thermal Polyakov blocks.
    }
    \label{fig:DensityPlots_SigmaSigma}
\end{figure}

In order to compare the MC results with the expansion~\eqref{eq:3dIsing_SS_GMI}, we introduce an overall constant via:\footnote{Note that we also shift the MC data by a constant $c\sim 0.00408$ to ensure that the correlator is positive everywhere.
This amounts to considering the connected correlator (or equivalently setting $\vev{\sigma}_\beta$ to zero).}
\begin{equation}
    g^{\vev{\sigma\sigma}}_\text{dr} (z,\zb)
    =
    \frac{C_{\vev{\sigma\sigma}}}{\beta^{2\Delta_\sigma}} \sum_{\Om=\mathds{1},\eps,T,\eps'}
    a^{\vev{\sigma\sigma}}_\Om\, F^{\vev{\sigma\sigma}
    }_\Om (z,\zb)\,, \nonumber
\end{equation}
and perform a fit in the limit $x=0$.
This provides the overall normalization of the correlator.
We obtain:
\begin{equation}
    C_{\vev{\sigma\sigma}}
    \sim
    0.3319\,. \nonumber
\end{equation}
The correlator in the OPE regime is plotted in Figure~\ref{fig:Plot3d_SigmaSigma}, while the discrepancy between the MC and the thermal Polyakov block resummation is shown in Figure~\ref{fig:DensityPlots_SigmaSigma}.
We observe a very good agreement ($< 5{\%}$) in a large portion of the OPE region.
The correlator at $\tau = 0$ is also plotted in Figure~\ref{fig:DensityPlots_SigmaSigma}, and we observe a very good agreement with the MC results in the OPE regime $x^2 < \beta^2$.
In particular, we observe that the analytic correlator is well within the error bars of the Monte Carlo simulation.

\paragraph{Asymptotic OPE coefficients for heavy operators.}
As explained in Section~\ref{subsec:AsymptoticOPECoefficientsForHeavyOperators}, the expansion in thermal Polyakov blocks provides an approximation for the OPE coefficients of heavy operators in terms of the light operators.
For instance, for the double-twist family $[\sigma\sigma]_0$ we obtain from~\eqref{eq:Tauberian_RecursionRelation} the large-spin expansion
\begin{align}
    \hat{a}_{[\sigma\sigma]_0} (J)
    =
    (1 + (-1)^J)
    \biggl[ &
    \frac{1}{J^{1/2-\Delta_\sigma}} \left( 1.0355 + \frac{0.0001705}{J} + O (J^{-2}) \right) \notag \\
    &+ \frac{a_T}{J^{1-\Delta_\sigma}} \left( 0.012186 + \frac{0.006562}{J} + O (J^{-2}) \right) \notag \\
    &+ \frac{a_\eps}{J^{1/2+\Delta_\eps/2-\Delta_\sigma}} \left( - 0.28971 - \frac{0.0686}{J} + O (J^{-2}) \right) \notag \\
    &+ \ldots
    \biggr]\,.
    \label{eq:LargeJ_Us}
\end{align}
The first line corresponds to the identity contribution, for which the terms exact in $J$ are presented in Table~\ref{tab:LIFvsGMI}.
This can be compared to Equation~(6.78) in~\cite{Iliesiu:2018fao}, which was obtained using the Lorentzian inversion formula.
We observe a perfect agreement.\footnote{As a note to the diligent reader, we observe a disagreement for the second term associated with $a_T$. 
However, if one uses Equation~(6.18) in~\cite{Iliesiu:2018fao}, one obtains $0.006563$ which is very close to our value.}
As explained below~\eqref{eq:AsymptoticCoefficients_DSD}, the main difference between this formula and ours is the presence of zeta functions reflecting periodicity.
At large $J$ the zeta functions tend quickly towards $1$ and this explains why we observe such a good agreement in the large-spin regime.

It is important to note that this approximation for the heavy operators does not provide OPE coefficients for the \textit{physical} operators.
Instead the coefficients should be seen as \textit{averages} of weighted OPE coefficients (in the sense explained in \cite{Marchetto:2023xap}), whose peaks lie close to the classical dimensions $\Delta = \Delta_{[\sigma\sigma]_{n,J}}$.
The physical OPE coefficients for the three dominant families $[\sigma\sigma]_{n=0,1,J}$ and $[\eps\eps]_{0,J}$ were studied in~\cite{Iliesiu:2018zlz}.
It was found that, similarly to zero temperature, the family $[\sigma\sigma]_{0,J}$ appears mostly unmixed and thus the real OPE coefficients will be close to their asymptotic values.
However, the families $[\sigma\sigma]_{1,J}$ and $[\eps\eps]_{0,J}$ are mixed.
In order to extract their values from~\eqref{eq:Tauberian_RecursionRelation}, we would need to perform an unmixing procedure, as done for instance in Section~4.4 of~\cite{Iliesiu:2018zlz}.
This is a rather involved procedure, in which both correlators $\vev{\sigma\sigma}_\beta$ and $\vev{\eps\eps}_\beta$ must be included, and we do not repeat it here.
Since our asymptotic coefficients are close to the asymptotic coefficients of~\cite{Iliesiu:2018zlz} for large $J$ (see Table~\ref{tab:LIFvsGMI}), the results are not expected to change significantly for most of the operators.\footnote{Interestingly we observe an increased mixed value of the OPE coefficient for the operator $T = [\sigma\sigma]_{0,2}$, which appears to be underestimated in comparison to MC data in Figure~4 of~\cite{Iliesiu:2018zlz}.
It would be interesting to understand whether our improved asymptotics could explain and resolve this discrepancy.
We leave this study for future work.}

\begin{table}[t]
    \renewcommand{\arraystretch}{1.3}
    \centering
    \caption{Comparison of asymptotic values of the OPE coefficients for the families $[\sigma\sigma]_{0,J}$ and $[\sigma\sigma]_{1,J}$ associated with the identity operator $\mathds{1}$, using the Lorentzian inversion formula (LIF) of~\cite{Iliesiu:2018fao,Iliesiu:2018zlz} and the associated thermal Polyakov block (TPB).
    We observe that the two methods approach one another at large spin $J$. }
    \begin{tabular}{ccc}
        \hline 
        $J$ & $\hat{a}_{[\sigma\sigma]_{0,J}}^{\vev{\sigma\sigma}}$ (LIF) & $\hat{a}_{[\sigma\sigma]_{0,J}}^{\vev{\sigma\sigma}}$ (TPB) \\ \hline
        $2$ & $1.94836$ & $2.50676$ \\
        $4$ & $2.09419$ & $2.20012$ \\
        $6$ & $2.13310$ & $2.15686$ \\
        $8$ & $2.14921$ & $2.15483$ \\
        $10$ & $2.15901$ & $2.16037$ \\ 
        $12$ & $2.16641$ & $2.16674$ \\ \hline
    \end{tabular}
    \qquad
    \begin{tabular}{ccc}
        \hline 
        $J$ & $\hat{a}_{[\sigma\sigma]_{1,J}}^{\vev{\sigma\sigma}}$ (LIF) & $\hat{a}_{[\sigma\sigma]_{1,J}}^{\vev{\sigma\sigma}}$ (TPB) \\ \hline
        $2$ & $-0.042586$ & $0.014984$ \\
        $4$ & $-0.010563$ & $0.028374$ \\
        $6$ & $0.022499$ & $0.031922$ \\
        $8$ & $0.031206$ & $0.033812$ \\
        $10$ & $0.034068$ & $0.035014$ \\
        $12$ & $0.036450$ & $0.036461$ \\ \hline
    \end{tabular}
    \label{tab:LIFvsGMI}
\end{table}

\begin{figure}
    \centering
    \begin{minipage}{.5\textwidth}
        \centering
        \includegraphics[width=0.95\linewidth]{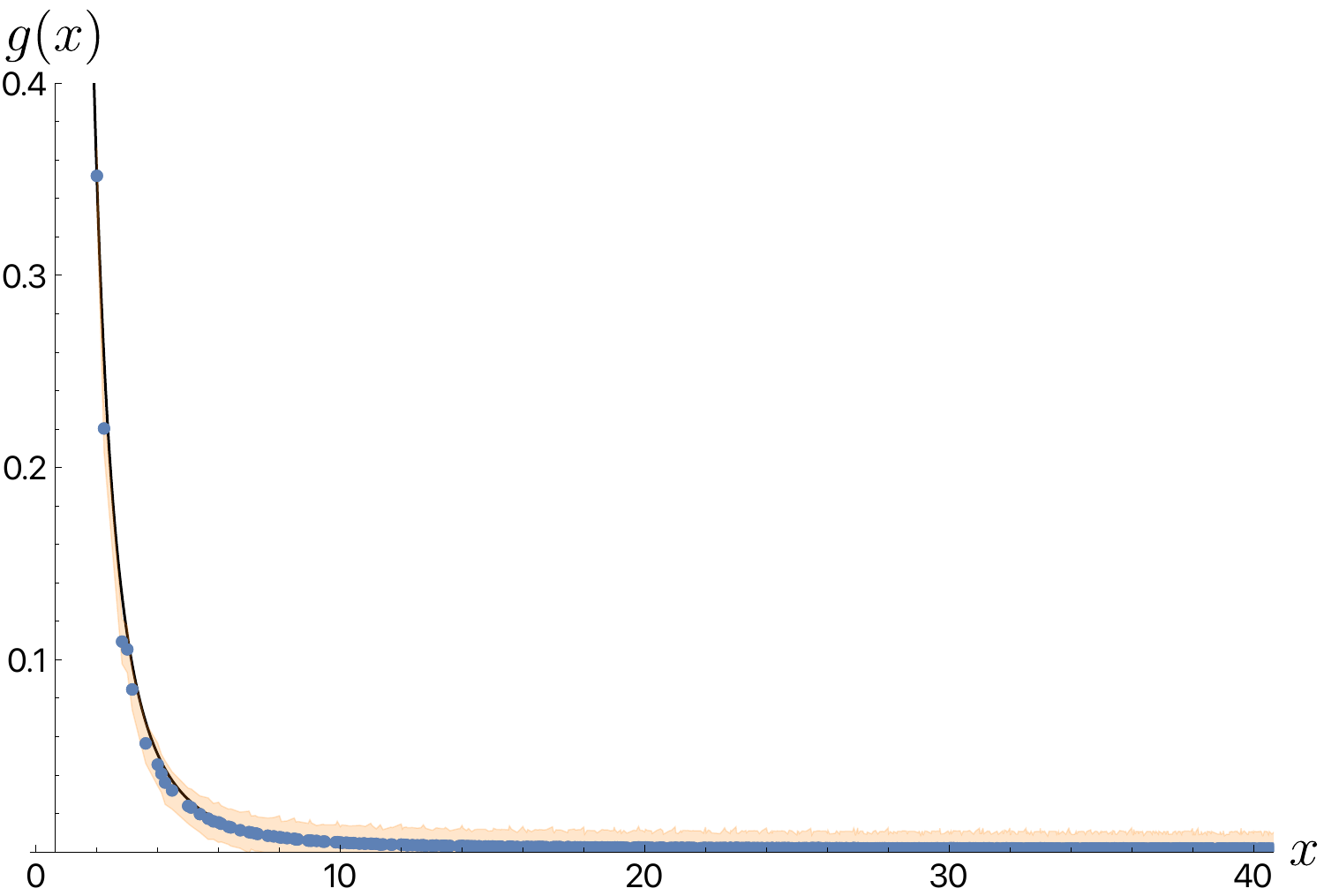}
    \end{minipage}%
    \begin{minipage}{0.5\textwidth}
        \centering
        \includegraphics[width=0.95\linewidth]{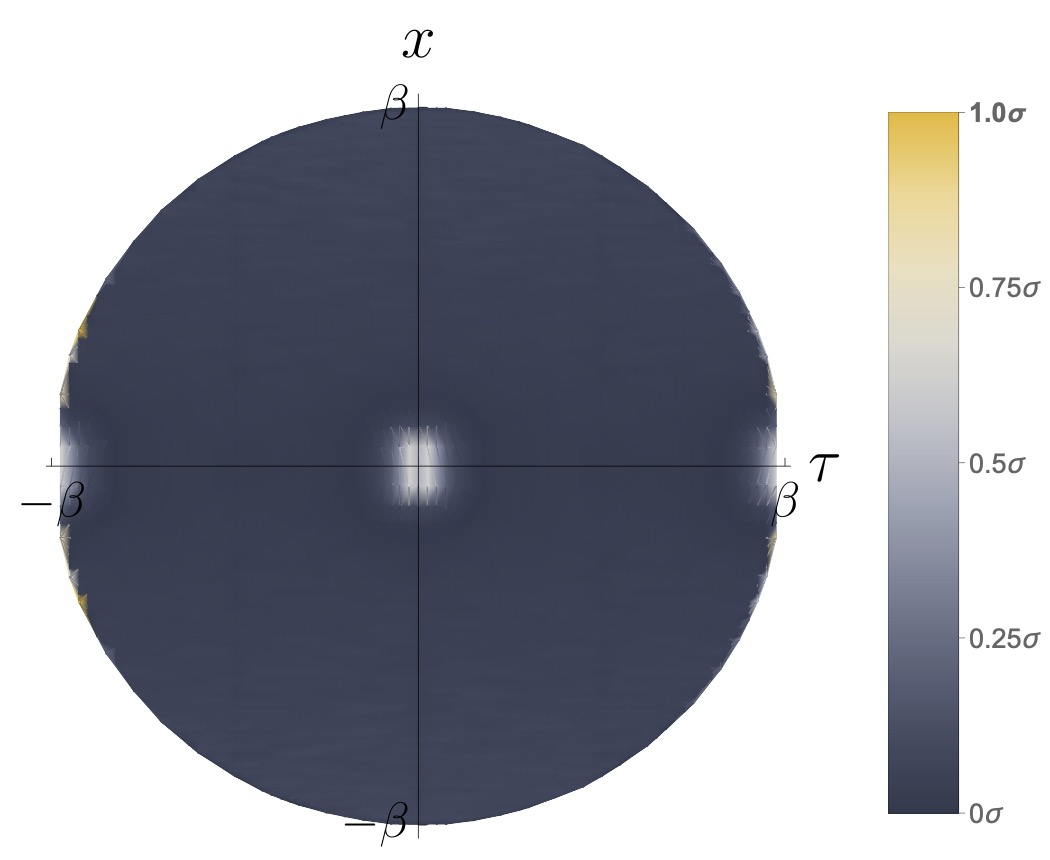}
    \end{minipage}
    \caption{
    \textbf{Left}: Plot of the Monte Carlo simulations (blue points with orange error bands) against the truncated expansion in thermal Polyakov blocks (black solid line) for the correlator $g(x)=\vev{\eps(0,x)\eps(0,0)}_\beta$ in the spatial direction $x$ for $\tau=0$.
    \textbf{Right}: Discrepancy between the analytic prediction of the truncated expansion in thermal Polyakov blocks and the Monte Carlo data for the correlator $\vev{\eps(\tau,x)\eps(0,0)}_\beta$, measured in units of the Monte Carlo statistical uncertainty $\sigma$.
    As for $\vev{\sigma(\tau,x)\sigma(0,0)}_\beta$ we observe a very good agreement.
    }
    \label{fig:DensityPlots_EpsEps}
\end{figure}

\paragraph{Correlator $\vev{\eps\eps}_\beta$.}
We now consider the correlator $\vev{\eps\eps}_\beta$.
The expansion in thermal Polyakov blocks is:
\begin{equation}
    g^{\vev{\eps\eps}}_\text{dr} (z,\zb)
    =
    \frac{C_{\vev{\eps\eps}}}{\beta^{2\Delta_\eps}}
    \sum_{\Om=\mathds{1},\eps,T}
    a^{\vev{\eps\eps}}_\Om\, F^{\vev{\eps\eps}}_\Om (z,\zb)\,.
    \label{eq:3dIsing_EE_GMI}
\end{equation}
We adapt the input for $\vev{\sigma\sigma}_\beta$ provided in~\eqref{eq:3dIsing_SS_Input1}-\eqref{eq:3dIsing_SS_Input4} by the simple conversion:
\begin{equation}
    a^{\vev{\eps\eps}}_\Om
    =
    \frac{f_{\eps\eps\Om}}{f_{\sigma\sigma\Om}}
    a^{\vev{\sigma\sigma}}_\Om\,,
    \label{eq:3dIsing_ConvertSSToEE}
\end{equation}
using the zero-temperature three-point functions obtained from the zero-temperature bootstrap~\cite{Simmons-Duffin:2016wlq}.
This yields:
\begin{align}
    a^{\vev{\eps\eps}}_\mathds{1} &= 1\,, \label{eq:3dIsing_EE_Input1} \\
    a^{\vev{\eps\eps}}_\eps &= 1.09(22)\,, \label{eq:3dIsing_EE_Input2} \\
    a^{\vev{\eps\eps}}_T &= 5.37(19)\,. \label{eq:3dIsing_EE_Input3}
\end{align}
We now fit the expression~\eqref{eq:3dIsing_EE_GMI} against the MC data to obtain the overall normalization constant, which reads:\footnote{As for the $\vev{\sigma\sigma}_\beta$ case, there is also a shift $c \sim 0.001676$.}
\begin{equation}
    C_{\vev{\eps\eps}}
    \sim
    2.4394\,. \nonumber
\end{equation}
The comparison between the truncated sum of thermal Polyakov blocks and MC data is presented in Figure~\ref{fig:DensityPlots_EpsEps}.
We observe again a very good agreement.

\section{Thermal Polyakov blocks in momentum space}
\label{sec:OPEInMomentumSpace}

In this section we use the decomposition of the thermal correlator into thermal Polyakov blocks derived in the previous section and we transform this decomposition to momentum space. We start by deriving the thermal Polyakov blocks for the Euclidean correlator in momentum space.
By using the direct connection between the Euclidean correlator and the retarded correlator we obtain the corresponding thermal Polyakov blocks for the retarded correlator and the decomposition of the momentum-space correlator in the complex-$\omega$ plane.
Our results apply in two dimensions and for $d>2$.
We consider special limits of the momentum-space correlator such as $k=0$ to connect with previous works. We conclude this section by reconstructing the exact $2d$ correlator for a Virasoro primary with $\Delta_\phi=1$.
We show that, although the expansion of the correlator in terms of thermal Polyakov blocks in momentum-space is asymptotic, one can apply a Borel transform to obtain the exact result.

\subsection{Fourier transform of thermal Polyakov blocks}
\label{subsubsec:GeneralizedMethodOfImagesInMomentumSpace}

The formula given in~\eqref{eq:GMI_maindef} has a very simple interpretation: it is the KMS-symmetric inversion of a single OPE block of the thermal two-point function.
As a consequence, the decomposition of the correlator into thermal Polyakov blocks is \emph{globally} defined in position space, unlike the OPE, which has a finite radius of convergence.
This allows us to perform a Fourier transform of the thermal Polyakov blocks and to obtain \emph{momentum-space blocks} globally defined in $(\omega,k)$. This result represents a novelty with respect to previous works which considered asymptotic decompositions in Fourier space at large frequency and momentum.\footnote{Note that this does not imply that the OPE is convergent in momentum space.
We later show examples that illustrate the fact that it is an asymptotic series in most cases.} 

In the following, we consider the Fourier transform of the sum over thermal images of a generic function $f(\tau,x)$:
\begin{equation}
    \Fm [f] (\omega_n,k)
    =
    \int_0^{\beta} \mathrm{d}\tau \int \mathrm{d}^{d-1}x \,
    e^{i (\tau\omega_n + k \cdot x)} \sum_{m=-\infty}^{\infty}f(\tau+m \beta, x)\,, \label{eq:fouriertransform}
\end{equation}
where we introduced the Matsubara frequencies $\omega_n=\frac{2 \pi n}{\beta}$. By swapping the integration with the sum, we can make use of Poisson resummation to obtain:
\begin{equation}
    \sum_{m=-\infty}^\infty
    \int_0^{\beta} \mathrm{d}\tau\int \mathrm{d}^{d-1}x\,
    e^{i (\tau\omega_n+k \cdot x)}\,
    f(\tau+m\beta,x)
    =
    \int_{-\infty}^\infty \mathrm{d}\tau \int \mathrm{d}^{d-1}x\,
    e^{i (\tau\omega_n+k \cdot x)}\,
    f(\tau,x) \ . \nonumber
\end{equation}
We use this result to perform the Fourier transform of the thermal Polyakov blocks given in~\eqref{eq:GMI_maindef}, which boils down to determining the Fourier transform of the Epstein-Hurwitz zeta function~\eqref{eq:EHZeta_Definition}:
\begin{equation}
    \Fm \left[ \zeta_{EH} \right] (s,\omega_n,k)
    =
    2^{\,d-2s}\,\pi^{d/2}\,
    \frac{\Gamma\!\left(d/2-s\right)}{\Gamma(s)}
    \left(k^2+\omega_n^2\right)^{\,s-d/2}\,,
    \label{eq:EH-momentum}
\end{equation}
This can be generalized to obtain the Fourier transform of the thermal Polyakov blocks:
\begin{equation}
    \Fm \left[ F_\Om \right] (\omega_n,k)
    =
    \pi^{d/2} (-1)^{J/2} 2^{d-2s}
    \frac{
    \Gamma\!\left(\frac{d+J-2s}{2}\right)
    }{
    \Gamma\!\left(\frac{J+2s}{2}\right)
    }
    (\omega_n^2 + k^2)^{\,s-\frac{d}{2}}
    \,
    C_J^{(\nu)}
    \!\left(\frac{\omega_n}{\sqrt{\omega_n^2 + k^2}}\right)\,,
    \label{eq:GMI-momentum}
\end{equation}
where $s=\Delta_\phi-\Delta_\Om/2$; the result only depends on $|\vec{k}|=k$, due to spatial rotational symmetry. As a consequence of the position space dispersion relation given in \eqref{eq:GMI_FullCorrelator} and its decomposition given in \eqref{eq:GMI_PolyakovBlocks}, we now obtain the decomposition of the momentum-space Euclidean correlator into thermal Polyakov blocks:
\begin{align}
    g_E(\omega_n,k)
    =
    \sum_\Om \frac{a_\Om}{\beta^{\Delta_\Om}} \Fm \left[ F_\Om \right] (\omega_n,k)+\tilde g_\text{arcs}(\omega_n,k)\; ,
    \label{eq:mom-decomposition}
\end{align}
where, for simplicity, we use the notation, $\tilde g_\text{arcs}(\omega_n,k) = \mathcal{F}[g_\text{arcs}](\omega_n,k)$. It should be emphasized that the decomposition into thermal Polyakov blocks in momentum space need not converge.
In fact, in most cases it should be understood as an asymptotic series.
This is expected due to the presence of infinitely many poles in the complex frequency plane, which are known as QNMs.
The novelty of~\eqref{eq:GMI-momentum} is that it provides these asymptotic expansions in different regimes of validity, such as finite $k$ or $k=0$, starting from a single decomposition valid for arbitrary $\omega$ and $k$. 

\paragraph{Generalized free field theory.} As the simplest example we consider a GFF theory with scaling dimension $\Delta_\phi$ in $d$ dimensions. Since only the identity operator contributes to the dispersion relation given in \eqref{eq:GMI} and there are no arc contributions, the Euclidean correlator takes the form:
\begin{align} \label{eq: GFF momentum}
    g_{\text{GFF}} (\omega_n,k)
    =
    \pi^{d/2}  2^{d-2\Delta_\phi}
    \frac{
    \Gamma\!\left(\frac{d-2\Delta_\phi}{2}\right)
    }{
    \Gamma\!\left(\Delta_\phi\right)
    }
    (\omega_n^2 + k^2)^{\,\Delta_\phi-\frac{d}{2}}
    \, ,
\end{align}
which agrees with the exact correlator up to overall numerical factors. Note that when: 
\[
\Delta_\phi = \frac{d}{2} + n, 
\qquad n = 0,1,2,\dots,
\]
the correlator develops a pole. Thus the momentum-space correlator is singular and requires regularization/renormalization. Physically, this signals conformal anomalies, and the usual power-law behavior is replaced by a logarithmic term which we will discuss further in this section.

The check \eqref{eq: GFF momentum} is also useful because it suggests a convenient normalization for the thermal Polyakov blocks in momentum space. In fact, in the current normalization the Euclidean free scalar propagator ($\Delta_\phi=\nu$) would read:
\begin{equation} \nonumber
    g_{\text{free}} (\omega_n,k)
    =
    \frac{
    4\pi^{\nu+1}
    }{
    \Gamma\!\left(\nu\right)
    }
    \frac{1}{\omega_n^2 + k^2}
    \,.
\end{equation}
By requiring the propagator to be unit-normalised, the normalization of the thermal Polyakov blocks becomes:
\begin{equation} \nonumber
    \Fm \left[ F_\Om \right] (\omega_n,k)
    =
     (-1)^{J/2} 4^{\nu-s}
    \frac{\Gamma\!\left(\nu\right)
    \Gamma\!\left(\nu+1-s+J/2\right)
    }{
    \Gamma\!\left(s+J/2\right)
    }
    (\omega_n^2 + k^2)^{\,s-\nu-1}
    \,
    C_J^{(\nu)}
    \!\left(\frac{\omega_n}{\sqrt{\omega_n^2 + k^2}}\right)\,,
\end{equation}
and the non-kinematical factors can be encoded into an overall factor: 
\begin{equation} \nonumber
    \mathcal{N}_{\mathcal{O}}^{(\nu)}=(-1)^{J/2} 4^{\nu-s}
    \frac{
    \Gamma\!\left(\nu+1-s+J/2\right)
    }{
    \Gamma\!\left(s+J/2\right)
    } \ .
\end{equation} 
If we define a \emph{momentum-space} OPE coefficient for each operator in the OPE: 
\begin{equation} \label{eq: momopeco}
    \tilde{a}_{\mathcal{O}}=a_{\mathcal{O}} \; \Gamma\!\left(\nu\right) \,
    \mathcal{N}_{\mathcal{O}}^{(\nu)} =  
    (-1)^{J/2} 4^{\nu-s}
    \frac{\Gamma\!\left(\nu\right)
    \Gamma\!\left(\nu+1-s+J/2\right)
    }{
    \Gamma\!\left(s+J/2\right)
    } \frac{J!}{2^{J} (\nu)_J}\frac{f_{\phi \phi \mathcal O} b_{\mathcal O}}{c_{\mathcal O}}\ ,
\end{equation}
we can produce the final clean form for the thermal Polyakov blocks in momentum space:
\begin{equation} \label{eq:finalblock}
    \tilde{F}_\Om (\omega_n,k)
    =
    \frac{\Fm \left[ F_\Om \right] (\omega_n,k)}{\Gamma(\nu) \Nm_\Om^{(\nu)}}
    =
    (\omega_n^2 + k^2)^{\,s-\nu-1}
    \,
    C_J^{(\nu)}
    \!\left(\frac{\omega_n}{\sqrt{\omega_n^2 + k^2}}\right)\,.
\end{equation}
Using this notation, the decomposition of the Euclidean correlator into thermal Polyakov blocks becomes
\begin{equation}
    g_E(\omega_n,k)
    =
    \sum_\Om \frac{\tilde{a}_\Om}{\beta^{\Delta_\Om}} \tilde{F}_\Om (\omega_n,k)+ \tilde{g}_\text{arcs} (\omega_n,k)\;.\label{eq:mom-decomposition2}
\end{equation}
It should be appreciated that, although the definition \eqref{eq: momopeco} associates a momentum-space OPE coefficient to every position-space OPE coefficient, the $\mathcal{N}_{\mathcal{O}}^{(\nu)}$ will vanish whenever $\Delta=2\Delta_\phi+2\mathbb{Z}^{\geq 0}$, i.e., in correspondence with classical double-twist conformal dimensions. This feature encodes the fact that only operators contributing to the discontinuity of the Euclidean correlator will contribute to the thermal Polyakov-block decomposition in momentum space.
\paragraph{Thermal Polyakov blocks in $2d$.}
Since in $d=2$ the thermal OPE blocks are different from those in~\eqref{eq:ThermalBlocks}, this case needs to be handled separately. The regularization in the limit $\nu \to 0$ leads to the following OPE:

\begin{align}
    g(\tau,x)
    =
    \sum_{\mathcal{O}} \frac{a^{2d}_{\mathcal{O}}}{\beta^{\Delta_\Om}}\left(\tau^2+x^2\right)^{-s}\, T_J\left(\frac{\tau}{\sqrt{\tau^2+x^2}}\right)\,,
    \label{eq:2dOPE}
\end{align}
where $T_n(x)$ denotes the Chebyshev polynomial of the first kind and, in contrast to the general definition~\eqref{eq:aO_Definition}, the OPE coefficients are defined as follows:
\begin{equation} \label{eq: 2dao}
    a^{2d}_{\mathcal{O}}=\frac{1}{2^{J}}\frac{f_{\phi\phi\mathcal{O}}b_{\mathcal{O}}}{c_{\mathcal{O}}}\left(\frac{2}{1+\delta_{J,0}} \right) \ .
\end{equation}
Following the logic previously adopted, we can define the momentum-space OPE coefficients in $d=2$ as:
\begin{equation} \label{eq: def2datilde}
    \tilde{a}^{2d}_{\mathcal{O}}=a^{2d}_{\mathcal{O}} \; 4\pi \; \mathcal{N}_{\mathcal{O}}^{(0)}  \ .
\end{equation}
From the expansion \eqref{eq:2dOPE} we perform the Fourier transform of the KMS-invariant blocks and obtain:
\begin{equation}
    \tilde{F}_\Om^{2d} (\omega_n,k)
    =
\left(\omega_n^2+k^2\right)^{s-1}
T_J\!\left(
\frac{\omega_n}{\sqrt{\omega_n^2+k^2}}
\right)\,.
    \label{eq:GMI-momentum2d}
\end{equation}

We reproduce the same structure as observed in~\eqref{eq:finalblock}, namely the special polynomials contained in the thermal conformal blocks as given in \eqref{eq:2dOPE} resurface inside the momentum space dispersion relation.

\paragraph{Zero-momentum limit.}
We will extensively use the $k=0$ limit of~\eqref{eq:finalblock} in the following.
Since the spin $J$ is an even integer, in this limit the block simplifies to:
\begin{equation} 
    \tilde{F}_\Om (\omega_n, 0)
    =
    (\omega_n^2)^{\,s-\nu-1}
    \,
    C_J^{(\nu)}
    \!\left(1\right)\,.
    \label{eq:gmizeromomentum}
\end{equation}
We observe that in this limit the spin $J$ is completely disentangled from the kinematics, through a mechanism similar to that in the $x=0$ limit in position space, where the sum over operators of equal dimensions is weighted by a Gegenbauer polynomial \cite{Marchetto:2023xap}.
We denote the weighted sum as $\tilde{a}_{\Delta}$, and the formula is completely analogous to the position space version:
\begin{align}
    \tilde{a}_\Delta=\sum_{\Om, \Delta_\Om = \Delta} \tilde{a}_{\mathcal{O}}
    \,
    C_J^{(\nu)}
    \!\left(1\right)\;.
    \label{eq:aDelta-momentum}
\end{align}

\paragraph{Retarded correlator and the spectral function.} The presence of the thermal circle implies that the Euclidean correlator in momentum space is defined at discrete Matsubara frequencies.
We can transform the Euclidean correlator to the retarded one by making use of the relation \cite{LeBellac:1996,Meyer:2011gj,Kapusta:2023ftft}:
\begin{align}
g_E(\omega_n, k)=g_R(i\omega_n, k)\,,
\label{eq:euctoret}
\end{align}
for $\omega_n > 0$.
Then, we analytically continue $\omega + i 0=i\omega_n$ to obtain the retarded correlator defined in the complex $\omega$-plane.
We suppress the prescription $+ i 0$ in the following for compactness.
The remarkably simple relation between the Euclidean and retarded correlators allows us to directly adapt the momentum-space decomposition obtained in the previous section to the retarded correlator, 
\begin{equation}
    g_R(\omega,k)=
    \sum_{\mathcal O} \frac{\tilde{a}_{\mathcal O}}{\beta^{\Delta_\Om}}
    (k^2-\omega^2)^{\,s-\nu-1}
    \,
    C_J^{(\nu)}
    \!\left(\frac{i\omega}{\sqrt{k^2-\omega^2}}\right) + \tilde{g}_{\rm arcs}(\omega,k) \,.
    \label{eq:momentumOPE}
\end{equation}
where $\tilde{g}_{\text{arcs}}(\omega,k) = \mathcal F \left[g_{\text{arcs}} \right] (-i\omega,k)$ is the Fourier transform of the function $g_{\text{arcs}}$ as given in \eqref{eq:GMI_FullCorrelator}.

In $d=2$, the retarded correlator has a very similar form and it is given by
\begin{equation} \nonumber
    g_R(\omega,k)=
    \sum_{\mathcal O} \frac{\tilde{a}^{2d}_{\mathcal O}}{\beta^{\Delta_\Om}}
    (k^2-\omega^2)^{\,s-1}
    \,
    T_J
    \!\left(\frac{i\omega}{\sqrt{k^2-\omega^2}}\right) + \tilde{g}_{\rm arcs}(\omega,k) \,,
\end{equation}
with the convention~\eqref{eq: 2dao} for the OPE coefficients.

\paragraph{Spectral density.} Furthermore, we can decompose the spectral density in terms of the spectrum using the definition:\footnote{Note that if the retarded correlator does not fall off exponentially as we take $\omega\to\infty$ then one can add subtractions to make the integral convergent.}
\begin{equation}
    g_R(\omega,k) = \int_{-\infty}^{\infty} \frac{\text{d}\omega'}{2 \pi } \frac{\rho(\omega',k)}{\omega'-\omega-i 0}\label{eq:specdensity}
\end{equation}
or, equivalently:
\begin{equation} \nonumber
    \rho(\omega,k)= 2\operatorname{Im} g_R(\omega+i0,k)=-i \,\text{Disc}_\omega \; g_R(\omega, k) \ ,
\end{equation}
where the discontinuity is defined as:
\begin{equation} \nonumber
    \text{Disc}_\omega \; g_R(\omega, k)=g_R(\omega+i0,k)-g_R(\omega-i0,k) \ .
\end{equation}
Starting from the decomposition in~\eqref{eq:momentumOPE}, one can exchange the sum over operators in the OPE and the discontinuity on the real axis. Computing the spectral density term by term, this results in
\begin{equation} \nonumber
    \rho^{(\Om)}(\omega,k)
    =
    2\,\operatorname{Im}
    \left[
    (k^2-(\omega+i0)^2)^{\,s-\nu-1}
    \,
    C_J^{(\nu)}
    \!\left(\frac{i(\omega+i0)}{\sqrt{k^2-(\omega+i0)^2}}\right) 
    \right].
\end{equation}
In the time-like region $|\omega|>k$, and for even spin $J$, this becomes:
\begin{equation} \nonumber
    \rho^{(\Om)}(\omega,k)
    =
    -2\,\operatorname{sgn}(\omega)\,
    \sin\!\left[
    \pi\left(s-\frac d2\right)
    \right]
    \left(\omega^2-k^2\right)^{s-\frac d2}
    C_J^{(\nu)}
    \left(
    \frac{|\omega|}{\sqrt{\omega^2-k^2}}
    \right)
    \Theta(\omega^2-k^2)\,.
\end{equation}
Thus, schematically, the decomposition of the spectral density in terms of the spectrum is given by:
\begin{equation} \nonumber
    \rho(\omega,k)
    =
    \sum_{\Om} \frac{\tilde{a}_\Om}{\beta^{\Delta_\Om}} \;
    \rho^{(\Om)}(\omega,k)
    +
    \rho_{\text{arcs}}(\omega,k).
\end{equation}
The last term is present only if $\tilde g_{\text{arcs}}(\omega,k)$ has a non-trivial imaginary part on the real frequency axis:
\begin{equation} \nonumber
    \rho_{\text{arcs}}(\omega,k) = 2 \operatorname{Im}\left[ \tilde{g}_{\text{arcs}}(\omega,k)\right]\;.
\end{equation}
Although this work does not focus on the spectral density, we have defined it in terms of the retarded correlator and highlighted that it can be accessed from the thermal Polyakov block decomposition by analytically continuing the Euclidean correlator. Consequently, the Euclidean KMS-invariant block decomposition provides a direct route to several physically interesting regimes of the retarded correlator. Using this connection we hope to report on deeper investigations of Lorentzian dynamical information, such as dissipation and spectral response, to be studied using the same Euclidean data that organize the thermal block expansion.
\paragraph{Connection with previous works.}
Previously, a notion of momentum-space OPE was constructed by \cite{Caron-Huot:2009ypo,Manenti:2019wxs} starting from the OPE in position space as in~\eqref{eq:OPE}.
Additionally, in various asymptotic regimes in momentum space, contributions from the OPE were considered in \cite{Dodelson:2023vrw,Jia:2025jbi}.
In \cite{Manenti:2019wxs}, the momentum-space thermal correlator was decomposed into thermal OPE blocks, giving in the large-$k$ limit:
\begin{align}
    G^{(\Delta_\phi,d)}_{\Delta,J}(\omega_n,k)
    &=
    \sum_{j=0}^{J/2}
    c^{(\nu)}_{J,j}\,
    \frac{2^{\beta_-}\pi^{d/2}\Gamma(\beta_+)}{\Gamma(\alpha)}
    \,{}_2F_1\!\left(
    \begin{matrix}
        \beta_-,\,1-\beta_-\\
        1-\beta_+
    \end{matrix}
    ; \frac{k+i\omega_n}{2k}
    \right)
    \,k^{-\beta_-}(k+i\omega_n)^{-\beta_+} \notag
    \\&\phantom{=\ }
    + O(e^{-k})\,,
    \label{eq:manenti-block}
\end{align}
with the following auxiliary variables defined:
\begin{align} \nonumber
    \tilde{\jmath}=\frac{J}{2}-j,
    \qquad
    \alpha=\tilde{\jmath}+\Delta_\phi-\frac{\Delta}{2},
    \qquad
    \beta_\pm=\frac{d}{2}+\frac{\Delta}{2}-\Delta_\phi \pm \tilde{\jmath}\,.
\end{align}
This expression is valid up to non-perturbative corrections.
It turns out that these momentum-space OPE blocks consist of linear combinations of hypergeometric functions $_2F_1$, which are directly related to the momentum space blocks presented in~\eqref{eq:GMI-momentum}:
\begin{align} \nonumber
    G^{(\Delta_\phi,d)}_{\Delta,J} (\omega_n,k)
    =
    \int_0^1 \mathrm{d}\tau
    \Bigg[\int \mathrm{d}^{d-1}x\, C_J^{(\nu)}\left(\frac{\tau}{\sqrt{\tau^2+x^2}}\right)(\tau^2+x^2)^{-s}e^{i\tau\omega_n+i\vec{k}\cdot\vec{x}}\Bigg]_{k\to\infty}\,.
\end{align}
If we consider the blocks as given in~\eqref{eq:GMI-momentum} in the large-$k$ limit, we recover the expression~\eqref{eq:manenti-block}.
This means that a significant part of the non-perturbative corrections contained in the $O(e^{-k})$ terms, missing in~\eqref{eq:manenti-block}, is included in our blocks~\eqref{eq:GMI-momentum}.
This is due to the fact that the thermal Polyakov blocks are defined in all of position space, therefore attributing non-perturbative corrections of order $O(e^{-k})$ to each operator contributing to the OPE.

It should, however, be noted that there may still be non-perturbative corrections that are not captured by the thermal Polyakov blocks in momentum space.
They are encoded in the arc contributions $\tilde{g}_\text{arcs}(\omega_n,k)$.
We discuss this point further in Section~\ref{subsec:2dBorel} by focusing on the $2d$ case.

\paragraph{Anomalies and integer dimensions.}
It is well known that for scalar operators whose scaling dimensions satisfy $\Delta_{\phi}-d/2\in\mathbb Z^{\geq 0}$, the naive Fourier transform of conformal correlation functions requires regularization.
Logarithmic terms arise in momentum space after renormalization.
In the literature, this phenomenon is usually interpreted as a manifestation of conformal anomalies~\cite{Bzowski:2013sza,Bzowski:2015pba,Schwimmer:2018hdl}.\footnote{More precisely, these logarithms are associated with Type B conformal anomalies~\cite{Deser:1993yx,Osborn:1991gm,Schwimmer:2010za,Gomis:2015yaa,Niarchos:2019onf,Niarchos:2020nxk,Andriolo:2022lcb,Schwimmer:2023nzk,Baume:2024poj}.}

The mechanism can be understood already from the two-point function of a scalar primary operator at zero temperature,
\begin{equation} \nonumber
    \vev{\phi(x)\phi(0)}
    =
    \frac{C_{\phi}}{|x|^{2\Delta_{\phi}}}\,.
\end{equation}
The Fourier transform is formally given by:
\begin{equation} \nonumber
    \vev{\phi(p)\phi(-p)}
    \propto
    \Gamma\!\left(\frac d2-\Delta_\phi\right)
    \left(p^2\right)^{\Delta_\phi - d/2}.
\end{equation}
When $d$ is even and $\Delta_{\phi}-d/2$ is a non-negative integer, the $\Gamma$-function develops a pole and the Fourier transform is ill-defined.
A convenient regularization scheme consists in analytically continuing the scaling dimension $\Delta_{\phi}=d/2+n-\varepsilon$, where $n\in \mathbb Z^{\ge 0}$.
Using:
\begin{equation} \nonumber
    \Gamma(-n+\varepsilon)
    =
    \frac{(-1)^n}{n!}
    \left(
        \frac1\varepsilon
        +
        \psi(n+1)
        +O(\varepsilon)
    \right)\,,
\end{equation}
together with:
\begin{equation} \nonumber
    (p^2)^{n-\varepsilon}
    =
    (p^2)^n
    \left(
        1-\varepsilon\log p^2+O(\varepsilon^2)
    \right)\,,
\end{equation}
one finds that, after subtracting the divergent $1/\varepsilon$ contribution by means of a local counterterm, the renormalized correlator takes the form
\begin{equation} \nonumber
    \langle \phi(p)\phi(-p)\rangle_{\rm ren}
    \propto
    \left(p^2\right)^{\Delta_\phi-d/2}
    \left(
        \log\frac{p^2}{\mu^2}
        + c
    \right)\,.
\end{equation}
The logarithmic term is universal, whereas the constant $c$ is scheme-dependent since it can be modified by finite local counterterms.

The same phenomenon occurs at finite temperature.
In particular, the identity contribution to the thermal correlator coincides with the zero-temperature correlator and therefore inherits the same logarithmic structure whenever the corresponding scaling dimensions satisfy the condition above.
This explains the origin of the divergences appearing in~\eqref{eq:EH-momentum} and~\eqref{eq:GMI-momentum}, as well as their proper treatment. We illustrate the specific mechanism for a generic thermal Polyakov block in the case $k=0$: the block with explicit coefficients reads:
\begin{equation} \nonumber
    \Fm \left[ F_\Om \right] (\omega,0)
    =
     (-1)^{J/2} 4^{\nu-s}
    \frac{\Gamma\!\left(\nu\right)
    \Gamma\!\left(\nu+1-s+J/2\right)
    }{
    \Gamma\!\left(s+J/2\right)
    }
    (-\omega^2)^{\,s-\nu-1}
    \ .
\end{equation}
Whenever $\nu+1-s+J/2= -m$ with $m \in \mathbb{Z}^{\geq 0}$, the $\Gamma$-function in the numerator develops a pole. After a regularization, the scheme-independent part of the block reads:
\begin{equation} \nonumber
    \Fm \left[ F_\Om \right] (\omega,0)
    =
    - \frac{1}{4\, \Gamma(m+1)} \frac{\Gamma\!\left(\nu\right)
    }{
    \Gamma\!\left(s+J/2\right)
    }\left(\frac{\omega^2}{4}\right)^{J/2+m}
    \log\left(-\frac{\omega^2}{4}\right)
    \ .
\end{equation}

Throughout this paper we write all formulas assuming generic scaling dimensions.
Whenever $\Delta_{\phi}-d/2$ becomes a non-negative integer, the corresponding expressions should be understood in the renormalized sense, with the appearance of logarithmic terms encoding the associated conformal anomalies.

\subsection{Application: Virasoro primaries with $\Delta_\phi=1$}
\label{subsec:2dBorel}

As a simple two-dimensional example, let us consider the thermal two-point function of a Virasoro primary with scaling dimension
\(
\Delta_\phi=1
\)
on the Euclidean cylinder with $\beta = 1$. The exact finite-temperature correlator in position space is given by
\begin{equation}
g(\tau,x)
=
\langle
\phi(z,\bar z)\phi(0,0)
\rangle_\beta
=\pi^2
\csc\!\left(\pi z\right)
\csc\!\left(\pi\bar z\right),
\label{eq:2dexact}
\end{equation}
where
\(
z=\tau+ix
\)
and
\(
\bar z=\tau-ix
\). On the thermal cylinder, only operators in the vacuum module can acquire non-vanishing one-point functions and these can be extracted explicitly from \eqref{eq:2dexact} using the OPE given in \eqref{eq:2dOPE}. 

With respect to the decomposition in thermal Polyakov blocks, we consider only descendants of the vacuum with vanishing twist, i.e.,
\(\Delta = J\) \cite{Barrat:2025nvu}, since their contribution to the discontinuity is non-zero. In the case
\(
\Delta_\phi=1
\),
the OPE coefficients are given as:
\begin{equation}
a^{2d}_J
=
-4\,
\operatorname{Li}_J(-1)=4 \left(1-2^{1-J}\right) \zeta (J).\label{eq:2dopecoef}
\end{equation}
The Fourier transform of the exact Euclidean correlator, after analytic continuation to the complex-$\omega$ plane using~\eqref{eq:euctoret}, takes the form
\begin{equation}
g_R(\omega,k)
=
-\pi
\left[
\psi
\!\left(
\frac12
-
\frac{i\omega}{4\pi}
-
\frac{i k}{4\pi}
\right)
+
\psi
\!\left(
\frac12
-
\frac{i \omega}{4\pi}
+
\frac{i k}{4\pi}
\right)
\right],
\label{eq:2dexactmom}
\end{equation}
where $\psi(z)$ denotes the digamma function. The poles of the retarded correlator above form the exact quasinormal spectrum \cite{Son_2002}:\footnote{Note that in the following part of this work, we use $\omega_n$ to denote the QNM spectrum of the retarded correlator and avoid using this notation for Euclidean Matsubara frequencies.}
\begin{equation}
\omega_n^\pm(k)
=
\pm k
-
2\pi i (2n+1),
\qquad
n=0,1,2,\ldots\;.
\label{eq:qnm2d}
\end{equation}
This example is particularly instructive: since the thermal data are known exactly, it is an ideal testing ground for the thermal Polyakov block decomposition \eqref{eq:mom-decomposition} developed in the previous section. Similarly to what we did for the large $N$ $\mathrm{O}(N)$ model in Section \ref{subsec:ApplicationLargeNONModel}, we can explicitly identify the non-perturbative contributions $\tilde{g}_{\text{arcs}}(\omega,k)$ that are not captured by thermal Polyakov blocks.

We recall the thermal Polyakov blocks expansion for $2d$ correlators:

\begin{equation} \nonumber
    g_R(\omega,k)=
    \sum_{J} \tilde{a}^{2d}_{J}  \,
    (k^2-\omega^2)^{\,s-1}
    \,
    T_J
    \!\left(\frac{i\omega}{\sqrt{k^2-\omega^2}}\right) + \tilde{g}_{\rm arcs}(\omega,k) \,, \quad \tilde{a}^{2d}_{J}=a^{2d}_{J} \; 4\pi \; \mathcal{N}_{\mathcal{O}}^{(0)} \ .
\end{equation}
Once we open the coefficients and we set $\Delta=J=2n$, the sum over thermal Polyakov blocks reads:
\begin{align}
    \tilde{g}_{\text{dr}}(\omega,k)
    &=
    \sum_{n=1}^{\infty} 4\pi
    (-1)^n
    \left(4^n-2\right)
    \zeta_{2n}\,
    \Gamma(2n)\,
    (k^2-\omega^2)^{-n}
    T_{2n}
    \!\left(
    \frac{i\omega}{\sqrt{k^2-\omega^2}}
    \right)\notag\\
    &\phantom{=\ }-\pi \log \left(\frac{k^2-\omega^2}{\mu^2}\right) \ ,
    \label{eq: deltaphi1blocks}
\end{align}
where the $n=0$ term, corresponding to the identity block, was regularized following the procedure described in the previous paragraph.

The sum over thermal Polyakov blocks diverges due to the factorial growth originating from the 
\(
\Gamma(2n)
\) factor above. Therefore this expansion is only asymptotic. Physically, this behavior is directly related to the existence of the infinite tower of QNMs accumulating at large frequency. Since the divergence is entirely controlled by the factorial growth of the coefficients, the series is Borel transformable. For simplicity, we restrict to the case
\(
k=0
\)
and we focus on the asymptotic series appearing in the thermal Polyakov block decomposition \eqref{eq: deltaphi1blocks}:
\begin{equation} \label{eq: Fomega}
    F(\omega)=\sum_{n=1}^{\infty} 4\pi
\left(4^n-2\right)
\zeta_{2n}\,
\Gamma(2n)\,
\omega^{-2n} \ .
\end{equation}
We consider the generalized Borel transform of the series \eqref{eq: Fomega}:
\begin{equation} \label{eq: boreltr}
    \mathcal{B}[F](\xi)=\sum_{n=1}^{\infty} \frac{\tilde{a}^{2d}_{2n}}{\Gamma(2n)}\, \xi^{2n-1}
 =
\frac{
2\pi\left(
2\pi\xi\,\csc(2\pi\xi)-1
\right)
}{\xi}.
\end{equation}
The series $F(\omega)$ can then be determined by evaluating the Laplace transform:
\begin{equation} \nonumber
    F(\omega)
=
\int_0^{\infty}
\text{d}\xi\,
e^{-\omega\xi}
\mathcal B[F](\xi) \,.
\end{equation}
This integral converges for $\text{Re} \omega > 0$ and can be analytically continued to other regions.
The function \eqref{eq: boreltr} possesses simple poles located at
\[
\xi=\frac{m}{2},
\qquad
m\in\mathbb Z\setminus\{0\},
\]
lying directly on the Borel contour. By adopting the $\pm i0$ prescription in Borel space, the integral can be written as follows:
\begin{equation} \nonumber
F(\omega)
=
\operatorname{PV}
\int_0^\infty
\text{d}\xi\,
e^{-\omega\xi}
\mathcal B[F](\xi)
\pm
2\pi^2 i\,
\frac{e^{-\omega/2}}
{1+e^{-\omega/2}},
\end{equation}
where the second term arises from the sum of the residues at the poles lying on the contour.
These residue contributions are known as Stokes jumps.
The principal-value Borel sum can be evaluated exactly:
\begin{equation} \nonumber
\operatorname{PV}
\int_0^\infty
\text{d}\xi\,
e^{-\omega\xi}
\mathcal B[F](\xi)
=
\pi\log \frac{\omega^2}{16 \pi^2}
-\pi
\left[\psi
\!\left(\frac12+\frac{i\omega}{4\pi}
\right)+\psi
\!\left(\frac12-\frac{i\omega}{4\pi}
\right)\right] \ ,
\end{equation}
and the full series $F(\omega)$ can be rewritten as follows: 
\begin{equation} \nonumber
    F(\omega)=\pi\log \left(\frac{\omega^2}{16 \pi^2}\right)
-\pi
\left[\psi
\!\left(\frac12+\frac{i\omega}{4\pi}
\right)+\psi
\!\left(\frac12-\frac{i\omega}{4\pi}
\right)\right]\pm
2\pi^2 i\,
\frac{e^{-\omega/2}}
{1+e^{-\omega/2}}
\end{equation}
Finally, we can use the identity:
\begin{equation} \nonumber
    \pi \psi 
\!\left(\frac12+\frac{i\omega}{4\pi}
\right)=\pi \psi
\!\left(\frac12-\frac{i\omega}{4\pi}
\right)+i \pi^2-2 \pi^2 i \frac{e^{-\omega/2}}{1+e^{-\omega/2}}
\end{equation} 
to finalise the expression for our Borel-resummed series: 
\begin{equation} \nonumber
    F(\omega)=\pi\log \left(-\frac{\omega^2}{16 \pi^2}\right)-2\pi \psi
\!\left(\frac12-\frac{i\omega}{4\pi}
\right)+2 (1\pm 1) \pi^2 i \frac{e^{-\omega/2}}{1+e^{-\omega/2}} \ .
\end{equation}
After adding the identity contribution ($n=0$) and an appropriate local counterterm to regularize it, the thermal Polyakov blocks decomposition~\eqref{eq: deltaphi1blocks} reads as follows:
\begin{equation} \nonumber
    \tilde{g}_{\text{dr}}(\omega,0)=-2\pi \psi
\!\left(\frac12-\frac{i\omega}{4\pi}
\right)+2 (1\pm 1) \pi^2 i \frac{e^{-\omega/2}}{1+e^{-\omega/2}} \ ,
\end{equation}
and we can obtain an exact expression for the $\tilde{g}_{\text{arcs}}$ contribution by subtracting this from the exact correlator \eqref{eq:2dexactmom} in the $k=0$ limit:
\begin{align} \nonumber
\tilde{g}_\text{arcs}(\omega,0) = -2 (1\pm 1) \pi^2 i \frac{e^{-\omega/2}}{1+e^{-\omega/2}}=-2 (1\pm 1) \pi^2 i\sum_{m=0}^{\infty} (-1)^m e^{-\frac{m+1}{2}\omega}
\end{align}
This is consistent with the initial definition of $g_\text{arcs}$ in the position-space dispersion relation given in \eqref{eq:GMI_FullCorrelator}, because up to a constant factor, it is exponentially suppressed at large frequency and therefore cannot be generated by the momentum-space OPE.
Note that the choice of integration prescription in Borel space changes the result: the $-i 0$ prescription yields exact agreement between the thermal Polyakov decomposition and the exact retarded correlator, while the $+i0$ prescription identifies a non-trivial arc term, necessary to compensate the Stokes jumps.

\section{Inversion formulae and complex $\omega$-plane: zero momentum}\label{sec:UVIR}

This section is devoted to the derivation and implications of the inversion formula following from Equation~\eqref{eq:momentumOPE}. We begin with the simpler case of vanishing spatial momentum, $k=0$. In this setting, we show that, under minimal assumptions on the analytic structure of the correlator in the complex $\omega$-plane, the inversion formula naturally leads to a bootstrap problem for the QNMs, expressing them in terms of a finite set of thermal OPE data. We then investigate the consequences of this construction. In particular, assuming a suitable asymptotic behavior of the QNM spectrum at large frequencies, we derive a complementary bootstrap problem in which the thermal OPE data determine the large-$n$ behavior of the QNM frequencies $\omega_n$, while the corresponding residues are constrained by an infinite set of sum rules.

Let us consider a generic CFT and specialize the momentum-space OPE in Equation ~\eqref{eq:momentumOPE} to the case of vanishing spatial momentum, $k=0$. The expansion then takes the form:\footnote{The phase is defined by choosing the principal branch for the logarithm, i.e., $\log(-z)=\log|z|+i \pi$. We will follow this convention throughout this section.} 
\begin{equation}\label{eq:OPEk = 0simply}
g_{R}(\omega)= g_R(\omega,0)
=\sum_{\Delta} \frac{\tilde{a}_\Delta}{\beta^{\Delta}} \, e^{ \pi i (s-d/2)} \, \omega^{2\Delta_\phi-\Delta-d}
+\tilde{g}_{\rm arcs}(\omega,0)\;,
\end{equation}
where $\tilde{a}_\Delta$ is the weighted OPE coefficient defined in momentum space in~\eqref{eq:aDelta-momentum}. The sum runs over all primary operators $\mathcal O_{\Delta,J}$ with scaling dimension $\Delta$. 
From now on we set $\beta = 1$.
Factors of $\beta$ are easy to reinstate by dimensional analysis.

We now make the minimal assumption that the retarded correlator is a meromorphic function of the complex frequency $\omega$. In particular, we assume that its singularities consist exclusively of isolated simple poles, with no branch cuts or essential singularities. This assumption implies that the retarded correlator admits the representation:
\begin{equation} \nonumber
g_R(\omega)
=\sum_n\frac{r_n}{\omega-\omega_n} \, \ ,
\end{equation}
where $\omega_n$ denote the QNM frequencies and $r_n$ their corresponding residues. The set of QNMs may be finite or, more generally, infinite (see discussion later in Section \ref{subsec:largeomegapert}).
We later consider the non-meromorphic case by including branch cuts in Section~\ref{subsec:beyondmeromorphicity}.

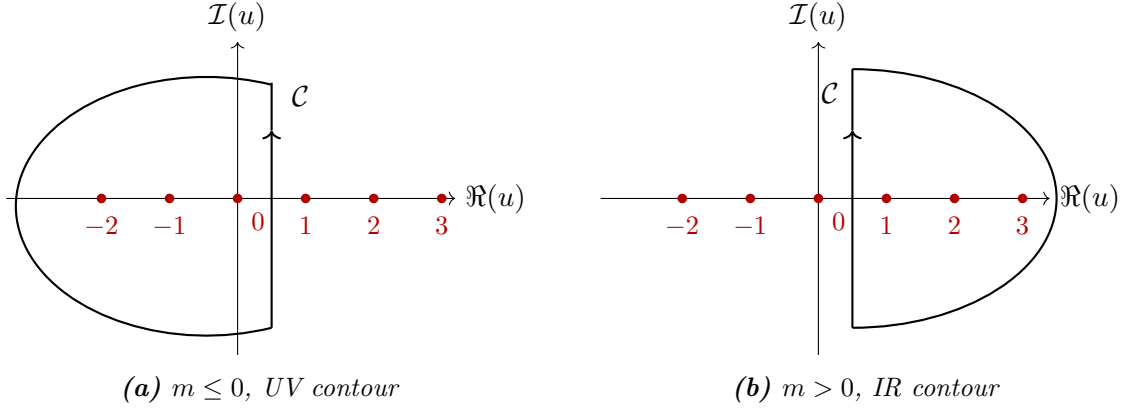
\begin{figure}[ht]
  \centering
  
  \begin{minipage}[t]{0.47\textwidth}
    \centering
    \begin{tikzpicture}[scale=0.9]
      \draw[->] (-3.4,0) -- (3.2,0) node[right] {$\Re(u)$};
      \draw[->] (0,-2.3) -- (0,2.3) node[above] {$\Im(u)$};

      \foreach \x/\lab in {-2/$-2$,-1/$-1$,1/$1$,2/$2$,3/$3$}
        \fill[red!70!black] (\x,0) circle (2pt) node[below=3pt] {\small \lab};
      \fill[red!70!black] (0,0) circle (2pt) node[below right=2pt] {\small $0$};
      \node[left] at (1.2,1.5) { $\mathcal C$};

      \draw[thick] (0.5,-1.9) arc[start angle=-70,end angle=-290,x radius=2.8,y radius=1.9];

      \draw[thick] (0.5,1) -- (0.5,1.7) ;

      \draw[thick,->] (0.5,-1.9) -- (0.5,1) ;
    \end{tikzpicture}
    \subcaption{$m\leq0$, UV contour}
  \end{minipage}
  \hfill
  \begin{minipage}[t]{0.47\textwidth}
    \centering
    \begin{tikzpicture}[scale=0.9]
      \draw[->] (-3.2,0) -- (3.4,0) node[right] {$\Re(u)$};
      \draw[->] (0,-2.3) -- (0,2.3) node[above] {$\Im(u)$};

      \foreach \x/\lab in {-2/$-2$,-1/$-1$,1/$1$,2/$2$,3/$3$}
        \fill[red!70!black] (\x,0) circle (2pt) node[below=3pt] {\small \lab};
      \fill[red!70!black] (0,0) circle (2pt) node[below right=2pt] {\small $0$};

      \draw[thick,->] (0.5,-1.9) -- (0.5,1);
      \draw[thick] (0.5,1) -- (0.5,1.9);
      \node[left] at (0.45,1.55) { $\mathcal C$};

      \draw[thick] (0.5,1.9) arc[start angle=90,end angle=-90,x radius=3.0,y radius=1.9];
      
    \end{tikzpicture}
    \subcaption{$m>0$, IR contour}
  \end{minipage}
  \caption{The same Mellin--Barnes contour in the complex $u$-plane can be deformed in two different ways. The red dots are the kinematic poles of $\pi/\sin(\pi u)$ at integer $u$. Closing to the right gives the Taylor expansion around $\omega=0$, while closing to the left gives the large-$\omega$ asymptotic expansion and, when present, the OPE poles coming from $\mathcal Z(u)$.}
  \label{fig:u-plane-closures}
\end{figure}

The latter expression can be conveniently rewritten in terms of a Mellin--Barnes representation:  
\begin{equation}\label{eq:MellinBarnsk=0}
g_R(\omega)
=
\int_{\mathcal C}\frac{\text{d}u}{2\pi i}\,
\Gamma(u)\Gamma(1-u) \,
\omega^{u-1}
\sum_n r_n^{\vphantom{-u}} (-\omega_n)^{-u}\,,
\end{equation}
where $\mathcal C$ is a vertical contour lying inside the fundamental strip \(0<\operatorname{Re}(u)<1\). The fundamental strip is determined with respect to the Gamma functions that constitute the kernel of the Mellin--Barnes integral. Depending on whether we consider the large- or small-$\omega$ expansion, we close the contour on the left or on the right, as shown in Figure~\ref{fig:u-plane-closures}.

Comparing the OPE expansion in Equation~\eqref{eq:OPEk = 0simply} with Equation~\eqref{eq:MellinBarnsk=0}, it follows immediately that a term proportional to
$\omega^{2\Delta_\phi-\Delta-d}$ arises upon deforming the contour to the left and picking up the residue at:
\begin{equation} \nonumber
u=2\Delta_\phi-\Delta-d+1.
\end{equation}
Closing the contour on the left, as shown in Figure \ref{fig:u-plane-closures}, we therefore obtain the inversion formula:
\begin{equation}
\tilde a_\Delta
= e^{ -i \pi  (s-d/2)} \operatorname*{Res}_{u=2s-d+1}
\left[
\frac{\pi e^{- i \pi u}}{\sin(\pi u)}
\,\mathcal Z(u)
\right] \ ,
\quad \text{with}\quad
\mathcal Z(u)
=
\sum_n r^{\vphantom{u}}_n\,\omega_n^{-u}, \label{eq:kzeroinversion}
\end{equation}
where the overall phase is inherited from \eqref{eq:OPEk = 0simply}. In \cite{NewProjectCristo} an equivalent expression is derived directly in position space by requiring the OPE to admit a decomposition in terms of power laws. 

The contribution from the arc term $\tilde{g}_{\rm arcs}$ does not affect the inversion formula. This is the case because, by construction, it consists of exponentially suppressed terms, which do not generate poles in the Mellin variable $u$ and therefore do not contribute to the residues appearing in the inversion formula. For this reason, we will omit the arc contribution throughout the present discussion. An analogous statement holds for the inversion formula at generic spatial momentum $k$, although the corresponding argument is more subtle.

The function $\mathcal Z(u)$ plays a distinguished role in the inversion formula. Indeed, it naturally packages the complete information carried by the quasinormal-mode spectrum and its residues into a single analytic object. This suggests that $\mathcal Z(u)$ is the fundamental quantity associated with the QNM spectrum, and motivates a detailed study of its analytic properties. In the remainder of this section we will investigate these properties and show how they encode the thermal OPE data.

The inversion formula immediately implies a set of non-trivial constraints on the function $\mathcal Z(u)$. Indeed, the prefactor $1/\sin(\pi u)$ possesses simple poles at every integer value $u=m\in\mathbb Z$. As we will now see, the poles with $m<0$ and those with $m\ge0$ have completely different physical interpretations. This distinction reflects the fact that closing the Mellin contour to the left or to the right probes different kinematic regimes of the correlator (see Figure \ref{fig:u-plane-closures}). Closing the contour to the left reproduces the ultraviolet expansion, namely the thermal OPE, whereas closing it to the right generates the infrared expansion, corresponding to the Taylor series around $\omega=0$.

\paragraph{$m<0$, UV contour.} Let us first consider the left half-plane, corresponding to negative integers $m<0$. There are two qualitatively distinct possibilities.

(1) The first possibility is that there exists a pole at $u=m<0$, associated with a primary operator whose scaling dimension satisfies:
\begin{equation} \nonumber
\Delta=2\Delta_\phi-d+1-m \ ,
\end{equation}
for $m\in\mathbb Z^{<0}$. In this case the pole is physical, as it corresponds to an operator appearing in the thermal OPE. Matching the inversion formula with the OPE immediately fixes the value of $\mathcal Z(u)$ at the corresponding integer:
\begin{equation} \nonumber
\mathcal Z(m)
=
e^{i\pi \frac{m-1}{2}}
\tilde{a}_{2\Delta_\phi-d+1-m} \ .
\end{equation}
Thus, whenever an operator contributes at an integer value of the Mellin variable, the corresponding value of $\mathcal Z(u)$ is entirely determined by its thermal OPE coefficient. 

There is an important special case of this discussion in which the inversion formula requires a slight modification.
Suppose that an operator contributes at a negative integer value \[
u=m \ ,\qquad
m\in\mathbb Z_{<0} \ , \]
and that, at the same point, the function $\mathcal Z(u)$ itself develops a simple pole. Since the universal kernel:
\[
\frac{\pi}{\sin(\pi u)}
\]
already has a simple pole at every integer, the integrand of the Mellin--Barnes representation develops a double pole.

From the point of view of the inversion formula, this situation appears singular. However, it has a clear physical interpretation. Indeed, performing explicitly the Mellin--Barnes transform of a double pole shows that the corresponding contribution to the retarded correlator is:
\begin{equation} \nonumber
g_R(\omega)
\supset
\omega^{m-1}\log\omega.
\end{equation}
Thus, the double pole generates logarithmic corrections to the large-frequency expansion. These terms are not captured by the generic OPE expansion \eqref{eq:OPEk = 0simply}, which consists only of pure powers of $\omega$. Their appearance is therefore not a failure of the inversion formula but rather a signal that the asymptotic expansion must be enlarged to include logarithmic terms.

From the CFT point of view, the emergence of such logarithms is expected. They are the momentum-space manifestation of conformal anomalies, discussed also in Section~\ref{sec:OPEInMomentumSpace}, which occur whenever scale invariance is broken by anomalous terms in the renormalization of composite operators. The double poles of the Mellin--Barnes integrand therefore provide a simple analytic criterion for identifying anomalous logarithms directly from the quasinormal-mode generating function $\mathcal Z(u)$.

(2) The second possibility is that no operator satisfies:
\begin{equation} \nonumber
\Delta=2\Delta_\phi-d+1-m \ ,
\end{equation}
for $m\in\mathbb Z^{<0}$. In this case the pole generated by the Mellin kernel is unphysical and must therefore be cancelled. This requires the residue to vanish, implying:
\begin{equation} \nonumber
\mathcal Z(m)=0 \ .
\end{equation}
Equivalently, one obtains the infinite family of sum rules:
\begin{equation} \nonumber
\sum_k r_k^{\vphantom{-m}}\,\omega_k^{-m}=0 \ ,
\end{equation}
which impose highly non-trivial constraints on the spectrum of quasinormal modes and their residues. In this way, the absence of operators at specific scaling dimensions is translated into exact relations satisfied by the quasinormal mode data.
Exploring this relation would be very interesting, and we leave it for future work.

The last possible scenario we should consider for the UV contour is when there is a pole at $u \notin \mathbb{Z}$, associated with an operator for which the quantity $2\Delta_\phi-\Delta-d+1$  is not an integer: the $\sin(\pi u)$ factor in the denominator does not produce any pole, therefore, the function $\mathcal Z(u)$ must supply it.
We conclude that $\mathcal Z(u)$ contains the following term:
\begin{equation} \nonumber
    \mathcal Z(u) \supset \frac{\sin(\pi u)}{\pi}  e^{i \pi (s-d/2+u)}  \frac{\tilde{a}_\Delta}{u-\left(2\Delta_\phi-\Delta-d+1\right)} \ . 
\end{equation}
This formula directly relates OPE data to the spectrum of QNMs. Such a relation must exist, since under the assumptions above both sets of data completely determine the correlator. Nevertheless, the way in which the OPE data are reorganized into the QNM expansion is highly non-trivial.
In the following, we will study this expression in a few examples.

\paragraph{$m\geq0$, IR contour.} Let us now turn to the opposite case, namely positive integers $m\ge0$, corresponding to closing the contour in the right half-plane. In contrast to the previous case, the residues no longer reconstruct the ultraviolet OPE but instead generate the infrared expansion of the correlator around $\omega=0$. Indeed, assuming that the Taylor expansion exists, one finds:
\begin{equation}\label{eq:Taylorexpansionk=0}
g(\omega)
\overset{\omega \ll 1}{=}
-\sum_{p=0}^{\infty}\omega^p
\sum_n r_n^{\vphantom{-p}}\,\omega_n^{-p-1}
=
\sum_{p=0}^{\infty}\mu_p\,\omega^p \ ,
\end{equation}
from which it immediately follows that:
\begin{equation} \nonumber
\mu_p=-\mathcal Z(p+1) \ .
\end{equation}
The same conclusion follows directly from the Mellin--Barnes representation \eqref{eq:MellinBarnsk=0}: closing the contour to the right picks up precisely the poles at positive integers and reconstructs the Taylor expansion of the correlator. Therefore, the inversion formula provides a unified framework in which the left half-plane encodes the ultraviolet OPE data, while the right half-plane determines the infrared moments of the retarded correlator.
\paragraph{Summary: $\mathcal Z(u)$ properties.} Before proceeding, let us briefly summarize the main properties of the function $\mathcal Z(u)$:
\begin{itemize}
    \item[$\star$] (\textit{IR expansion}) The retarded correlator admits a Taylor expansion around the origin:
    \begin{equation} \nonumber
        g_R(\omega)
        =
        -\sum_{p=0}^{\infty}
        \mathcal Z(p+1)\,\omega^p \ ,
    \end{equation}
    which exists under the assumption that $g_R(\omega)$ is meromorphic. Its radius of convergence is expected to be finite and determined by the quasinormal mode closest to the origin in the complex $\omega$-plane.

    \item[$\star$](\textit{UV-expansion/sum rules})  If there is no primary operator with scaling dimension
    \begin{equation} \nonumber
        \Delta=2\Delta_\phi-d+1-m \ , \quad m \in \mathbb{Z}^{<0} \ ,
    \end{equation}
    this produces a zero:
    \begin{equation} \nonumber
        \mathcal Z(m)=0 \ .
    \end{equation}
    Conversely, whenever such an operator is present in the spectrum, $\mathcal{Z}$ evaluates to a thermal OPE coefficient:
    \begin{equation} \nonumber
        \mathcal Z(m)
=
e^{i\pi \frac{m-1}{2}}
\tilde{a}_{2\Delta_\phi-d+1-m} \ .
    \end{equation}
    In the presence of logarithms in the OPE blocks, due to conformal anomalies, $\mathcal Z(u)$ has a pole in $u=m<0$ whose residue is fixed by the thermal OPE coefficient.

    \item[$\star$](\textit{OPE limit})  Every primary operator with non-integer scaling dimension $\Delta$ gives rise to a simple pole of $\mathcal Z(u)$ located at:
    \begin{equation} \nonumber
        u=2\Delta_\phi-\Delta-d+1 \ ,
    \end{equation}
    with residue fixed by the corresponding thermal OPE coefficient,
    \begin{equation} \nonumber
        \mathcal Z(u) \supset \frac{\sin(\pi u)}{\pi}  e^{i \pi (s-d/2+u)} \frac{\tilde{a}_\Delta}{u-\left(2\Delta_\phi-\Delta-d+1\right)}  \  .
    \end{equation}
\end{itemize}
At fixed $k \neq 0 $, the analysis above can be performed as well by considering the OPE and expanding the kinematical structure in a series to obtain a power-series in $\omega$ at large $\omega$ \cite{NewProjectCristo}.

\subsection{Finite numbers of QNMs}
Let us first consider the simplest scenario, in which the retarded correlator possesses only a finite number $M$ of quasinormal modes. In this case the inversion formula reduces to:
\begin{equation} \nonumber
\tilde a_\Delta
= e^{ -\pi i (s-d/2)} \operatorname*{Res}_{u=2s-d+1}
\left[
\frac{\pi e^{-i \pi u}}{\sin(\pi u)}
\,\sum_{n=1}^{M} r^{\vphantom{u}}_n\,\omega_n^{-u}
\right] \,.
\end{equation}
Since the sum
\begin{equation} \nonumber
\mathcal Z(u)=\sum_{n=1}^{M} r^{\vphantom{u}}_n\,\omega_n^{-u}
\end{equation}
contains only finitely many terms, it defines an entire function of $u$. Consequently, the only singularities of the integrand originate from the universal Mellin kernel $1/\sin(\pi u)$, whose poles are located at integer values of $u$. It follows immediately that the thermal OPE can contain only operators whose scaling dimensions satisfy:
\begin{equation} \nonumber
2\Delta_\phi-\Delta-d+1\in\mathbb Z \ .
\end{equation}

It is important to emphasize that the converse statement does not hold. While a finite number of quasinormal modes necessarily implies an integer spectrum of scaling dimensions, the existence of such a spectrum does not imply that the number of quasinormal modes is finite. Indeed, an infinite collection of quasinormal modes may combine into an entire function $\mathcal Z(u)$ while still reproducing an OPE consisting exclusively of operators at integer values of the Mellin variable.

In the case under study the thermal OPE coefficients are completely determined by the quasinormal mode data:
\begin{equation} \label{eq: k=0 inversion}
\tilde a_\Delta
= e^{ -\pi i (s-d/2)} 
\,\sum_{n=1}^{M} r^{\vphantom{-u}}_n\,\omega_n^{-2\Delta_\phi+\Delta+d-1} \ .
\end{equation}
Conversely, since the unknown data consist only of the finite set of frequencies $\omega_n$ and residues $r_n$, a finite number of thermal OPE coefficients is sufficient, at least in principle, to reconstruct the complete quasinormal mode spectrum.

The infrared expansion is equally simple. The Taylor coefficients of the retarded correlator around $\omega=0$ are given in Equation~\eqref{eq:Taylorexpansionk=0}, with
\begin{equation} \nonumber
\mu_p
=
-\sum_{n=1}^{M}
r_n\,\omega_n^{-p-1}
=
-\mathcal Z(p+1)  \ .
\end{equation}
Thus, in the finite-QNM case, both the ultraviolet data (the thermal OPE coefficients) and the infrared data (the Taylor coefficients) are determined by the same finite set of quasinormal modes.

Although this situation is rather special, it is nevertheless realized in interesting examples. In particular, the $\mathrm O(N)$ model at large $N$ provides an explicit realization of a retarded correlator with only finitely many quasinormal modes. 

\subsection{Application: $\mathrm{O}(N)$ model at large $N$} To test the finite-QNM scenario, we consider the $3d$ $\mathrm{O}(N)$ model at large $N$ in the $k=0$ limit. In this case, the thermal mass of the theory corresponds to a pole on the real positive axis in the $\omega$-plane. In fact, the thermal mass and its negative counterpart are the only two QNMs of the $\mathrm{O}(N)$ model at large $N$. The momentum-space retarded correlator is given for $k=0$ by:
\begin{equation}
    g_{R}(\omega)=\frac{1}{-\omega^2+m_{\text{th}}^2}=\frac{1}{2m_{\mathrm{th}}} \left(\frac{1}{\omega-m_{\mathrm{th}}}
-
\frac{1}{\omega+m_{\mathrm{th}}} \right) \ .
\label{eq:ONkzero}
\end{equation}
From this expression we can read off $r_n$ and $\omega_n$:
\begin{equation}
\omega_\pm
=
\pm m_{\mathrm{th}} \ , \qquad  r_\pm
=
\mp\frac{1}{2 m_{\mathrm{th}}} \ .
\end{equation}
In particular, they obey the reality condition $\omega_\pm^*=-\omega_\mp$. Using the $k=0$ inversion formula \eqref{eq: k=0 inversion} derived above we conclude that
\begin{align} \nonumber
    \tilde a_\Delta=\frac{1+(-1)^{\Delta}}{2} \ (-1)^{\frac{\Delta}{2}}m_{\text{th}}^{\Delta}\;.
\end{align}
The selection rule ensures that $\Delta=2m$, with $m \in \mathbb Z^{\geq 0}$. Hence, the final result reads
\begin{align} \nonumber
    \tilde a_{2m}=(-1)^{m}m_{\mathrm{th}}^{2m}\;,
\end{align}
which are precisely the momentum-space OPE coefficients of the operators $[\sigma^m]$ contributing to the discontinuity of the Euclidean correlator.

We will continue our investigation of the $\mathrm{O}(N)$ model at large $N$ in Section \ref{subsec:ONlargeN}.

\subsection{Large $\omega$ perturbation theory for QNMs}
\label{subsec:largeomegapert}
Since, as we discussed above, a finite number of QNMs would imply an integer spectrum together with very special OPE coefficients, the most generic situation is that the number of QNMs is infinite. This is indeed what we expect in the vast majority of cases. Conformal dimensions are generically non-integer, so a finite QNM expansion would be too restrictive. Even when the spectrum happens to be integer, as in some holographic theories, the number of QNMs is still typically infinite.

 If the number of QNMs is infinite and the spectrum contains operators of non-integer dimension, the main constraints come from the fact that $\mathcal Z(u)$ has a pole for each operator.
 In order to proceed, we first make a simplifying assumption which will help us show the consequences of the inversion formula. 
 
\paragraph{Asymptotics of QNMs.}
In general, very little is known about the asymptotic distribution of quasinormal modes in generic QFTs or CFTs. Nevertheless, all known examples exhibit a remarkably simple pattern: at large mode number $n$, the quasinormal frequencies tend to organize themselves into one or more asymptotic trajectories in the complex frequency plane. Motivated by these observations, we shall make the minimal assumption that the high-energy quasinormal modes lie on a single asymptotic trajectory. More precisely, we assume that:
\begin{equation}\label{eq:omeganasymp}
\omega_n
\sim
r\,e^{i\theta}\,n^\delta \left(1+\frac{r^{(1)}e^{i \theta^{(1)}}}{n}+\ldots \right) \ ,
\qquad n\rightarrow\infty \ ,
\end{equation}
where $r>0$, $\theta$ and $\delta$ are constants. Such an asymptotic behavior is ubiquitous in holographic examples and in the few exactly solvable cases currently available (see, e.g., \cite{Dodelson:2023vrw}). 

To characterize the large-$n$ behavior completely, we also need an assumption on the residues. For simplicity, we assume that they follow a power-law scaling:
\begin{equation}\label{eq:residueasympto}
r_n
\sim
\,n^\gamma \left(\alpha+\frac{\alpha^{(1)}}{n}+\ldots \right),
\qquad n\rightarrow\infty \ ,
\end{equation}
where $\alpha$, $\alpha^{(1)}$, and $\gamma$ are constants.
This represents the simplest possible ansatz compatible with the expected asymptotic behavior of the correlator.

More general scenarios can certainly be considered. For instance, the quasinormal modes may split into several asymptotic trajectories, each characterized by its own parameters $(r,\theta,\delta)$ and residue scaling. Since all the explicit examples discussed in this paper involve a single trajectory, we restrict ourselves to this case and leave the multi-trajectory analysis for future work.

\paragraph{UV perturbation theory.}

Let us now turn to the asymptotic regime of large mode number. A natural question is whether the inversion formula can be used to develop a systematic large-$n$ expansion of the quasinormal mode contribution. The answer is affirmative, at least at the level of a formal asymptotic expansion. Throughout this section we assume the asymptotic behavior introduced in Equations~\eqref{eq:omeganasymp} and~\eqref{eq:residueasympto}. The corresponding expansion coefficients are, of course, theory dependent.

Expanding the quasinormal frequencies for large $n$, one finds:
\begin{align}
\omega_n^{-u}
&=
r^{-u}e^{-i\theta u}\,
n^{-u \, \delta}
\left(
1+\frac{r^{(1)}e^{i\theta^{(1)}}}{n}
+\cdots
\right)^{-u}
\nonumber\\
&=
r^{-u}e^{-i\theta u}\,
n^{-u \, \delta }
\left[
1
-\frac{u\,r^{(1)}e^{i\theta^{(1)}}}{n}
+\ldots 
\right] \ , \nonumber
\end{align}
where the second line follows from a large-$n$ expansion.
Multiplying by the corresponding asymptotic expansion of the residues gives:
\begin{align}
r_n^{\vphantom{-u}}\,\omega_n^{-u}
=
r^{-u}e^{-i\theta u}
n^{\gamma-u \,\delta }
\left[
A_0(u)
+\frac{A_1(u)}{n}
+\cdots
\right], \nonumber
\end{align}
where the coefficients $A_k(u)$ are polynomials in $u$ determined by the asymptotic expansions of both $\omega_n$ and $r_n$. The first two coefficients are
\begin{align}
A_0(u)&=\alpha, \nonumber\\ \nonumber
A_1(u)&=\alpha^{(1)}
-u\,\alpha\, r^{(1)}e^{i\theta^{(1)}} \ .
\end{align}
Reality of the retarded function implies $g_R(-\omega^\star) =g_R^\star(\omega)$, which means that the QNMs appear in pairs $(\omega_n,-\omega_n^\star)$, with residues $(r_n, -r_n^{\star})$. Therefore, we also need to consider the contributions of the reflected tail.
We first consider the conjugate QNMs in the large-$n$ limit:
\begin{equation} \nonumber
    -\omega^{\star}_n
\sim
r\,e^{i(\pi -\theta)}\,n^\delta \left(1+\frac{\left(r^{(1)}\right)^{\star}e^{-i \theta^{(1)}}}{n}+\ldots \right) \ ,
\end{equation}
where $q \in 2\mathbb{Z}+1$ selects the branch. The choice becomes important when we consider the generic term in the $\mathcal{Z}(u)$ function associated with the conjugate QNMs:
\begin{equation} \nonumber
    -r_n^\star (-\omega_n^\star)^{-u} =- r^{-u} e^{-i (\pi -\theta) u }n^{\gamma-u \, \delta} \left[\tilde A_0(u) + \frac{\tilde A_1(u)}{n}+\ldots \right] \ ,
\end{equation}
where $\tilde A_n(u) = A_n^\star(u^\star)$ are obtained by complex-conjugating all coefficients while leaving $u$ unchanged, explicitly:
\begin{align}
\tilde A_0(u)&=\alpha^\star, \nonumber\\
\tilde A_1(u)&=\left(\alpha^{(1)}\right)^\star
-u\,\alpha^\star \left(r^{(1)}\right)^\star e^{-i \theta^{(1)}}\ . \nonumber
\end{align}
Summing term by term over the large-$n$ tail naturally produces an expansion in terms of Riemann zeta functions:
\begin{equation}
\mathcal Z(u)
\sim 
r^{-u}
\sum_{\ell}
\left[A_\ell(u) e^{-i\theta u}- \tilde A_\ell(u) e^{-i(\pi -\theta) u}\right]\,
\zeta(u \, \delta-\gamma+\ell) \ ,
\label{eq:zetaexpansion}
\end{equation}
which should be regarded as a formal asymptotic expansion. More precisely, Equation~\eqref{eq:zetaexpansion} captures only the contribution of the asymptotic quasinormal modes, while the finite number of low-lying modes must be treated separately.

Whether the series over $\ell$ converges or should instead be interpreted only asymptotically depends on the analytic properties of the large-$n$ expansion of the quasinormal frequencies and residues. In particular, if $\omega_n$ and $r_n$ admit convergent expansions in powers of $1/n$, the series above is expected to converge in the corresponding domain. Otherwise, it should be interpreted as an asymptotic expansion, in the same spirit as many semiclassical or large-order expansions encountered in quantum field theory.

It is now straightforward to identify the physical origin of the poles appearing in Equation~\eqref{eq:zetaexpansion}. Since the Riemann zeta function possesses a single pole at unit argument, each term in the expansion contributes a pole whenever:
\begin{equation}
u \, \delta-\gamma+\ell=1 \ . \nonumber
\end{equation}
These singularities must coincide with the poles of $\mathcal Z(u)$ associated with primary operators of non-integer scaling dimension. Therefore, the asymptotic distribution of quasinormal modes is directly constrained by the spectrum of the thermal OPE.
Keeping only the leading contribution ($\ell=0$), the first two operators with \emph{non-integer} dimensions $\Delta_0'$ and $\Delta_1'$ already determine the asymptotic exponents, 
\begin{equation}
\delta
=
\frac{1}{\Delta_1'-\Delta_0'},
\qquad
\gamma
=
\frac{2\Delta_\phi-2\nu-1-\Delta_0'}
{\Delta_1'-\Delta_0'}
-1\ , \nonumber
\end{equation}
where the prime indicates that these are the operators with non-integer conformal dimensions associated with the poles of $\mathcal Z(u)$.\footnote{An operator may also be associated with a pole of $\Zm (u)$ through a conformal anomaly, as discussed above.}
In other words, the spacing of the leading non-integer operators fixes the asymptotic growth of the quasinormal frequencies, while the location of the first operator determines the overall power governing the residues.

Once $\delta$ and $\gamma$ have been determined, the residues of the poles of $\mathcal Z(u)$ provide further constraints on the remaining parameters, namely $r$, $\alpha$, $\theta$, and the subleading coefficients entering the large-$n$ expansions. However, the leading asymptotic analysis alone is not sufficient to determine all of these quantities uniquely.

A natural strategy is therefore to regard Equation~\eqref{eq:zetaexpansion} as an asymptotic rather than an exact representation. One truncates the expansion at a given order in $1/n$ and matches the resulting expression to as many thermal OPE data as are available. In this way the unknown asymptotic parameters can be determined order by order, systematically improving the approximation. We shall illustrate this procedure explicitly in the case of the $\mathcal N=4$ SYM $R$-current retarded correlator in the following.

It is important, however, to emphasize the limitations of the large-$n$ analysis. By construction, the $1/n$ expansion determines only the asymptotic tail of the quasinormal mode spectrum. It does not reconstruct the low-lying modes, nor does it automatically satisfy the exact sum rules:
\begin{equation}\label{eq:sumrulesform}
\sum_{n=1}^{\infty}r_n^{\vphantom{-m}}\,\omega_n^{-m}=0\ , \quad m \in \mathbb{Z}^{<0} \ ,
\end{equation}
which depend on the complete spectrum and are therefore sensitive to non-asymptotic data. In other words, the large-$n$ expansion provides local information in mode number, whereas the cancellation of the poles of the Mellin kernel imposes global constraints on the entire spectrum.

A natural strategy to overcome this limitation is inspired by the procedure introduced in \cite{Barrat:2024fwq}. Rather than neglecting the contribution of the asymptotic tail in Equation~\eqref{eq:sumrulesform}, one separates the spectrum into a finite number of low-lying modes and an asymptotic tail. More precisely, one keeps the first $N$ quasinormal modes exact, while replacing all modes with $n>N$ by their asymptotic expansions, Equations~\eqref{eq:omeganasymp} and \eqref{eq:residueasympto}.
The sum rules then become, for negative integer $m$, 
\begin{equation} \nonumber
\sum_{n=1}^{N}
r_n^{\vphantom{-m}}\,\omega_n^{-m}
+
r^{m}
\sum_{\ell=0}^{\infty}
\left[
A_{\ell}(m)e^{-im\theta}
-
\tilde A_{\ell}(m)e^{-i(\pi -\theta)m}
\right]
\zeta_H(m \, \delta-\gamma+\ell,N+1)
\approx 0 \ ,
\end{equation}
where $\zeta_H(x,y)$ is the Hurwitz $\zeta$-function already introduced above.

The advantage of this formulation is that the infinitely many unknown quasinormal modes are replaced by a finite set of asymptotic parameters together with the first $N$ exact modes. As a consequence, the infinite system of sum rules is reduced to a finite-dimensional problem that can be solved, at least approximately, by matching a sufficiently large number of OPE constraints. Increasing $N$ or including higher orders in the asymptotic expansion systematically improves the approximation.

\paragraph{Generalization to multiple asymptotic trajectories.}

The assumption of a single asymptotic family of quasinormal modes is highly restrictive. As we have seen, it implies that the poles of $\mathcal Z(u)$ generated by the asymptotic tail are organized into a single arithmetic progression:
\begin{equation} \nonumber
u=\frac{\gamma+1-\ell}{\delta},
\qquad \ell=0,1,\ldots,
\end{equation}
which translates into an equally rigid pattern for the non-integer scaling dimensions appearing in the thermal OPE. Therefore, if the spectrum of a given theory does not exhibit such a structure, one should conclude that the underlying quasinormal mode spectrum cannot be described by a single asymptotic trajectory.

There are several possible ways in which this simple picture may fail. The most natural possibility is that the quasinormal modes split into several asymptotic families, each characterized by its own asymptotic parameters. Alternatively, the large-$n$ expansion itself may have to be generalized, for example by including logarithmic or exponentially small corrections that are invisible in a pure $1/n$ expansion.

Assuming the existence of several asymptotic families labeled by an index $\lambda$, the previous derivation generalizes straightforwardly to:
\begin{equation} \nonumber
    \mathcal Z(u)
\sim
\sum_{\lambda}r_{\lambda}^{-u}
\sum_{\ell}
\left[A_{\lambda,\ell}(u) e^{-i\theta_\lambda u}- \tilde A_{\lambda,\ell}(u) e^{-i(\pi -\theta_{\lambda}) u}\right]\,
\zeta(u \, \delta_\lambda-\gamma_\lambda+\ell) \ ,
\end{equation}
so that the singularity structure of $\mathcal Z(u)$ becomes the superposition of the singularities generated by each individual trajectory. Consequently, the non-integer operators appearing in the thermal OPE are naturally interpreted as arising from the union of several asymptotic families of quasinormal modes, each contributing its own arithmetic sequence of poles.

Although we shall not pursue this direction here, the generalization above suggests that the decomposition of the OPE spectrum into different asymptotic sequences may provide a direct way of identifying the distinct families of quasinormal modes present in a given theory.

\subsection{Beyond meromorphicity}
\label{subsec:beyondmeromorphicity}

So far we have worked under the simplest possible assumption, namely that the retarded correlator is a meromorphic function of the complex frequency. While this assumption is realized in many holographic examples and provides a remarkably simple framework, it is certainly not the most general situation encountered in quantum field theory. In generic interacting theories, one expects branch cuts to arise from multiparticle continua or other non-perturbative effects. Extending the inversion formula to this more general setting is conceptually straightforward. The main obstacle is practical rather than conceptual: in a generic theory, neither the location of the branch cuts nor the corresponding discontinuities are known a priori.

In the presence of branch cuts, the Cauchy representation becomes: 
\begin{equation}
g_R(\omega)
=
\sum_n
\frac{r_n}{\omega-\omega_n}
+
\int_{\mathcal C_{\mathrm{cut}}}
\frac{\text{d}\omega'}{2\pi i}
\frac{\mathrm{Disc}\,g_R(\omega')}
{\omega'-\omega},
\label{eq:poles-plus-cut}
\end{equation}
where $\mathcal C_{\mathrm{cut}}$ runs along the branch cuts and all powers of $\omega'$ are understood on the same Riemann sheet as the original correlator.

Applying the Mellin--Barnes representation of the Cauchy kernel and exchanging the order of integration, one finds:
\begin{equation}
g_R(\omega)
=\int_\mathcal{C}
\frac{\text{d}u}{2\pi i}\,
\Gamma(u)\Gamma(1-u)
e^{-i\pi u}
\omega^{u-1}
\mathcal Z(u),
\label{eq:mb-cut}
\end{equation}
where the generalized Mellin transform is:
\begin{equation}
\mathcal Z(u)
=
\sum_n
r_n\omega_n^{-u}
+
\int_{\mathcal C_{\mathrm{cut}}}
\frac{\text{d}\omega'}{2\pi i}
\,
\mathrm{Disc}\,g_R(\omega')
(\omega')^{-u}.
\label{eq:Zfull-general}
\end{equation}
The momentum-space inversion formula therefore becomes:
\begin{equation}
\tilde a_\Delta
=
 e^{ -i \pi  (s-d/2)} \operatorname*{Res}_{u=2s-d+1}
\left[
\frac{\pi e^{-i\pi u}}{\sin(\pi u)}
\,
\mathcal Z(u)
\right].
\label{eq:inversion-with-cut}
\end{equation}
The structure of the inversion formula is unchanged: the only modification is that the discrete Mellin transform $\mathcal Z(u)$ is replaced by a generalized transform receiving contributions both from isolated quasinormal modes and the continuous spectral density associated with the branch cuts.

From this perspective, the analytic structure of $\mathcal Z(u)$ directly encodes the large-frequency expansion of the correlator. Simple poles reproduce the familiar power-law contributions to the thermal OPE, regardless of whether they originate from isolated quasinormal modes or from the Mellin transform of the discontinuity. More generally, branch points or higher-order singularities of $\mathcal Z(u)$ generate logarithmic corrections or other non-analytic terms in the asymptotic expansion. In this way, the inversion formula naturally extends from a discrete spectral decomposition to the most general analytic structure allowed by the retarded correlator.

It would be very interesting to pursue this direction in concrete interacting examples. This is particularly relevant since the vast majority of physically interesting critical theories, including the $3d$ Ising universality class, the $\mathrm O(N)$ models, and the Gross--Neveu--Yukawa models, are expected to exhibit branch cuts in addition to isolated poles in the complex frequency plane. These theories describe a wide range of condensed matter systems, including quantum antiferromagnets probed by neutron scattering experiments \cite{Sachdev:2011fcc,Wen:2007} and quantum critical points associated with ferromagnetic phase transitions \cite{Sachdev:1995bf}.

\subsection{Application: $\mathcal N = 4$ SYM $R$-current}

It is useful to illustrate the previous discussion in an explicit holographic example. At zero spatial momentum $k=0$, the retarded correlator of the $SU(4)_R$ currents in strongly coupled $\mathcal N=4$ SYM at finite temperature reads \cite{Myers:2007we}: 
\begin{equation}
g_{R}(\omega)
=
\frac{N_c^2
}{8}\frac{i \omega}{2\pi}
+
\frac{N_c^2
}{8}\left(\frac{\omega}{2\pi}\right)^2
\left[
\psi\!\left(-\frac{i-1}{2}\frac{\omega}{2\pi}\right)
+
\psi\!\left(-\frac{i+1}{2}\frac{\omega}{2\pi}\right)
\right] \ .
\label{eq:Rcurrent-exact}
\end{equation}
Since the digamma function $\psi(z)$ has simple poles at:
\[
z=0,-1,-2,\ldots,
\]
the retarded correlator possesses two towers of quasinormal modes:
\begin{equation}
\omega_n^{(+)}
=
-2\pi (1-i)\,n\  ,
\qquad
\omega_n^{(-)}
=
2\pi (1+i)\,n \ ,
\qquad
n=1,2,\ldots \ ,
\label{eq:Rcurrent-qnms}
\end{equation}
which satisfy the reality condition $\omega_n^{(-)}=-\left(\omega_n^{(+)}\right)^\star$.
Taking residues with respect to the frequency $\omega$ gives:
\begin{equation}
r_n^{(+)}
=
-\frac{N_c^2}{8\pi}(1+i)n^2 \ ,
\qquad
r_n^{(-)}
=
\frac{N_c^2}{8\pi}(1-i)n^2 \ ,
\label{eq:Rcurrent-residues}
\end{equation}
which indeed obey \( r_n^{(-)}=-\left(r_n^{(+)}\right)^{\star}\),
as required for a retarded correlator.
The Mellin-like QNM generating function introduced previously is given in \eqref{eq:kzeroinversion}: here we substitute the exact spectrum and denote the function as $\mathcal Z_R(u)$:
\begin{equation}
\mathcal Z_R(u)
=
\frac{N_c^2}{4\pi}
(2\pi)^{-u} \left[(1-i) e^{-i\frac{\pi}{4}u}-(1+i)e^{-i\frac{3\pi}{4}u} \right]
\zeta(u-2) \ ,
\label{eq:Rcurrent-Z-closed}
\end{equation}
whose structure is also reproduced by the $\ell=0$ term of Equation~\eqref{eq:zetaexpansion}.
For the conserved current one has:
\[
\Delta_\phi=3,
\qquad
d=4,
\]
which gives the pole locations
\begin{equation}
u
=
2\Delta_\phi-\Delta-d+1
=
3-\Delta.
\label{eq:Rcurrent-uDelta}
\end{equation}
The inversion formula therefore takes the form:
\begin{equation}
\tilde a_\Delta
= -i \, e^{ i \pi \Delta} \operatorname*{Res}_{u=3-\Delta}
\left[
\frac{\pi e^{- i \pi u}}{\sin(\pi u)}
\,\mathcal Z_{R}(u)
\right] \ .
\label{eq:Rcurrent-inversion}
\end{equation}
It is straightforward to check that this formula matches the OPE in Equation~\eqref{eq:Rcurrent-exact}.
We now interpret the main singularities of the expression above in terms of operators appearing in the OPE of the conformal field theory:

\begin{itemize}

\item[i)] \textit{Identity and the double pole}:

The identity operator corresponds to \(\Delta=0,\) and therefore \(u = 3\).
Both the Riemann zeta function $\zeta(u-2)$ and the universal kernel:
\[
\frac{\pi e^{-i\pi u}}{\sin(\pi u)}
\]
have a simple pole at $u=3$.

Consequently, the integrand
of the inversion formula develops a \emph{double pole} at $u=3$. This is the Mellin-space manifestation of the familiar $\omega^2\log\omega$ term in the large-frequency expansion of the exact retarded correlator.
More precisely, we directly match the flavour anomaly coefficient from the coefficient of the logarithmic term
\begin{equation} \nonumber
    g_R(\omega)\supset \frac{N_c^2}{ 16 \pi^2} \omega^2 \log \omega  \ .
\end{equation}
This is precisely the expected Type B anomaly associated with the $R$-current of $\mathcal N = 4$ SYM, in the usual normalization of the two-point function  
\cite{Osborn:1993cr,Erdmenger:1996yc,Henningson:1998gx,Binder:2019jwn} \begin{equation} \nonumber
    C_J= \frac{N_c^2-1}{4 \pi^2} \overset{N_c \gg 1}{\sim} \frac{N_c^2}{4 \pi^2} \ .
\end{equation}

\item[ii)] \textit{Stress-tensor coefficient}:

The conformal dimension of the stress tensor in $d=4$ is 
\(\Delta = 4\), which corresponds to
\( u=-1\).
Since $u=-1$ is a negative integer, only the simple pole of the universal
kernel contributes, and the residue reduces to the value of
$\mathcal Z_R(u)$:
\begin{equation} \nonumber
\tilde a_\Delta
= -i  \, \mathcal Z_{R}(-1) \ .
\end{equation}
Using the closed expression for $\mathcal Z_R(u)$, one finds:
\begin{equation}
\tilde a_{\Delta=4}
=
-\frac{N_c^2}{120}.
\label{eq:Rcurrent-a4}
\end{equation}
This reproduces the unit-normalized thermal stress-tensor OPE coefficient directly from the
exact quasinormal-mode spectrum.\footnote{This is the exact coefficient expected for $\mathcal N =4 $ SYM \cite{Gubser:1996de}. In fact the coefficient $\tilde{a}_{\Delta = 4}$ at zero coupling is $\tilde{a}_{\Delta = 4}^{\rm free} =- N_c^2/90$, which means that:
\begin{equation} \nonumber
    \frac{\tilde{a}_{\Delta = 4}^{\rm strong }}{\tilde{a}_{\Delta =4}^{\rm free}} = \frac{3}{4} \ ,
\end{equation} as expected. }

\item[iii)] \textit{Zeros and selection rules}:

This example also illustrates that zeros of $\mathcal Z_R(u)$ arise from two
independent mechanisms.

First, the trivial zeros of the Riemann zeta function imply:
\begin{equation}
\mathcal Z_R(0)
=
\mathcal Z_R(-2)
=
\mathcal Z_R(-4)
=
\cdots
=
0.
\label{eq:Rcurrent-zeta-zeros}
\end{equation}
Second, the oscillating factor
vanishes whenever:
\[
\left[(1-i) e^{-i\frac{\pi}{4}u}-(1+i)e^{-i\frac{3\pi}{4}u} \right]=0 \quad \longrightarrow \quad 
u=1+4m,
\quad
m\in\mathbb Z \ .
\]
In particular,
\begin{equation}
\mathcal Z_R(-3)
=
\mathcal Z_R(-7)
=
\cdots
=
0.
\label{eq:Rcurrent-interference-zeros}
\end{equation}
Hence the absence of certain OPE coefficients is controlled partly by the
trivial zeros of the zeta function and partly by destructive interference
between the two conjugate quasinormal-mode towers.

\item[iv)] \textit{Small-frequency expansions}:

Finally, the Taylor coefficients around $\omega=0$ are obtained from:
\[
\mu_p=-\mathcal Z_R(p+1) \ .
\]
As an example, it is easy to see that:
\[
\mathcal Z_R(1)=0 \ ,
\]
which implies that the constant term in the
small-frequency expansion vanishes. Higher moments are encoded in the same
closed expression \eqref{eq:Rcurrent-Z-closed}, providing a concrete
realization of the general relation between low-frequency moments and the
quasinormal-mode spectrum.

\end{itemize}

Let us use this example to make an important comment. In these holographic models, the inversion formula captures only the contribution of the multi-stress-tensor sector. The reason is that the retarded correlator is obtained from the Euclidean correlator by taking its discontinuity across the appropriate branch cut. Double-trace operators, although they contribute to the Euclidean correlator, do not contribute to this discontinuity because their conformal dimensions are protected at the classical level. As a consequence, the inversion formulae presented here do not directly probe the double-trace sector, whose contribution has been studied in detail in Refs.~\cite{Buric:2025fye,Buric:2025anb,Niarchos:2026wsw,Barrat:2025twb,Arnaudo:2026tcy}. Nevertheless, these contributions can be reconstructed by combining our results with the techniques developed in Ref.~\cite{Barrat:2025twb}.

Beyond the $R$-current case, deriving exact expressions for holographic thermal correlators becomes prohibitively difficult. Nevertheless, further examples and consistency checks can be carried out numerically; see, e.g., \cite{NewProjectCristo} for explicit tests.

\subsection{Application: BTZ black brane}
In this section we consider the two-dimensional Virasoro-primary example with $\Delta_\phi=1$, for which the exact correlator is given in~\eqref{eq:2dexactmom}.
When we set $k=0$, we observe that there is only one QNM tower, given in~\eqref{eq:qnm2d}, and its frequencies obey the reality condition for $k=0$:
\begin{align} \nonumber
    \omega_n^{\vphantom{\star}}=-\omega_n^\star\;.
\end{align}
Additionally, we can read off the coefficients from the exact correlator in \eqref{eq:2dexactmom}:
\begin{align} \nonumber
    r_n=8 i \pi^2 \;,
\end{align}
and using these we can define the function $\mathcal{Z}(u)$: 
\begin{align} \nonumber
    \mathcal{Z}(u)=\sum_{n=0}^\infty r_n^{\vphantom{-u}} \omega_n^{-u}
=
8i\pi^2(2\pi)^{-u} e^{-i\pi u/2}
\left(1-2^{-u}\right)\zeta(u) \ .
\end{align}
The $k=0$ inversion yields:
\begin{equation} \nonumber
    \tilde{a}^{2d}_{\Delta=1-m}=8 \pi^2 (2 \pi)^{-m} e^{-i \pi m}(1-2^{-m}) \zeta(m) \ , \quad m \in \mathbb{Z}^{<0} \ .
\end{equation}
The zeta function automatically allows only the odd values of $m$ to be non-zero, returning the correct result:
\begin{equation} \nonumber
     \tilde{a}^{2d}_{\Delta=2n}
=\frac{\pi}{2n} \left(4\pi\right)^{2n}(2^{1-2n}-1)B_{2n} \ , \quad n\in \mathbb{Z}^{>0} \,,
\end{equation}
where $B_2n$ are the Bernoulli numbers. This result precisely matches the OPE coefficients of the two-dimensional theory, given in \eqref{eq:2dopecoef}, together with the weights arising from the thermal Polyakov blocks in momentum space, as defined in \eqref{eq:aDelta-momentum} and specialized to the two-dimensional setup. This provides the final check of the $k=0$ inversion formula. In the next section, we generalize the analysis to $k\neq 0$.

Note that the function $\mathcal Z(u)$, as expected, is written in terms of a zeta function in $u$.
As noted above, this function correctly reproduces the OPE coefficients. In addition, the sum rules for missing operators (negative integer dimensions) are automatically satisfied since these zeros coincide with the zeros of the $\zeta$-function: $\mathcal Z(-2m) = 0$. Finally, it is also straightforward to see that the function $\mathcal Z(p+1)$ matches the small-$\omega$ expansion of the correlator in Equation~\eqref{eq:2dexactmom}.

\section{Inversion formulae and complex $\omega$-plane: non-zero momentum}
\label{sec:non-zeromom}

We now turn to the more general case of non-vanishing spatial momentum, $k\neq0$. We derive the complete inversion formula, which refines the previous construction by resolving the spin dependence of the exchanged operators. The case of two spacetime dimensions requires a separate treatment, and we derive the corresponding inversion formula in this setting as well. Finally, under the same minimal assumptions on the analytic structure of the retarded correlator in the complex $\omega$-plane, we discuss how the inversion formula can be used to extract non-perturbative information about the QNM spectrum. Compared to the $k=0$ case, the main new feature is that both the QNM frequencies $\omega_n(k)$ and their residues become momentum dependent. Nevertheless, we show that the inversion formula still provides useful constraints and, in particular, allows one to predict their scaling behavior in the large-$k$ regime.

Since we have obtained an explicit momentum-space OPE, the next natural step is to extract the corresponding OPE coefficients by inverting the conformal blocks. This construction may be regarded as the momentum-space analogue of the Euclidean inversion formula for thermal two-point functions developed in \cite{Iliesiu:2018fao}. The key observation is that, after an appropriate change of variables, the momentum-space blocks factorize into a radial part and an angular part, allowing the inversion to proceed in close analogy with the position-space case.

Let us start from Equation~\eqref{eq:momentumOPE}. It is convenient to introduce the variables:
\begin{equation}
q=\sqrt{k^2-\omega^2},
\qquad
\eta=\frac{i\omega}{q} \ .
\label{eq:q_eta_variables}
\end{equation}
The inverse change of variables is given by:
\begin{equation}
\omega=-iq\eta,
\qquad
k=q\sqrt{1-\eta^2}.
\label{eq:omega_k_q_eta}
\end{equation}
On the Euclidean section, $\eta$ is real and satisfies:
\begin{equation} \nonumber
-1\leq\eta\leq1.
\end{equation}
Let us discuss the low-momentum and high-momentum limits. On the Euclidean section, $\omega_E=q \eta$ and $k=q\sqrt{1-\eta^2}$: if $\eta$ is generic and fixed, then the limit $q\to \infty$ requires $\omega_E, k\to \infty$; conversely, if $q\to 0$, then $\omega_E, k\to 0$; similar conclusions are reached when $\eta=0,\pm 1$. For this reason, $q\to \infty$ can be considered a UV limit, while $q \to 0$ can be considered an IR limit.\footnote{There is another limit worth considering: $q \to 0$, paired with $\omega_E, k$ fixed (or both large), corresponding to the lightcone limit. Nevertheless, such limit requires $|\eta| \to \infty$ and therefore lies outside the regime explored in this work.}

For later convenience, we define:
\begin{equation}
G(q,\eta)
\equiv
g_R\!\left(-iq\eta,q\sqrt{1-\eta^2}\right),
\qquad
G_{\rm arcs}(q,\eta)
\equiv
\tilde{g}_{\rm arcs}\!\left(-iq\eta,q\sqrt{1-\eta^2}\right).
\label{eq:G_definitions}
\end{equation}
In these variables the momentum-space OPE takes the particularly simple form:
\begin{equation}
    G(q,\eta)=
    \sum_{\mathcal O} \frac{\tilde{a}_{\mathcal O}}{\beta^{\Delta_\Om}}
    q^{2s-d}
    \,
    C_J^{(\nu)}
    \!\left(\eta\right) + G_{\rm arcs}(q,\eta) \ .
    \label{eq:OPE_q_eta}
\end{equation}
Subtracting the arc contribution gives
\begin{equation}
G(q,\eta)-G_{\rm arcs}(q,\eta)
=
\sum_{\mathcal O} \frac{\tilde{a}_{\mathcal O}}{\beta^{\Delta_\Om}}  \,
    q^{2s-d}
    \,
    C_J^{(\nu)}
    \!\left(\eta\right) \ .
\label{eq:subtracted_OPE_q_eta}
\end{equation}
From now on we set $\beta = 1$.
We use the orthogonality of Gegenbauer polynomials to project onto a specific spin $J$. The orthogonality relation is given by:
\begin{equation} \nonumber
    \int_{-1}^{1} \text{d}\eta \, (1-\eta^2)^{\nu-\frac12}
    C_J^{(\nu)}(\eta)C_{J'}^{(\nu)}(\eta)=\delta_{J,J'} \frac{\pi 2^{1-2\nu}}{J+\nu} \frac{\Gamma(J+2\nu)}{\Gamma(J+1)\Gamma(\nu)^2}= \delta_{J,J'}^{\vphantom{(\nu)}} h_{J}^{(\nu)}
\end{equation}
and it can be used to project the retarded correlator onto the spin $J$:
\begin{align}
    B_J(q)=\frac{1}{h_{J}^{(\nu)}}\int_{-1}^{1}
\text{d}\eta\,
(1-\eta^2)^{\nu-\frac12}
C_J^{(\nu)}(\eta)
\left[
G(q,\eta)-G_{\rm arcs}(q,\eta)
\right]\label{eq:BJ_projection}\;.
\end{align}
Having projected onto a fixed spin, the remaining task is to isolate the individual powers of the radial variable \(q\). This is naturally achieved by taking a Mellin transform with respect to \(q\). Since the expansion is only expected to hold asymptotically at large momentum, we introduce a UV Mellin transform:
\begin{equation}
\mathcal M_J^{\rm UV}(\alpha)
=
\int_{\Lambda}^{\infty}
\frac{\text{d}q}{q}\,
q^{-\alpha}
B_J(q),
\label{eq:UV_Mellin_transform}
\end{equation}
where the cutoff \(\Lambda\) is chosen sufficiently large so that the asymptotic OPE is valid. As we now show, the precise value of \(\Lambda\) is irrelevant for extracting the OPE data. Indeed, consider a single contribution to the asymptotic expansion:
\begin{equation} \nonumber
B_J(q)
\supset \; \tilde{a}_{\mathcal O}  \,
    q^{2s-d} \ .
\end{equation}
Its Mellin transform is simply given by:
\begin{equation}
\tilde{a}_{\mathcal O}
\int_{\Lambda}^{\infty}
\frac{\text{d}q}{q}\,
q^{2s-d-\alpha}
=
\tilde{a}_{\mathcal O}
\frac{\Lambda^{2s-d-\alpha}}
{\alpha-(2s-d)},
\qquad
\mathrm{Re}(\alpha)>2s-d.
\label{eq:single_power_Mellin}
\end{equation}
Thus every power of \(q\) is mapped into a simple pole in Mellin space, whose location determines the scaling dimension while its residue gives the corresponding OPE coefficient. Explicitly,
\begin{equation}
\operatorname*{Res}_{\alpha=2s-d}
\mathcal M_J^{\rm UV}(\alpha)
=
\tilde{a}_{\mathcal O} \ .
\label{eq:residue_N}
\end{equation}
Substituting the definitions of the Mellin transform and of the spin projection, Equations~\eqref{eq:UV_Mellin_transform} and~\eqref{eq:BJ_projection}, returns the inversion formula:
\begin{equation}
\tilde{a}_{\mathcal O}
=
\frac{1}
{
h_J^{(\nu)}}
\operatorname*{Res}_{\alpha=2s-d}
\int_{\Lambda}^{\infty}
\frac{\text{d}q}{q}\,
q^{-\alpha}
\int_{-1}^{1}
\text{d}\eta\,
(1-\eta^2)^{\nu-\frac12}
C_J^{(\nu)}(\eta)
\left[
G(q,\eta)-G_{\rm arcs}(q,\eta)
\right].
\label{eq:Euclidean_inversion_formula}
\end{equation}
The logic underlying this formula is remarkably simple. The Gegenbauer orthogonality projects onto fixed spin, while the Mellin transform projects onto fixed scaling dimension by converting powers of the radial momentum into poles. The result is therefore a momentum-space Euclidean inversion formula for the asymptotic OPE data. Note that its derivation relies only on the existence of the asymptotic large-momentum expansion and does not require any assumption on the analytic structure of the retarded correlator in the complex frequency plane.

Let us emphasize an important feature of the inversion formula. The presence of the cutoff $\Lambda$ makes it manifest that the OPE coefficients are extracted entirely from the UV behavior of the correlator, namely from arbitrarily large momenta. This statement holds independently of whether the OPE is a genuinely convergent expansion or merely an asymptotic one.

Indeed, the cutoff $\Lambda$ only affects the regular part of the Mellin transform. The singular part, and in particular the residues from which the OPE coefficients are extracted, are independent of the precise value of $\Lambda$. As long as $\Lambda$ is chosen within the asymptotic regime where the large-momentum expansion is valid, all dependence on the cutoff drops out of the inversion formula. Therefore, the OPE coefficients are genuine UV observables, determined entirely by the asymptotic behavior of the correlator.

\paragraph{Contribution of the arc terms.}

While the OPE contribution is under complete kinematical control, the same is not true for the arc contribution. Nevertheless, we now argue that $G_{\rm arcs}(q,\eta)$ is not expected to contribute to the inversion formula. The physical reason is simple: the inversion formula has been constructed precisely to project onto the momentum-space OPE blocks. Since the arc contribution is, by construction, not captured by the OPE expansion, one naturally expects it to contribute only to the regular part of the Mellin transform and therefore not to the residues that determine the OPE coefficients.

This expectation can be made precise under very mild assumptions. Consider:
\[
I(\alpha)
=
\int_{\Lambda}^{\infty}
\frac{\text{d}q}{q}\,
q^{-\alpha}
\int_{-1}^{1}
\text{d}\eta\,
(1-\eta^2)^{\nu-\frac12}
C_J^{(\nu)}(\eta)\,
G_{\rm arcs}(q,\eta) \ .
\]
The arc function is, by definition, non-polynomial in momentum at high energy, specifically, it takes the form:
\[
G_{\rm arcs}(q,\eta)
=
e^{-\mu q}
H(q,\eta),
\qquad
\operatorname{Re}(\mu)>0 \ ,
\]
where $H(q,\eta)$ grows at most polynomially as $q\rightarrow\infty$. Physically, this assumption simply states that the arc contribution is exponentially suppressed at large momentum and therefore does not contribute to the asymptotic power-law expansion generated by the OPE.

 We now show that $I(\alpha)$ and all of its derivatives are regular.
 First of all, we observe that there exist constants $C,N>0$ such that, uniformly for complex $\alpha$:
\[
\left|
q^{-\alpha-1}
G_{\rm arcs}(q,\eta)
\right|
\le
C\,
q^{N-\operatorname{Re}(\alpha)-1}
e^{-\operatorname{Re}(\mu)q}.
\]
Since the exponential suppression dominates every polynomial growth, the integral defining $I(\alpha)$ converges absolutely for every complex $\alpha$. Moreover, the same estimate holds after differentiating with respect to $\alpha$, since:
\[
\partial_\alpha^n I(\alpha)
=
(-1)^n
\int_{\Lambda}^{\infty}
\frac{\text{d}q}{q}
(\log q)^n
q^{-\alpha}
\int_{-1}^{1}
\text{d}\eta\,
(1-\eta^2)^{\nu-\frac12}
C_J^{(\nu)}(\eta)\,
G_{\rm arcs}(q,\eta),
\]
and the additional logarithmic factors are dominated by the exponential decay.

Standard differentiation-under-the-integral arguments therefore imply that $I(\alpha)$ is an entire function of $\alpha$. Consequently, the arc contribution cannot generate poles in the Mellin transform and therefore does not contribute to the residues entering the inversion formula. Under the assumptions above, all OPE data are entirely determined by the asymptotic power-law part of the correlator. We shall verify this expectation explicitly in the examples discussed below.

Since the arc terms produce only regular contributions and therefore do not affect the residues, i.e., the thermal OPE coefficients, the inversion formula can be written as 
\begin{equation}
\tilde{a}_{\mathcal O}
=
\frac{1}
{
h_J^{(\nu)}}
\operatorname*{Res}_{\alpha=2s-d}
\int_{\Lambda}^{\infty}
\frac{\text{d}q}{q}\,
q^{-\alpha}
\int_{-1}^{1}
\text{d}\eta\,
(1-\eta^2)^{\nu-\frac12}
C_J^{(\nu)}(\eta)
G(q,\eta) \ .
\label{eq:final_inversion_formula}
\end{equation}

\paragraph{IR inversion.}

The inversion formula discussed above projects onto the UV/OPE expansion at large momentum. An analogous construction can be carried out for other kinematic regimes. In particular, one may consider the deep infrared limit, i.e., $|\omega|,|k|\ll1$, where the two-point function admits the expansion \cite{Delacretaz:2026owo}:
\begin{equation}
\label{eq:IRexpansion_general}
g(\omega,k)=\sum_{n,m}\mu_{n,m}\,\omega^n k^m \, .
\end{equation}
The coefficients $\mu_{n,m}$ generalize the low-frequency moments discussed previously in the $k=0$ case. Extracting these coefficients is straightforward and follows directly from Cauchy's integral formula:
\begin{equation} \nonumber
\mu_{n,m}
=
\oint_{\omega\in B_\omega}\frac{\text{d}\omega}{2\pi i}
\oint_{k\in B_k}\frac{\text{d}k}{2\pi i}
\frac{g(\omega,k)}{\omega^{n+1}k^{m+1}} \, ,
\end{equation}
where $B_\omega$ and $B_k$ are closed contours around the origin in the complex $\omega$ and $k$-planes, respectively. Their radius (analogous to $\Lambda$ in the UV inversion) is chosen sufficiently small so that the contours lie entirely within the domain of convergence of the expansion \eqref{eq:IRexpansion_general}.

Compared to the UV inversion formula, this construction is considerably simpler since it relies solely on Cauchy's theorem. We leave a detailed investigation of its applications and possible implications for future work.
\subsection{Application: Free scalar theory}
We test the inversion formula~\eqref{eq:final_inversion_formula} in the simple case of free scalar theory. In the $q,\eta$ variables, the free retarded correlator takes the particularly simple form:
\begin{equation} \nonumber
G(q,\eta)=\frac{1}{q^2},
\end{equation}
so that the inversion formula reads:
\begin{equation} \nonumber
\tilde{a}_{\mathcal O}
=
\frac{1}{
h_J^{(\nu)}
}
\operatorname*{Res}_{\alpha=2s-d}
\int_{\Lambda}^{\infty}
\frac{\text{d}q}{q}\,
q^{-\alpha-2}
\int_{-1}^{1}
\text{d}\eta\,
(1-\eta^2)^{\nu-\frac12}
C_J^{(\nu)}(\eta).
\end{equation}
The $\eta$ integral can be evaluated exactly using the orthogonality properties of the Gegenbauer polynomials:
\begin{equation} \nonumber
\frac{1}{
h_J^{(\nu)}
}\int_{-1}^{1}
\text{d}\eta\,
(1-\eta^2)^{\nu-\frac12}
C_J^{(\nu)}(\eta)
=\delta_{J,0} \ ,
\end{equation}
showing that only $J=0$ operators contribute. The inversion formula simplifies to:
\begin{equation} \nonumber
\tilde{a}_{\mathcal O}
=
\delta_{J,0}
\operatorname*{Res}_{\alpha=2s-d}
\int_{\Lambda}^{\infty}
\text{d}q\,
q^{-\alpha-3}.
\end{equation}
The $q$ integral converges for $\mathrm{Re}(\alpha)>-2$, where it evaluates to:
\begin{equation} \nonumber
\int_{\Lambda}^{\infty}
\text{d}q\,
q^{-\alpha-3}
=
\frac{\Lambda^{-\alpha-2}}{\alpha+2}.
\end{equation}
The analytic continuation of the right-hand side has a pole at $\alpha=-2$, which translates into a pole at $\Delta=0$. Thus, the inversion formula correctly identifies the identity as the only operator contributing to the thermal Polyakov block decomposition:
\begin{equation} \nonumber
    \tilde{a}_{\mathcal O}=\delta_{J,0} \delta_{\Delta,0} \ .
\end{equation}

\subsection{Application: $\mathrm O(N)$ model in $d=4-\varepsilon$}

We test the inversion formula \eqref{eq:final_inversion_formula} not only on free theories, but also on weakly interacting ones. The perfect playground is provided by the $\mathrm O(N)$ model in $d=4-\varepsilon$, since in this theory only one operator, $\phi^2$, contributes at order $O(\varepsilon)$ to the discontinuity. At the same time, we will assume the absence of arc contributions \cite{Barrat:2025nvu}. 

The first step is to build the retarded correlator that will later be inverted. Following the decomposition in $q,\eta$ coordinates \eqref{eq:OPE_q_eta}, we start by considering the contribution from the identity operator:
\begin{equation} \nonumber
    G^{\mathds{1}}(q,\eta)=\frac{1}{q^2} \;.
\end{equation}
The contribution of the $\phi^2$ operator reads
\begin{equation} \nonumber
    G^{\phi^2}(q,\eta)=\frac{\tilde{a}^{(0)}_{\phi^2}+\varepsilon \tilde{a}^{(1)}_{\phi^2}}{q^{4+\varepsilon (\gamma_{\phi^2}-1)}} +O(\varepsilon^2)\,,
\end{equation}
where $\Delta_{\phi^2}=2\Delta_\phi+\varepsilon \gamma_{\phi^2}$, with $\gamma_{\phi^2}=\frac{N+2}{N+8}$.

The full retarded correlator at order $O(\varepsilon)$ reads:
\begin{equation} \label{eq: expcorreps}
    G(q,\eta)=\frac{1}{q^2}+\frac{\tilde{a}^{(0)}_{\phi^2}+\varepsilon \tilde{a}^{(1)}_{\phi^2}}{q^4}+\frac{\varepsilon \tilde{a}^{(0)}_{\phi^2}(1-\gamma_{\phi^2})}{q^4}\log q +O(\varepsilon^2)\;.
\end{equation}

We can substitute this expression into the inversion formula given in \eqref{eq:final_inversion_formula} and expand the final result to order $O(\varepsilon)$. Since the retarded correlator is not a function of $\eta$, the $\eta$ integral is trivial and the inversion formula simplifies:
\begin{equation} \nonumber
\tilde{a}_{\mathcal O}
=\delta_{J,0}
\operatorname*{Res}_{\alpha=2s-d}
\int_{\Lambda}^{\infty}
\frac{\text{d}q}{q}\,
q^{-\alpha} \left( \frac{1}{q^2}+\frac{\tilde{a}^{(0)}_{\phi^2}+\varepsilon \tilde{a}^{(1)}_{\phi^2}}{q^4}+\frac{\varepsilon \tilde{a}^{(0)}_{\phi^2}(1-\gamma_{\phi^2})}{q^4}\log q \right)+O(\varepsilon^2)
\ .
\end{equation}
The $q$ integral can be easily performed and it gives
\begin{multline} \nonumber
    \tilde{a}_{\mathcal O}
=\delta_{J,0}
\operatorname*{Res}_{\alpha=2s-d} \bigg[ \frac{\Lambda^{-\alpha-2}}{\alpha+2}
+
\left(
\tilde{a}_{\phi^2}^{(0)}
+
\varepsilon\,\tilde{a}_{\phi^2}^{(1)}
\right)
\frac{\Lambda^{-\alpha-4}}{\alpha+4}
\\+
\varepsilon\,\tilde{a}_{\phi^2}^{(0)}
\left(1-\gamma_{\phi^2}\right)
\Lambda^{-\alpha-4}
\left(
\frac{\log\Lambda}{\alpha+4}
+
\frac{1}{(\alpha+4)^2}
\right) \bigg]
+O(\varepsilon^2) \ .
\end{multline}
Naively evaluating the residues at this stage would lead to a non-trivial dependence on the cutoff $\Lambda$, which is not allowed by construction. Indeed, the correct way to proceed is to redefine the cutoff:
\begin{equation} \nonumber
    \log \Lambda = \lambda \ , \qquad \Lambda^{-\alpha-n}=1-(\alpha+n)\lambda+\frac{1}{2}(\alpha+n)^2 \lambda^2 + \dots 
\end{equation}
Under this redefinition, a precise cancellation makes the pole structure of the integral clearer:
\begin{equation} \nonumber
    \tilde{a}_{\mathcal O}
=\delta_{J,0}
\operatorname*{Res}_{\alpha=2s-d} \left[ \frac{1}{\alpha+2}
+
\left(
\tilde{a}_{\phi^2}^{(0)}
+
\varepsilon\,\tilde{a}_{\phi^2}^{(1)}
\right)
\frac{1}{\alpha+4}
\\+
\varepsilon\,\tilde{a}_{\phi^2}^{(0)}
\left(1-\gamma_{\phi^2}\right)
\frac{1}{(\alpha+4)^2}\right]
+O(\varepsilon^2) \ .
\end{equation}
Since we are working perturbatively at order $O(\varepsilon)$, it is possible to eliminate the second-order pole by shifting one of the simple poles:
\begin{equation} \nonumber
    \tilde{a}_{\mathcal O}
=\delta_{J,0}
\operatorname*{Res}_{\alpha=2s-d} \left[ \frac{1}{\alpha+2}
+
\frac{
\tilde{a}_{\phi^2}^{(0)}
+
\varepsilon\,\tilde{a}_{\phi^2}^{(1)}}{\alpha+4-\varepsilon 
\left(1-\gamma_{\phi^2}\right)}
\right]
+O(\varepsilon^2) \ .
\end{equation}
In conclusion, we recover exactly the same CFT data we started from at the beginning of this section. Although this exercise may seem trivial, let us highlight the fact that the correlator~\eqref{eq: expcorreps} could be replaced by any perturbative computation in a weakly-coupled theory: requiring the final result to be independent of the cutoff and absorbing double poles through pole-shifting, we would have recovered non-trivial thermal CFT data.

\subsection{Inversion formula in $2d$} \label{ssec: 2dinversion}
In the previous sections we discussed the inversion formula for $d>2$. In two dimensions, for kinematical reasons, the thermal OPE blocks are different and as a consequence their KMS-symmetric inversion is different. In fact, the expansion from which we start in two spacetime dimensions is given in~\eqref{eq:GMI-momentum2d}.
Making use of the orthogonality condition of Chebyshev polynomials, we can straightforwardly derive the $2d$ version of the inversion formula \eqref{eq:Euclidean_inversion_formula}:
\begin{equation}
\tilde{a}^{2d}_{\mathcal O}
=
\frac{1}{\pi}\frac{2}{1+\delta_{J,0}}
\operatorname*{Res}_{\alpha=2s-2}
\int_{\Lambda}^{\infty}
\frac{\text{d}q}{q}\,
q^{-\alpha}
\int_{-1}^{1} \text{d}\eta\,
(1-\eta^2)^{-\frac12}
T_{J}(\eta)
G(q,\eta) \ .
\label{eq:2dEuclidean_inversion_formula}
\end{equation}

We can use the exact correlator we considered in Section~\ref{subsec:2dBorel} to test this formula. In $q,\eta$ coordinates it reads:

\begin{equation} \nonumber
G(q,\eta)
=
-\pi
\left[
\psi
\!\left(
\frac12
-
\frac{q}{4\pi}
\left(\eta+i \sqrt{1-\eta^2} \right)
\right)
+
\psi
\!\left(
\frac12
-
\frac{q}{4\pi}
\left(\eta-i \sqrt{1-\eta^2} \right)
\right)
\right] \ .
\end{equation}
We verify that the inversion formula \eqref{eq:2dEuclidean_inversion_formula} returns the correct coefficients. In this case, it is convenient to set $\eta=\cos \vartheta$ and change variables: the inversion formula reads:
\begin{equation} \nonumber
\tilde{a}^{2d}_{\mathcal O}
=
\frac{1}{\pi}\frac{2}{1+\delta_{J,0}}
\operatorname*{Res}_{\alpha=2s-2}
\int_{\Lambda}^{\infty}
\frac{\text{d}q}{q}\,
q^{-\alpha}
\int_{0}^{\pi} \text{d}\vartheta\,
\cos (J \vartheta)
G(q,\vartheta) \ ,
\end{equation}
while the retarded correlator simplifies to:
\begin{equation} \nonumber
G(q,\eta)
=
-\pi
\left[
\psi
\!\left(
\frac12
-
\frac{q e^{i \vartheta}}{4\pi}
\right)
+
\psi
\!\left(
\frac12
-
\frac{q  e^{-i \vartheta}}{4\pi}
\right)
\right] \ .
\end{equation}
We start with the evaluation of the $\vartheta$ integral, which is denoted by $I_J(q)$ in the following.
It can be simplified by introducing the complex variable $z=e^{i \vartheta}$: 
\begin{equation} \label{eq: complexint}
   I_{J}(q)= 
-\pi^2 \oint_{|z|=1} \frac{\text{d}z}{2 \pi i} \,
\left(
\frac{1}{z^{1+J}}
+
\frac{1}{z^{1-J}}
\right)
\psi
\!\left(
\frac12
-
\frac{q z}{4\pi}
\right) \ .
\end{equation}
Evaluating these integrals is not straightforward since the pole locations are $q$-dependent. We then focus on performing the $q$ integral first:
\begin{equation} \nonumber
   H_{J}(z)=\int_{\Lambda}^{\infty}
\frac{\text{d}q}{q}\,
q^{-\alpha} \psi
\!\left(
\frac12
-
\frac{q z}{4\pi}
\right) \ .
\end{equation}
Since the integration goes from $\Lambda$ to $\infty$, it is convenient to use the large-$q$ asymptotic expansion of the digamma function:
\begin{equation} \nonumber
    \psi
\!\left(
\frac12
-
\frac{q z}{\mu}
\right) \sim \log\left(-
\frac{q z}{4\pi} \right) + \sum_{n=1}^{\infty}  \left(\frac{4\pi}{q}\right)^{2n}\frac{(1-2^{1-2n})B_{2n}}{2n \; z^{2n}} \ ,
\end{equation}
where $\mu$ is a regulator and $B_{2n}$ is the $2n$-th Bernoulli number. Substituting it into the $H_{J}(z)$ integral, we find the analytic continuation:
\begin{equation} \nonumber
    H_{J}(z)=\frac{\Lambda^{-\alpha}}{\alpha^2}+\frac{\Lambda^{-\alpha}}{\alpha}\log\left(-\frac{z \Lambda}{4 \pi} \right) + \sum_{n=1}^{\infty} \frac{\Lambda^{-\alpha-2n}}{\alpha+2n}\left(4\pi\right)^{2n}\frac{(1-2^{1-2n})B_{2n}}{2n \; z^{2n}} \ .
\end{equation}
Applying the contour integral in Equation~\eqref{eq: complexint}, we can easily pick up the residues in $z$ (or perform a contour integration, in the presence of the logarithm), ending up with the result:
\begin{multline} \nonumber
    \tilde{a}^{2d}_{\mathcal O}
=
-\frac{2 \pi}{1+\delta_{J,0}}
\operatorname*{Res}_{\alpha=-\Delta} \bigg[\frac{\Lambda^{-\alpha}}{\alpha^2} \delta_{J,0}+ \frac{\Lambda^{-\alpha}}{\alpha} \log\left(\frac{ \Lambda}{\mu} \right) \delta_{J,0}\\+\sum_{n=1}^{\infty}  \frac{\Lambda^{-\alpha-2n}}{\alpha+2n}\frac{\left(4\pi\right)^{2n}}{4n}(1-2^{1-2n})B_{2n} (\delta_{J, 2n}+\delta_{J, -2n}) \bigg] \ .
\end{multline}
To show explicitly that the result is independent of the cutoff $\Lambda$, we relabel:
\begin{equation} \nonumber
    \log\left(\frac{\Lambda}{\mu}\right)=\lambda \ , \quad \Lambda^{-\alpha-2n}=\left(1-(\alpha+2n)\log \mu-(\alpha+2n)\lambda+\dots\right) \ ,
\end{equation}
and this greatly simplifies the pole structure:
\begin{equation} \nonumber
     \tilde{a}^{2d}_{\mathcal O}
=
-\frac{4 \pi}{1+\delta_{J,0}}
\operatorname*{Res}_{\alpha=-\Delta} \bigg[-\frac{\log \mu}{\alpha} \delta_{J,0}+\sum_{n=1}^{\infty}  \frac{1}{\alpha+2n}\frac{\left(4\pi\right)^{2n}}{4n}(1-2^{1-2n})B_{2n} (\delta_{J, 2n}+\delta_{J, -2n}) \bigg] \ .
\end{equation}
We first treat carefully the pole at $\Delta=0$, which corresponds to the identity operator; the coefficient reads:
\begin{equation} \nonumber
    \tilde{a}^{2d}_{\mathds{1}}=2 \pi  \log \mu \ .
\end{equation}
Despite the appearance of the regulator $\mu$, this is simply due to the fact that by applying the definition \eqref{eq: def2datilde} we end up with a divergent $\Gamma(0)$, which is regularized by $\mu.$ The remaining poles are simpler to handle and yield, when selecting $J>0$,
\begin{equation} \nonumber
     \tilde{a}^{2d}_{J}
=\frac{\pi}{n} \left(4\pi\right)^{2n}(2^{1-2n}-1)B_{2n} \; \delta_{J, 2n} \; \delta_{\Delta,2n} \ ,
\end{equation}
which is in perfect agreement with the known value of these coefficients. It is important to note that the inversion formula automatically generates the selection rules for spin and conformal dimension, reproducing the correct spectrum of operators associated with thermal Polyakov blocks.

\subsection{QNMs at large momentum and the inversion formula}
In this section we investigate the general consequences of applying the momentum-space inversion formula under the same assumption adopted in the case \(k=0\), namely that the retarded correlator is meromorphic in the complex \(\omega\)-plane. More precisely, we assume that the subtracted correlator admits the representation:
\begin{equation}
g_R(\omega,k)
=
\sum_n
\frac{r_n(k)}
{\omega-\omega_n(k)},
\label{eq:meromorphic-ansatz}
\end{equation}
where both the quasinormal frequencies and their residues are now allowed to depend on the spatial momentum. In terms of the variables \((q,\eta)\), this becomes: 
\begin{equation}
G(q,\eta)
=
\sum_n
\frac{r_n\!\left(q \, \chi (\eta)\right)}
{-iq\eta-\omega_n\!\left(q\, \chi(\eta)\right)},
\label{eq:qeta-meromorphic}
\end{equation}
where \(\chi(\eta)=\sqrt{1-\eta^2}\). Unlike the case \(k=0\), the main difficulty is that the functional dependence of the quasinormal frequencies and residues on the spatial momentum is, in general, unknown. Consequently, the inversion formula cannot be applied directly without further input.

The Euclidean inversion formula derived in the previous section, however, provides precisely the information needed to overcome this difficulty. Since the OPE coefficients are extracted from the ultraviolet region, as emphasized by the presence of the cutoff \(\Lambda\), only the large-momentum behavior of the correlator is relevant.\footnote{The fact that only large momentum is relevant does not imply that an expansion around low momentum would not give any information. We will explore this direction in Section~\ref{sec:deepir}.} This suggests that the detailed low-energy dependence of \(\omega_n(k)\) and \(r_n(k)\) is largely irrelevant for the inversion procedure. Motivated by this observation, we develop a systematic large-\(k\) expansion of the quasinormal frequencies and residues, in complete analogy with the large-mode-number expansion discussed in Section~\ref{sec:UVIR}.

We now assume that the quasinormal frequencies admit a large-momentum expansion of the form:
\begin{equation}
\omega_n(k)
=
v_n\,k
+
\mu_n\,k^{1-\delta}
+
O(k^{1-2\delta}) \ ,
\qquad
\delta>0 \ ,
\label{eq:omega-large-k}
\end{equation}
together with a similar expansion for the residues,
\begin{equation}
r_n(k)
=
\rho_n\,k^\sigma
+
\rho_n^{(1)}\,k^{\sigma-\delta}
+
O(k^{\sigma-2\delta})\ .
\label{eq:r-large-k}
\end{equation}
The coefficients $v_n$, $\mu_n$, $\rho_n,\ldots$ are independent of $k$, although they may depend on the mode number $n$ and on the quantum numbers of the channel. Introducing the variables \(q,\eta\), it is convenient to define:
\begin{equation}
A_n(\eta)
\equiv
i\eta+v_n\chi(\eta) \ .
\label{eq:An-def}
\end{equation}
The denominator can then be rewritten as follows:
\begin{align}
-iq\eta-\omega_n(q \, \chi(\eta))
&=
-qA_n(\eta)
\left[
1+
\frac{\mu_n\chi(\eta)^{1-\delta}}
{A_n(\eta)}
q^{-\delta}
+
O(q^{-2\delta})
\right] \ ,
\label{eq:denominator-expand}
\end{align}
which immediately gives
\begin{equation}
\frac{1}
{-iq\eta-\omega_n(q \, \chi(\eta))}
=
-\frac{1}{qA_n(\eta)}
+
\frac{\mu_n\chi(\eta)^{1-\delta}}
{q^{1+\delta}A_n(\eta)^2}
+ O(q^{-1-2\delta}) \ .
\label{eq:denominator-leading}
\end{equation}
We can expand the numerator with a similar procedure:
\begin{equation}
r_n(q \,\chi(\eta))
=
\rho_n
q^\sigma
\chi(\eta)^\sigma
+
\rho_n^{(1)}
q^{\sigma-\delta}
\chi(\eta)^{\sigma-\delta}
+
O(q^{\sigma-2\delta}) \ .
\label{eq:residue-leading}
\end{equation}
Finally, multiplying the two expansions, one finds the full expansion of the correlator at large $k$:
\begin{align}
\frac{r_n(q \, \chi(\eta))}
{-iq\eta-\omega_n(q\, \chi(\eta))}
&=
-\rho_n
q^{\sigma-1}
\frac{\chi(\eta)^\sigma}
{A_n(\eta)}
\nonumber\\
&\quad
+
q^{\sigma-1-\delta}
\left[
\rho_n\mu_n
\frac{\chi(\eta)^{\sigma+1-\delta}}
{A_n(\eta)^2}
-
\rho_n^{(1)}
\frac{\chi(\eta)^{\sigma-\delta}}
{A_n(\eta)}
\right]
+
 O(q^{\sigma-1-2\delta}) \ .
\label{eq:single-mode-subleading}
\end{align}
Therefore the leading contribution of each quasinormal mode factorizes into a power of the radial momentum multiplied by a purely angular function:
\begin{equation}
\frac{r_n(q \, \chi(\eta))}
{-iq\eta-\omega_n(q\, \chi(\eta))}
=
-\rho_n
q^{\sigma-1}
\frac{\chi(\eta)^\sigma}
{i\eta+v_n\chi(\eta)}
+
 O(q^{\sigma-1-\delta}) \ .
\label{eq:single-mode-leading}
\end{equation}
The crucial observation is that the leading contribution has exactly the same structure as the momentum-space OPE: a power of the radial variable multiplied by a function of the angular variable alone. Consequently, the Euclidean inversion formula derived in the previous section can be applied term by term to the large-$k$ expansion of each quasinormal mode, relating its asymptotic data directly to the thermal OPE coefficients.

\paragraph{Consequences of the inversion formula.}

Equation \eqref{eq:single-mode-leading} is the direct analogue of the large-\(\omega\) expansion at \(k=0\): every simple pole contributes a universal
factor \(1/q\), so the large-\(q\) power is fixed by the large-\(k\) growth of
the residue. Since the inversion formula selects the coefficient of
\(
  q^{2s-d}
\),
consistency requires \( \sigma - 1 = 2s-d\) or, equivalently:
\begin{equation}
  r_n(k)\sim k^{2s-d+1}
  \ .
  \label{eq:main-scaling}
\end{equation}
Thus the residues grow by one extra power of \(k\) relative to the correlator
itself. This is the generic non-zero-momentum counterpart of the \(k=0\)
statement that the UV/OPE coefficient is extracted from the large-frequency
asymptotics of the pole sum. More explicitly, the minimal UV ansatz is:
\begin{equation}
  \omega_n(k)=v_n k + O(k^{1-\delta}),
  \qquad
  r_n(k)=\rho_n^{(s)}\,k^{2s-d+1} +  O(k^{2s-d+1-\delta}).
  \label{eq:minimal-uv-ansatz}
\end{equation}
The superscript on \(\rho_n^{(s)}\) emphasizes that different OPE powers are
associated with different large-\(k\) coefficients in the residue expansion. The leading power associated with the
identity is:
\begin{equation}
  r_n(k)
  =
  \rho_n\,k^{2\Delta_\phi-d+1}
  +
  O \!\left(k^{2\Delta_\phi-d+1-\delta}\right).
  \label{eq:identity-large-k}
\end{equation}
So, just as in the one-dimensional discussion, the identity fixes the first
allowed asymptotic growth of the QNM residues.

We now examine what the inversion formula predicts from this large-$k$ expansion.
Substituting \eqref{eq:minimal-uv-ansatz} into \eqref{eq:qeta-meromorphic} gives:
\begin{equation}
  G(q,\eta)
  \approx
  -\,q^{2s-d}
  \sum_n
  \rho_n^{(s)}
  \frac{\chi(\eta)^{2s-d+1}}{i\eta + v_n \chi(\eta)}
  + O\!\left(q^{2s-d-\delta}\right).
  \label{eq:full-leading}
\end{equation}
Inserting this into~\eqref{eq:final_inversion_formula}, focusing
on the pole \(\alpha=2s-d\), and using $\chi(\eta)=\sqrt{1-\eta^2}$, we find
\begin{equation}
    \tilde{a}_{\mathcal O}
=
-\frac{1}
{
h_J^{(\nu)}}
\sum_n
  \rho_n^{(s)}
\int_{-1}^{1}
\text{d}\eta\,
  \frac{(1-\eta^2)^{s-1}
C_J^{(\nu)}(\eta)}{i\eta + v_n \sqrt{1-\eta^2}} \ .
  \label{eq:final-uv}
\end{equation}

The radial Mellin integral has disappeared completely. The only remnant of the
QNM data is the angular kernel determined by the asymptotic slopes \(v_n\) and
the corresponding residue coefficients \(\rho_n^{(s)}\). 

To proceed, note that there are no poles on the integration path as long as $\text{Re}(v_n) \neq 0$.\footnote{If instead $\text{Re}(v_n) = 0$, there is always a pole on the integration path, and the computation requires a $+i0$ prescription. This corresponds to QNMs that are very close to the imaginary axis in the large-$k$ limit. In this section, we will not consider this case, but an example of $+i0$ prescription in the computation of the inversion formula can be found in Section~\ref{sec:deepir}.}
Thus, the integral converges and can be evaluated.
Assuming $J$ even,\footnote{The case $J$ odd is not automatically ruled out by the inversion formula, but it should be regarded as unphysical.} the integral reads:
\begin{multline} \nonumber
    \tilde{a}_{\mathcal O}
  =
  -\,\frac{1}{h_J^{(\nu)}}
  \sum_n \rho_n^{(s)} v_n \sum_{m = 0}^{J/2} \frac{(-1)^m 2^{J-2m}(\nu)_{J-m}}{m! (J-2m)!} \\\times \frac{\Gamma(J/2-m+1/2)\Gamma(s+1/2)}{\Gamma(J/2-m+s+1)} {}_2F_1\!\left(
\begin{matrix}
1,s+\frac12\\
s-m+1+\frac{J}{2}
\end{matrix}
;1-v_{n}^2
\right)
   \ .
\end{multline}

As a first application, let us consider the contribution of the identity operator. In this case $\tilde{a}_{\mathbb 1} =1$, $J = 0$ and $s = \Delta_\phi$,
while \(C_0^{(\nu)}(\eta)=1\). The general inversion formula \eqref{eq:final-uv} therefore reduces to:
\begin{equation}
1
=
-\frac{1}{h_0^{(\nu)}}
\sum_n
\rho_n^{(\Delta_\phi)} v_n \frac{\Gamma(\Delta_\phi+1/2)\sqrt{\pi}}{\Gamma(\Delta_\phi+1)} {}_2F_1\!\left(
\begin{matrix}
1,\Delta_\phi+\frac12\\
\Delta_\phi+1
\end{matrix}
;1-v_{n}^2
\right) \ 
.
\label{eq:identity-sumrule}
\end{equation}
This relation provides the leading constraint on the asymptotic quasinormal mode data. In particular, it fixes a weighted sum of the leading residue coefficients in terms of the asymptotic velocities \(v_n\), independently of the detailed structure of the low-lying spectrum.

Equation~\eqref{eq:identity-sumrule} is the natural non-zero momentum counterpart of the \(k=0\) identity sum rule. Just as in the zero-momentum case, the identity operator completely determines the leading ultraviolet behavior of the pole decomposition. Higher-dimensional operators then provide systematic corrections, constraining the subleading coefficients in the large-\(k\) expansion of the quasinormal frequencies and residues.

\paragraph{Generalization to multiple asymptotic families.}

The discussion above extends straightforwardly to the more realistic situation in which the quasinormal modes organize themselves into several asymptotic families, labeled by an index $\lambda$. The natural large-$k$ ansatz is:
\begin{equation}
\omega_{n,\lambda}(k)
=
v_{n,\lambda}\,k
+
O(k^{1-\delta}),
\qquad
r_{n,\lambda}(k)
=
\sum_{s'}
\rho_{n,\lambda}^{(s')}
k^{2s'-d+1}
+\cdots,
\label{eq:families}
\end{equation}
where each family is characterized by its own asymptotic slope and residue expansion.

Applying the inversion formula at the pole $\alpha=2s-d$ immediately projects onto the coefficient multiplying the corresponding power of $k$. One therefore finds
\begin{equation}
    \tilde{a}_{\mathcal O}
=
-\frac{1}
{
h_J^{(\nu)}}
\sum_{n,\lambda}
  \rho_{n,\lambda}^{(s)}
\int_{-1}^{1}
\text{d}\eta\,
  \frac{(1-\eta^2)^{s-1}
C_J^{(\nu)}(\eta)}{i\eta + v_{n,\lambda} \sqrt{1-\eta^2}} \ .
  \label{eq:families-final}
\end{equation}
As in the case of a single QNM family, the inversion formula automatically isolates the coefficient associated with the desired OPE power. Different terms in the large-$k$ residue expansion never mix, and each family contributes additively to the thermal OPE coefficient. In this sense, the large-$k$ expansion of the residues is naturally dual to the ultraviolet OPE expansion: every power in the OPE corresponds to a unique power in the asymptotic residue expansion.

A further simplification occurs whenever the asymptotic velocity becomes independent of the mode number within a given family:
\begin{equation}
v_{n,\lambda}=v_\lambda \ .
\label{eq:constant-v}
\end{equation}
In this case the angular dependence factorizes and Equation~\eqref{eq:families-final} reduces to:
\begin{equation}
    \tilde{a}_{\mathcal O}
=
-\frac{1}
{
h_J^{(\nu)}}\sum_{\lambda} \mathcal I_{J,s}^{(\nu)}(v_{\lambda})
\sum_{n,\lambda}
  \rho_{n,\lambda}^{(s)}
 \,,
 \label{eq:factorized-family}
\end{equation}
where we have denoted the previously computed $\eta$-integral by
\begin{equation}
\mathcal I_{J,s}^{(\nu)}(v_{\lambda})
=
\int_{-1}^{1}
\text{d}\eta\,
  \frac{(1-\eta^2)^{s-1}
C_J^{(\nu)}(\eta)}{i\eta + v_{\lambda} \sqrt{1-\eta^2}} \ .
\label{eq:kernel-def}
\end{equation}
Equation~\eqref{eq:factorized-family} provides the simplest realization of the momentum-space inversion formula. In this approximation, the entire dependence on the asymptotic kinematics is encoded in the universal kernel $\mathcal I_{J,s}^{(\nu)}(v_{\lambda})$, while the dynamical information is reduced to the sums of the asymptotic residue coefficients. This form is particularly convenient when fitting numerical or holographic quasinormal mode spectra.

\subsection{Application: $\mathrm O(N)$ model at large $N$}
\label{subsec:ONlargeN}
We now consider the case of the large-$N$ $\mathrm O(N)$ model in $d=3$, defined by the Lagrangian: 
\begin{equation} \nonumber
    \mathcal L = \frac{1}{2}(\partial \phi)^2+\frac{1}{2}\sigma \phi^2 \ ,
\end{equation}
where $\sigma$ is the Hubbard--Stratonovich field.
The retarded two-point function of identical fundamental scalars is known exactly and given by
\begin{equation} 
  g_R(\omega,k)
  =
  \frac{1}{k^2-\omega^2+m_{\mathrm{th}}^2} \ ,
  \label{eq:ON-E}
\end{equation}
with a thermal mass given in terms of the golden ratio:
\begin{equation}
  m_{\mathrm{th}}^2 = \langle \sigma\rangle_\beta
  =
  \left(\log\varphi^2\right)^2,
  \qquad
  \varphi= \frac{1+\sqrt5}{2}.
  \label{eq:ON-mth}
\end{equation}
\paragraph{Inverting the correlator.}
In $q,\eta$ coordinates, the thermal retarded correlator in momentum space takes a particularly simple form:
\begin{equation} \nonumber
     G(q,\eta)=\frac{1}{q^2 +m_{\text{th}}^2 } \,.
\end{equation}
Substituting it into the inversion formula, we obtain a factorization of the $q$ and $\eta$ integrals:
\begin{equation} \nonumber
\tilde{a}_{\mathcal O}
=
\frac{1}{
h_J^{(1/2)}
}
\operatorname*{Res}_{\alpha=2s-3}
\int_{\Lambda}^{\infty}
\text{d}q \,
\frac{q^{-\alpha-1}}{q^2 +m_{\text{th}}^2 }
\int_{-1}^{1}
\text{d}\eta\,
P_J(\eta) \ .
\end{equation}
The $\eta$-integral can be evaluated immediately and yields a selection rule for $J=0$, whereas the $q$-integral yields a hypergeometric function:
\begin{equation} \nonumber
\tilde{a}_{\mathcal O}
=
\delta_{J,0}
\operatorname*{Res}_{\alpha=2s-3}\left[\frac{\Lambda^{-2-\alpha}}{\alpha +2} \;  {}_2F_1\!\left(
\begin{matrix}
1,\alpha/2+1\\
\alpha/2+2
\end{matrix}
;-\frac{m_{\text{th}}^2}{\Lambda^2}
\right)\right] \ .
\end{equation}
The hypergeometric function's series representation exposes all the poles:
\begin{equation} \nonumber
    \tilde{a}_{\mathcal O}
=
\delta_{J,0}
\operatorname*{Res}_{\alpha=2s-3}\left[\sum_{m=0}^{\infty} (-m_{\text{th}}^2)^{m} \frac{\Lambda^{-\alpha-2-2m}}{\alpha+2+2m} \right] \ .
\end{equation}
From this expression, we extract the full set of thermal data contributing to the thermal Polyakov decomposition:
\begin{equation} \nonumber
    \tilde{a}_{[\sigma^m]}=(-1)^m m_{\text{th}}^{2m} \ , \quad \Delta=2m=0,2,4\dots 
\end{equation}

\paragraph{The large-$k$ limit from inversion.}
Having verified the microscopic correlator, we now illustrate the implications of the inversion formula applied to the large-$k$ limit.
The retarded correlator \eqref{eq:ON-E} is meromorphic and it trivially satisfies the assumptions adopted throughout this work. Its partial-fraction decomposition reads:
\begin{equation}
g_R(\omega,k)
=
\frac{r_+(k)}{\omega-\omega_+(k)}
+
\frac{r_-(k)}{\omega-\omega_-(k)} \ ,
\label{eq:ON-pole-decomp}
\end{equation}
with poles and residues:
\begin{equation}
\omega_\pm(k)
=
\pm\sqrt{k^2+m_{\mathrm{th}}^2} \ , \qquad  r_\pm(k)
=
\mp\frac{1}{2\sqrt{k^2+m_{\mathrm{th}}^2}} \ .
\label{eq:ON-poles}
\end{equation}

The large-$N$ $O(N)$ model therefore realizes the simplest possible version of our general framework: instead of an infinite tower of quasinormal modes, the retarded correlator contains only two stable poles related by the retarded reality condition. The large-momentum expansion of the poles reads:
\begin{equation}
\omega_\pm(k)
=
\pm k
\pm
\frac{m_{\mathrm{th}}^2}{2k}
+
O(k^{-3}) \ ,
\label{eq:ON-omega-expand}
\end{equation}
which agrees with the general large-$k$ ansatz given in \eqref{eq:omega-large-k} with parameters:
\begin{equation}
v_\pm=\pm1 \ ,
\qquad
\delta=2\  .
\label{eq:ON-vdelta}
\end{equation}
Similarly, the residues admit an expansion analogous to the one discussed in \eqref{eq:r-large-k}:
\begin{equation}
r_\pm(k)
=
\mp\frac{1}{2k}
\pm
\frac{ m_{\mathrm{th}}^2}{4k^3}
+
 O(k^{-5}) \ .
\label{eq:ON-r-expand}
\end{equation}
The leading behavior:
\begin{align} \nonumber
r_\pm(k)\sim k^{-1},
\end{align}
is precisely what is predicted by the general scaling law~\eqref{eq:main-scaling}. Indeed, for the identity operator, in the $3d$ large-$N$ $O(N)$ model one has $2\Delta_\phi-d+1=-1$,
exactly reproducing the observed scaling.

The first subleading correction scales as
\(
k^{-3},
\)
which is again in perfect agreement with the OPE. Indeed, the scalar operator
\(
\sigma
\)
has dimension
\(
\Delta_\sigma=2,
\)
and therefore
\(
2\Delta_\phi-\Delta_\sigma-d+1
=
-3 
\).
More generally, every successive term in the asymptotic expansion is naturally identified with the tower of scalar operators
\(
[\sigma^m],
\)
showing that the large-$k$ expansion of the residues precisely reproduces the thermal OPE.

We now explicitly check the sum rules \eqref{eq:final-uv} and in particular \eqref{eq:identity-sumrule}.
The compact inversion formula derived in the previous section can also be evaluated explicitly in this example.
For the scalar sector ($J=0$), we introduce the angular kernel
\begin{equation}
I_s(v)
=
\int_{-1}^{1}
\text{d}\eta\,
\frac{(1-\eta^2)^{s-1}}
{i\eta+v\sqrt{1-\eta^2}} \ .
\label{eq:Is-def}
\end{equation}
Adopting the change of variables
\[
\eta=\sin\vartheta \ ,
\qquad
\vartheta\in
\left[
-\frac{\pi}{2},
\frac{\pi}{2}
\right] \ ,
\]
one finds
\begin{equation}
I_s(v)
=
\int_{-\pi/2}^{\pi/2}
\text{d}\vartheta\,
\frac{\cos^{2s-1}\vartheta}
{v\cos\vartheta+i\sin\vartheta} \ .
\label{eq:Is-theta}
\end{equation}
For the two lightlike branches,
\(
v=\pm1,
\)
the denominator simplifies to a pure phase:
\[
\cos\vartheta+i\sin\vartheta=e^{i\vartheta} \ ,
\qquad
-\cos\vartheta+i\sin\vartheta=-e^{-i\vartheta} \ ,
\]
and the integrals can be evaluated exactly:
\begin{align*}
I_s(+1)
&=
\int_{-\pi/2}^{\pi/2}
\text{d}\vartheta\,
\cos^{2s-1}\vartheta\,
e^{-i\vartheta}
=
\sqrt{\pi} \;
\frac{\Gamma\!\left(s+\frac12\right)}
{\Gamma(s+1)} \ , \\
I_s(-1)
&=
-I_s(+1)\,,
\end{align*}
where the odd contribution vanishes by symmetry of the integration interval.

For the identity operator,
\(
s=\Delta_\phi=\frac12 \ ,
\)
and therefore
\begin{equation}
I_{1/2}(+1)=2 \ ,
\qquad
I_{1/2}(-1)=-2 \ .
\label{eq:ON-identity-kernels}
\end{equation}
Using the leading coefficients of the residue expansion,
\begin{equation} \nonumber
\rho_+
=
-1 \ ,
\qquad
\rho_-
=
1 \ ,
\end{equation}
the identity sum rule \eqref{eq:identity-sumrule} becomes:
\begin{equation} \nonumber
    1
=
-\sum_{n=\pm}
\rho_n v_n=-\left(-\frac{1}{2}-\frac{1}{2}\right) \ 
.
\end{equation}

This example provides a particularly transparent illustration of the power of the inversion formula. The angular kernel can be evaluated analytically, and the identity contribution is reconstructed exactly from the two asymptotic lightlike branches of the retarded correlator. The same strategy extends straightforwardly to the higher scalar operators in the large-$N$ $\mathrm{O}(N)$ spectrum, whose contributions are encoded in the successive terms of the large-momentum expansion of the residues.

\subsection{Deep IR and inversion formula}
\label{sec:deepir}

The OPE data for a thermal correlator in momentum space emerge naturally in the UV regime $(\omega, k \gg 1)$ \cite{Manenti:2019wxs,Caron-Huot:2009ypo}. In this work, by using the dispersion relation derived in position space \eqref{eq:GMI} and constructing the thermal correlator via thermal Polyakov blocks \eqref{eq:GMI_PolyakovBlocks}, we extend the validity of the OPE decomposition to the full $(\omega, k)$ plane. Furthermore, the decomposition into thermal Polyakov blocks in momentum space enables the formulation of the inversion formula presented in \eqref{eq:final_inversion_formula}. Given that this formula incorporates a built-in UV Mellin transform and a cutoff scale $\Lambda$, one might expect its sensitivity to be restricted to the UV regime. Nevertheless, it remains intriguing to investigate the sensitivity of the Euclidean inversion formula to the IR regime and to determine whether OPE coefficient contributions can be extracted from the deep-IR limit of the momentum-space correlator.

Since the thermal analytic bootstrap has, to date, been primarily formulated for scalar two-point functions, applying it to hydrodynamic correlators -- such as those associated with conserved currents or the stress-energy tensor -- remains outside the scope of this work. Instead, we focus on toy models in two and four dimensions and observe that the inversion formula is, to some extent, sensitive to IR data. Quantifying this sensitivity remains challenging, as we do not anticipate recovering exact OPE coefficients for specific theories. Nonetheless, we demonstrate in the following through explicit computations that the inversion formula yields non-vanishing results for operators with physically reasonable quantum numbers. This suggests that the Euclidean inversion formula may serve as a probe of the full correlator in different Lorentzian regimes, offering a promising avenue for future investigations into the thermal bootstrap’s reach beyond the high-momentum limit.

\paragraph{Deep-IR correlators in $2d$: truncated series.}
A precise definition of a deep-IR thermal correlator for scalar fields is not known; scalar densities, standard objects of interest in hydrodynamics, are to be interpreted as the time components of classically conserved currents $J^{\mu}$. Moreover, since CFT scalar operators can be decomposed into massive modes whose scale is set by the thermal mass, in the hydrodynamic regime they should decouple.
Keeping these points in mind, we use the exact correlator given in~\eqref{eq:2dexactmom} to probe the IR regime by performing an $\omega,k \ll 1$ expansion and retaining the leading contributions.
We start with the Taylor expansion of the thermal retarded correlator:
\begin{align}
    g(\omega,k)=\sum_{n,m=0}^\infty \mu_n^{(2m)}\omega^n k^{2m} \ . \label{eq:2dhydroext}
\end{align}
To perform an explicit computation, we use a set of $\mu_{n}^{2m}$ coefficients from the exact correlator for $\Delta_\phi=1$ Virasoro primaries given in \eqref{eq:2dexactmom}.
The Taylor expansion reads
\begin{align} \label{eq: coeffhydro}
g(\omega,k)
=
\mu_{0}^{(0)}
+
\sum_{\substack{n,m\ge 0\\ n+2m\ge 1}}
\left[\frac{
2\pi\, i^n\,(-1)^{m}
}{
(4\pi)^{n+2m}
}
\binom{n+2m}{2m}
\left(2^{n+2m+1}-1\right)
\zeta(n+2m+1)\right]
\omega^n k^{2m}.
\end{align}
The idea is to define the IR-correlator as a truncated sum:
\begin{align}
    g_{\text{IR}}(\omega,k)=\sum_{n,m=0}^M \mu_n^{(2m)}\omega^n k^{2m} 
    \label{eq:ir-expansion-2d}
\end{align}
and invert it through the inversion formula~\eqref{eq:2dEuclidean_inversion_formula}. We focus on a single monomial $\mu_n^{(2m)}\omega^n k^{2m} \subset g_{\text{IR}}(\omega,k)$: in $q, \eta$ coordinates it reads
\begin{equation} \nonumber
    G_{\text{IR}}(q,\eta) \supset \mu_n^{(2m)}(-i)^n q^{n+2m} \eta^n (1-\eta^2)^m \ .
\end{equation}
When substituted into the inversion formula, the $q$ and $\eta$ integrals factorize and can be solved separately.
The $q$ integral can be solved immediately and returns a physical pole at $\Delta=0$ and a finite tower of non-physical poles in $\Delta=-n-2m$, with $n+m \neq 0$:
\begin{equation} \nonumber
    \tilde{a}^{2d}_{\mathcal O}
=\frac{1}{\pi}\frac{2}{1+\delta_{J,0}}
\sum_{n,m=0}^{M} (-i)^n \mu_{n}^{(2m)}
\int_{-1}^{1} \text{d}\eta\,
\eta^n (1-\eta^2)^{m-\frac12}
T_{J}(\eta) \ , \quad \Delta=-n-2m \ .
\end{equation}
The $\eta$ integral can be simplified by switching to the angular coordinate $\eta= \cos \vartheta$:
\begin{equation} \nonumber
    I_{J}(n,m)=\int_{0}^{\pi} \text{d}\vartheta\,
(\cos \vartheta)^n (\sin \vartheta)^{2m}
\cos(J \vartheta)=\sum_{k=0}^{m} \binom{m}{k}(-1)^{k}\int_{0}^{\pi} \text{d}\vartheta\, \cos^{n+2k} \vartheta 
\cos(J \vartheta) \ .
\end{equation}
The remaining integral admits a closed form and it imposes  the selection rule: 
\begin{equation} \label{eq: selection}
    J+n \in 2 \mathbb Z^{\geq 0} \ .
\end{equation}
If the selection rule is satisfied, the integral returns:
\begin{equation} \nonumber
    I_{J}(n,m)=\sum_{k=0}^{m} (-1)^{k} \frac{\pi}{2^{n+2k}} \binom{m}{k} \binom{n+2k}{\frac{n+2k+J}{2}} \ .
\end{equation}
For computational purposes, it is probably easier to directly compute the integral over $\vartheta$, once $m,n$ and $J$ have been fixed.
The selection rules~\eqref{eq: selection} are nevertheless worth noting.
First of all, they impose an upper bound on the spin $J$; secondly, note that the coefficients \eqref{eq: coeffhydro} are real when $n$ is even, and purely imaginary when $n$ is odd: physical even spins are associated with real coefficients, while unphysical odd ones are associated with imaginary coefficients. 

\begin{figure}[t]
    \centering

    \begin{subfigure}[t]{0.4962\linewidth}
        \centering
        \includegraphics[width=\linewidth]{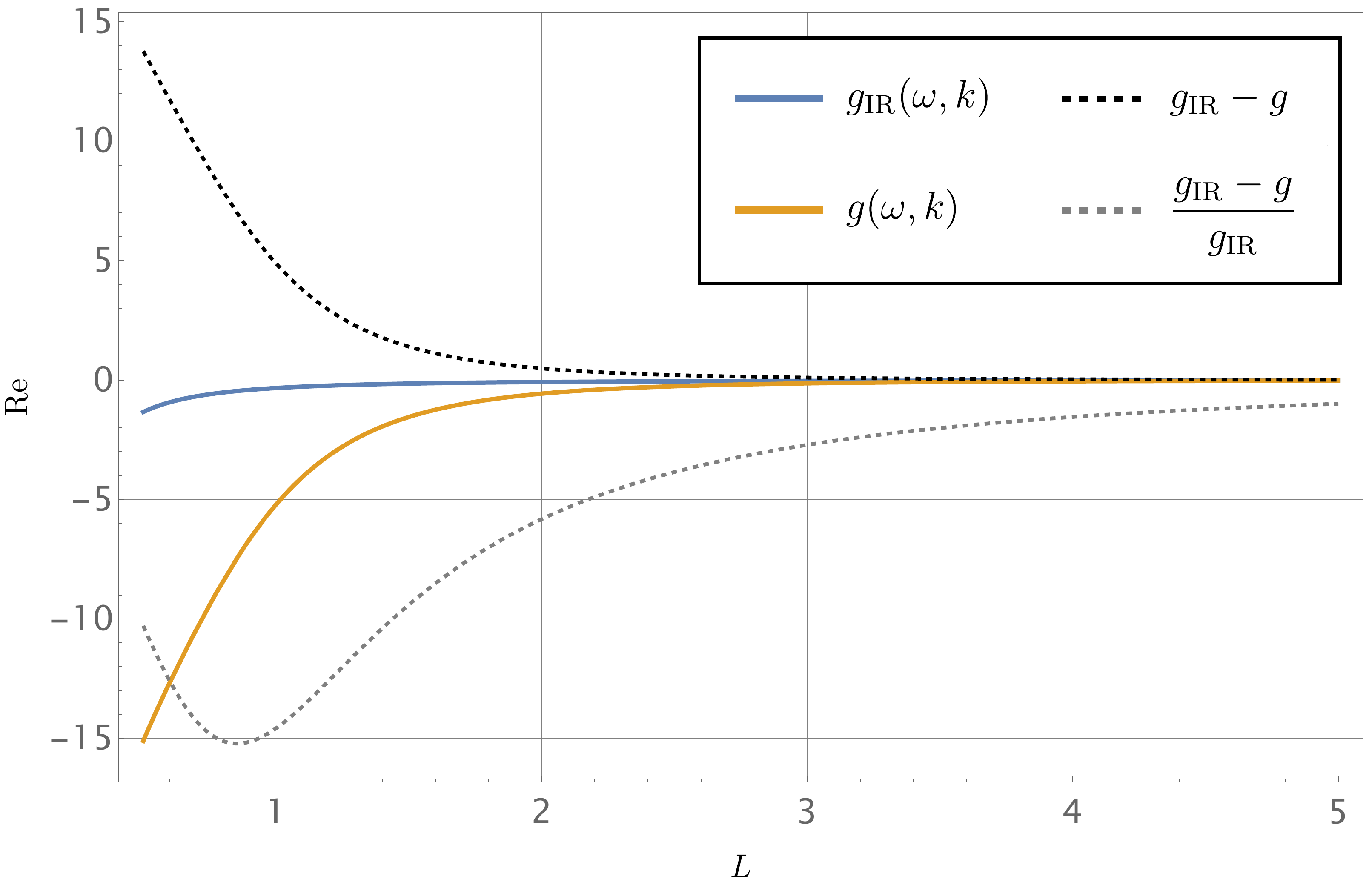}
        \label{fig:deepIR-real}
    \end{subfigure}
    \hfill
    \begin{subfigure}[t]{0.4962\linewidth}
        \centering
        \includegraphics[width=\linewidth]{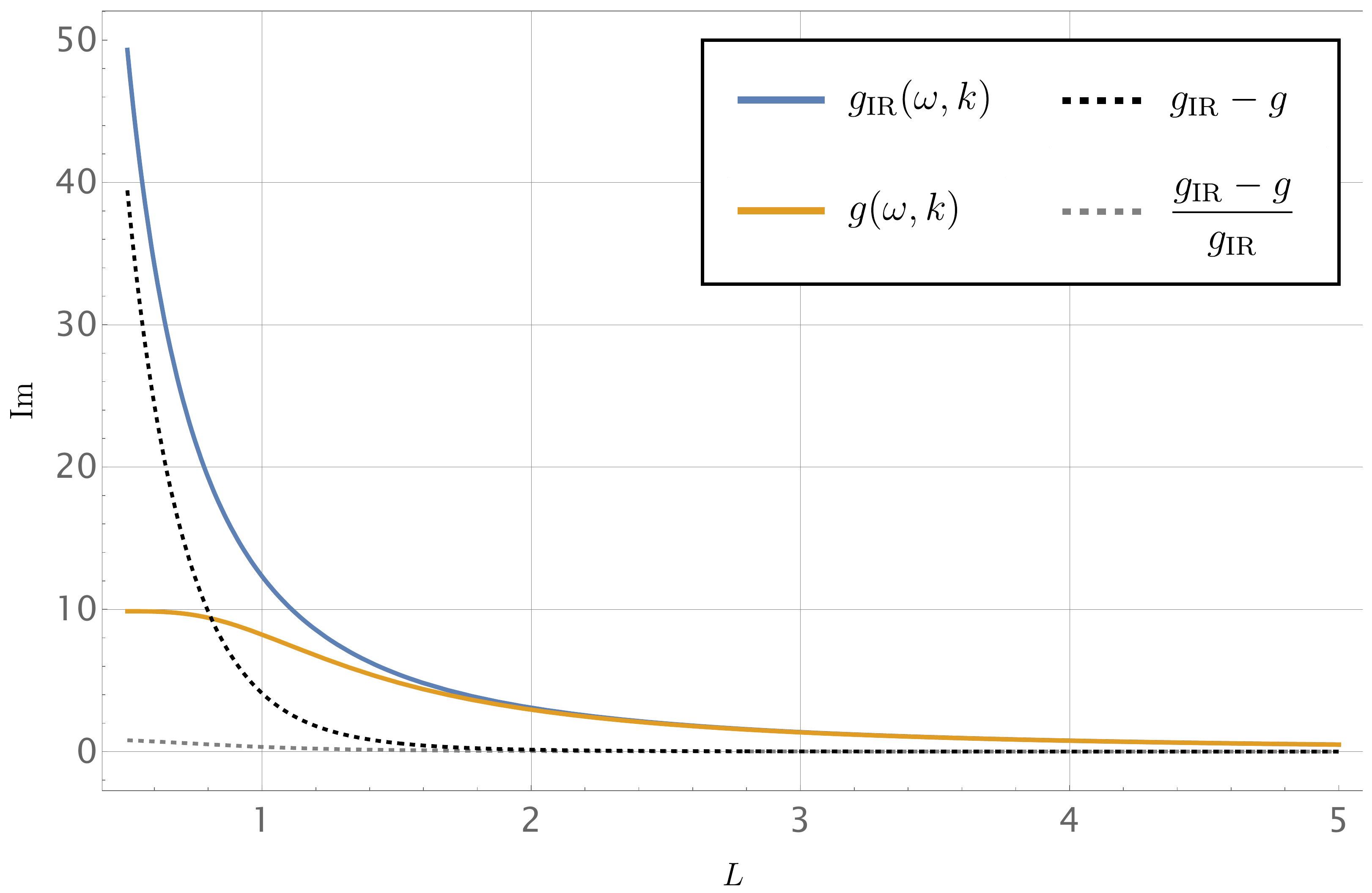}
        \label{fig:deepIR-imaginary}
    \end{subfigure}
    \caption{Real and imaginary part of the deep-IR Padé approximation $g_{\text{IR}}$ \eqref{eq: diffusiontoy2d} (blue line) and of the exact microscopic retarded correlator $g_{R}(\omega,k)-\mu_{0}^{0}$ \eqref{eq: coeffhydro} (orange line). We also plotted the absolute (black dashed) and relative (gray dashed) errors. The plots show the low-momentum limit $L \to \infty$, where $k=1/L$, $\omega=\lambda/L^2$, and $\lambda$ is big enough so that $|D/\lambda|\ll 1$. Here we chose $\lambda=5$, so that $|D/\lambda|\approx 0.027$.}
    \label{fig:deepIR}
\end{figure}

In conclusion, the inversion of the truncated IR expansion of a retarded correlator returns:
\begin{equation} \nonumber
    \tilde{a}^{2d}_{\mathcal O}
=\frac{1}{\pi}\frac{2}{1+\delta_{J,0}}
\sum_{n,m=0}^{M} (-i)^n \mu_{n}^{(2m)}
I_{J}(n,m) \ , \quad \Delta=-n-2m \ , \quad J+n \in 2 \mathbb Z^{\geq 0} \  .
\end{equation}
We observe that all the $\Delta$-poles are located at negative integer values. 
The only physical pole is at $\Delta=0$, which corresponds to the identity operator. This is a first indication that the deep-IR modes partially contribute to the OPE data.
All the other negative poles are not physical since they describe operators violating the unitarity bound; however, note that increasing $\Delta_\phi$ would allow us to obtain additional physical poles, even though we do not expect the associated coefficients $\tilde{a}^{2d}_{\mathcal{O}}$ to match their UV values. From an effective field theory point of view, we can see the interplay between $M$, the order of truncation in \eqref{eq:ir-expansion-2d}, and $\Delta\phi$, which is the scaling dimension of the external operator under study.
\paragraph{Deep-IR correlators in $2d$: Padé approximation.}
Another take on the deep-IR correlator is to truncate the Taylor expansion in~\eqref{eq:ir-expansion-2d} at order $M=2$, similarly to the previous paragraph, followed by a Padé approximation using an ansatz inspired by known hydrodynamic correlators.
For the $\Delta_\phi=1$ example, the constant $\mu_0^{(0)}$ comes from a zeta function regularization, hence we focus on $\tilde{g}_{\text{IR}}(\omega,k)=g_{\text{IR}}(\omega,k)-\mu_0^{(0)}$ instead. 
We consider the diffusion channel and take $\omega \sim k^2  \sim \varepsilon^2$, and we retain terms only through order $O(\varepsilon^2)$ for simplicity:
    \begin{equation} \nonumber
    \tilde{g}_{\text{IR}}(\omega, k)\approx \mu_0^{(2)}k^2+\mu_1^{(0)}\omega \ .
\end{equation}
When we ignore the microscopic origin of each deep-IR mode, we can provide a minimal Padé approximation for the diffusion-like behavior and approximate the correlator above with:
\begin{equation} \nonumber
    \tilde{g}_{\text{IR}}(\omega, k)\approx \mu_1^{(0)}\omega\left[1+\frac{\mu_0^{(2)}}{\mu_1^{(0)}}\left(\frac{k^2}{\omega}\right)\right] \approx \frac{\mu_1^{(0)}\omega}{1-\frac{\mu_0^{(2)}}{\mu_1^{(0)}}\left(\frac{k^2}{\omega}\right)} \ ,  \qquad \left|\frac{\mu_0^{(2)}}{\mu_1^{(0)}} \frac{k^2}{\omega}\right| \ll 1 \ ,
\end{equation}
which reproduces the correct Taylor expansion up to the chosen order. The combination above is reminiscent of a Dyson resummation of single IR modes. The ansatz can be reshaped as follows:
\begin{equation} \label{eq: diffusiontoy2d}
     \tilde{g}_{\text{IR}}(\omega, k)\approx \frac{i \chi D \, \omega^2}{\omega-i D k^2} \ , \qquad \left|D \frac{k^2}{\omega}\right|\ll 1 
\end{equation}
where we indicated the regime where the comparison with the microscopic correlator (see Figure~\ref{fig:deepIR}) makes sense and we introduced the coefficients $D$ and $\chi$, reminiscent of hydrodynamics transport coefficients,. In the $\Delta_\phi=1$ Virasoro primary microscopic model, they have the following values:
\begin{equation} \nonumber
    D=-\frac{i \mu_{0}^{(2)}}{\mu_{1}^{(0)}}=\frac{7 \zeta(3)}{2 \pi^3}\approx 0.13570\dots \ , \quad \chi=\frac{(\mu_{1}^{(0)})^2}{\mu_{0}^{(2)}}=\frac{\pi^5}{14 \zeta(3)}\approx 18.184\dots
\end{equation}
Note that the correlator \eqref{eq: diffusiontoy2d}, despite not being an hydrodynamic correlator, presents a diffusion pole.

We now explore the possibility to invert the correlator \eqref{eq: diffusiontoy2d}. In $q, \eta$ coordinates, this reads:
\begin{equation} \nonumber
     \tilde{G}_{\text{IR}}(q,\eta)\approx \frac{\chi D q \eta^2}{\eta+D q (1-\eta^2)} \ ,
\end{equation}
and the inversion formula is as follows, where we defined $D q = p$: 
\begin{equation} \label{eq: inversion app}
    \frac{\tilde{a}^{2d}_{\mathcal O}}{\chi  D^{\alpha}}
=
\frac{1}{\pi}\frac{2}{1+\delta_{J,0}}
\operatorname*{Res}_{\alpha=-\Delta}
\int_{D \Lambda}^{\infty}
\text{d}p\,
p^{-\alpha} 
\int_{-1}^{1} \text{d}\eta\;
\frac{\eta^2 (1-\eta^2)^{-\frac12}}{\eta+p (1-\eta^2)} \; T_{J}(\eta) \ .
\end{equation}
The computation of this inversion formula is rather involved, so we dedicate the appendix \ref{app: inte} to its detailed computation, and we provide the formulas to compute coefficients for an arbitrary large spin $J$ and conformal dimension $\Delta$. In the following, we limit ourselves to study the first few poles in the case $J=0$:
\begin{equation} \nonumber
     \tilde{a}^{2d}_{\Delta,0}
=\frac{\chi}{D^{\Delta}}
\operatorname*{Res}_{\alpha=-\Delta} \left[-\frac{1}{2}\,
\frac{1-i}{\alpha-\frac12}
+\frac{1}{\alpha}
-\frac{3}{8}\,
\frac{1+i}{\alpha+\frac12}-
\frac{5}{64}\,
\frac{1-i}{\alpha+\frac32}
+\dots \right] \ .
\end{equation}
We find one unphisical pole with $\Delta=-\frac{1}{2}$ and $\tilde{a}^{2d}_{-\frac{1}{2},0}=-\frac{1}{2}\chi \sqrt{D} (1-i)$, and a small collection of poles with conformal dimensions $\Delta=0,\frac{1}{2}, \frac{3}{2}, \dots$ and coefficients:
\begin{equation} \label{eq: coeffim}
    \tilde{a}^{2d}_{0,0}=\chi \ , \quad \tilde{a}^{2d}_{\frac{1}{2},0}=-\frac{3}{8}\frac{\chi}{\sqrt{D}}(1+i) \ , \quad \tilde{a}^{2d}_{\frac{3}{2},0}=-\frac{5}{64}\frac{\chi}{D^{3/2}}(1-i) \ , \dots \ .
\end{equation}
A few considerations are in order. From a technical perspective, spin $J=0$ returns a spectrum of half-integer dimensions, but it is possible that higher spins return also integer dimensions. If we interpret $\tilde{a}^{2d}_{0,0}=\chi$ as the coefficient of an identity operator in the deep IR, we can see that the transport coefficient $\chi$ plays the role of an overall normalization of all the other coefficients. Finally, the coefficients have both real and imaginary parts: the imaginary part is produced by a pole-residue contribution, which is ultimately tied to the Padé pole in the correlator \eqref{eq: diffusiontoy2d}.\footnote{Appendix \ref{app: inte} provides technical details on the appearance of this imaginary part.}

We can now interpret the result. These coefficients should not be interpreted in terms of the microscopic coefficients of Section~\ref{subsec:2dBorel}, since they are probing a deep-IR regime. Nevertheless, unlike the truncated IR expansion~\eqref{eq:ir-expansion-2d}, in its regime of validity (see Figure~\ref{fig:deepIR}) the correlator \eqref{eq: diffusiontoy2d} admits a thermal Polyakov block decomposition in terms of operators with realistic quantum numbers, as shown by the inversion formula: it is intriguing to think of deep-IR coefficients as EFT coefficients in the low-momentum limit. A more detailed study of this approach will be part of future work.

\acknowledgments

We are particularly grateful to  Paolo Arnaudo, Alejandra Castro, Sean Hartnoll, Cristoforo Iossa, Waltraut Knop, Zohar Komargodski, Robin Karlsson, Benjamin Withers, and Subir Sachdev.
JB and EP are supported by ERC-2021-CoG - BrokenSymmetries 101044226.
DB and EP have benefited from the German Research Foundation DFG under Germany’s Excellence Strategy – EXC 2121 Quantum Universe – 390833306.
EM and EP’s work is funded by the German Research Foundation DFG – SFB 1624 – “Higher structures, moduli spaces and integrability” – 506632645.

\appendix

\section{Derivation of thermal Polyakov blocks}
\label{app:DerivationOfTheGeneralizedMethodOfImages}

The starting point is the Euclidean two-point function:
\begin{equation} \nonumber
    g(r,w)
    =
    \vev{\phi(\tau,x) \phi(0,0)}_\beta\,,
\end{equation}
where we use the variables:
\begin{equation} \nonumber
    z = r w\,, \quad \zb = \frac{r}{w}\,.
\end{equation}
We wish to understand the analytic structure in the $w$-plane.
The first step is to notice that the Euclidean correlator has singularities at the collision points:
\begin{equation} \nonumber
    z = m\,, \qquad \zb = m\ \qquad \longleftrightarrow \qquad w = \frac{m}{r}\,, \qquad w = \frac{r}{m}\,.
\end{equation}
While these branch points are unambiguous, we need to choose branch cuts consistent with the Euclidean correlator.
The location of the latter in the complex $w$-plane is on the unit circle $|w|=1$, which the branch cuts are therefore not allowed to cross.

\begin{figure}
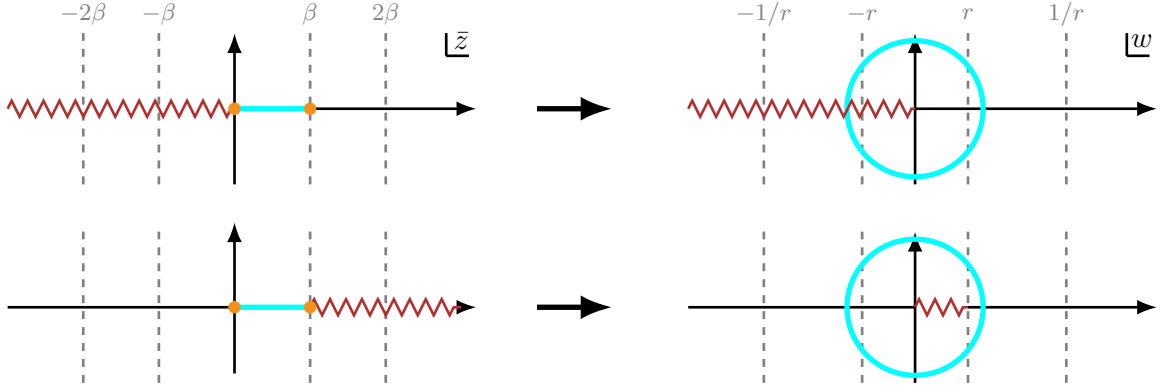

    \centering
    \GMIProofFig
    \caption{Translation of the choices of analytic structure from the $\zb$-plane to the $w$-plane.
    The first choice of branch cut, $\zb \in (-\infty,0)$ results in a cut crossing the Euclidean correlator (here represented by the cyan line and circle) in the $w$-plane.
    The second choice $\zb \in (1,\infty)$ produces a consistent cut for the $w$-plane.}
    \label{fig:AnalyticStructure1}
\end{figure}

Our strategy is to restrict ourselves to the main strip $0 < r <\beta$ and use the periodicity of the correlator (and thus the cuts) to introduce a sum of images.
Let us first consider fixed $z$.
In this domain, the singularities in the $\zb$-plane are located at $\zb=0,\beta$.
In the $w$-plane these points are located at $w=\frac{r}{\beta},+\infty$.
As mentioned above, the cut we choose cannot intersect the Euclidean correlator.
The only consistent and natural choice is therefore that the cut is located on:
\begin{equation} \nonumber
    \zb \in (\beta, \infty) \qquad \longleftrightarrow \qquad w \in (0,r)\,,
\end{equation}
as shown in Figure~\ref{fig:AnalyticStructure1}.
This coincides with the choice of analytic structure made in~\cite{Iliesiu:2018fao}.

\begin{figure}
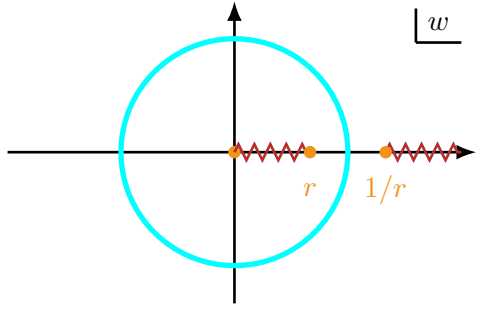

    \centering
    \AnalyticStructureThermal
    \caption{Choice of analytic structure of the correlator when restricted on the main strip $0 < r < \beta$.
    The symmetry $w \longleftrightarrow -w$ is restored by summing over images.}
    \label{fig:AnalyticStructure2}
\end{figure}

One can identify the cut in the $z$-plane in the same way keeping $\zb$ fixed.
We find that the consistent cut is:
\begin{equation} \nonumber
    z \in (\beta, \infty) \qquad \longleftrightarrow \qquad w \in (1/r,+\infty)\,.
\end{equation}
This is consistent with the symmetry $z \longleftrightarrow \zb$, i.e., $w \longleftrightarrow 1/w$.
All in all we obtain the analytic structure shown in Figure~\ref{fig:AnalyticStructure2}.

We are now ready to write the dispersion relation.
Since we consider only the cuts in the fundamental strip $0<r<\beta$, we need to sum the result over images to obtain the contributions from all the cuts.
Using Cauchy's integral formula this can be expressed as:
\begin{equation} \nonumber
    g(r,w)
    =
    \sum_{m=-\infty}^\infty \oint_{\Cm_1 + \Cm_2} \frac{\text{d}w'}{2\pi i} \frac{g(r_m, w')}{w' - w_m}\,,
\end{equation}
where we defined:
\begin{equation} \nonumber
    r_m^2 
    =
    (z+m\beta)(\zb +m\beta)\,, \qquad
    w_m^2
    =
    \frac{z+m\beta}{\zb +m\beta}\, ,
\end{equation}
and $\mathcal{C}_1$ and $\mathcal{C}_2$ are contours surrounding the branch cuts shown in Figure~\ref{fig:AnalyticStructure2}.
Using the symmetry $w \longleftrightarrow 1/w$ of $g(r,w)$ and rewriting the contour integral with a discontinuity, we arrive to the result:
\begin{align}
    g(r,w)
    &=
    \frac{1}{2}
    \sum_{m=-\infty}^\infty
    \int_{0}^{r_m} \frac{\text{d} w'}{2 \pi i}
    \frac{w_m^2 (1-w'^4)}{w' (w'-w_m)(w'+w_m) (1-w_m^2 w'^2)}
    \operatorname{Disc} g(r_m w',r_m/w') \notag \\
    &\phantom{=\ }
    + g_\text{arcs} (r,w)\,.
    \label{eq:GMI_FinalFormula}
\end{align}

We now wish to input the OPE in this formula.
Our goal here is to justify Equation~\eqref{eq:GMI_PolyakovBlocks}.
It turns out to be more convenient to start from the formulation as a Cauchy integral:
\begin{equation} \nonumber
    g(r,w)
    =
    \sum_{m=-\infty}^\infty \oint_{\Cm} \frac{\text{d}w'}{2\pi i} \frac{1}{w' - w_m} \sum_\Om a_\Om f_\Om (r_m,w')\,,
\end{equation}
where $\Cm$ is a contour around the point $(r,w)$ located inside the main strip $0 < r < \beta$, and where $f_\Om (r,w)$ is a thermal block in the $s$-channel.
In the $r, w$ variables the blocks take the form:
\begin{equation} \nonumber
    f_\Om (r,w)
    =
    r^{\Delta - 2\Delta_\phi} C_J^{(\nu)} \left( \frac{w + w^{-1}}{2} \right)\,.
\end{equation}
We now wish to commute the OPE sum with the integral and the sum of images.
This is a delicate operation which is justified only if the double sum is absolutely convergent.
In general we do not expect this condition to hold and we account for the error introduced by adding a homogeneous contribution we denote as ``arcs'':
\begin{equation} \nonumber
    g(r,w)
    =
    \frac{1}{\beta^{2\Delta_\phi}}
    \sum_\Om a_\Om \sum_{m=-\infty}^\infty \oint_{\Cm} \frac{\text{d}w'}{2\pi i} \frac{1}{w' - w_m} f_\Om (r_m,w')
    +
    \hat{g}_\text{arcs} (r,w)\,.
\end{equation}
The hat on the second term is here to emphasize that we do not expect these arcs to be the same function as the arcs in~\eqref{eq:GMI_FinalFormula}.
The next step is simply to use Cauchy's integral formula backwards.
Since $f_\Om (r,w')$ is holomorphic in $\Cm$, we simply obtain the block itself and the result is:
\begin{equation}
    g(r,w)
    =
    \frac{1}{\beta^{\Delta_\phi}}
    \sum_\Om a_\Om \sum_{m=-\infty}^\infty f_\Om (r_m,w_m)
    +
    \hat{g}_\text{arcs} (r,w)\,,
    \label{eq:GMI_SumOfImagesOfBlocks}
\end{equation}
Note that the first and second terms are periodic on their own.
In the first term, each OPE contribution is periodic independently.

It is possible to check this formula in a different, independent way by inputting the $t$-channel in~\eqref{eq:GMI_FinalFormula}.
The integral is challenging but one can check for a given choice of block that the result is precisely~\eqref{eq:GMI_SumOfImagesOfBlocks} after commuting the OPE sum with the integral and the sum of images.

\section{Evaluation of the inversion formula \eqref{eq: inversion app}} \label{app: inte}

In this appendix we compute explicitly the inversion formula. The object to evaluate reads:
\begin{equation} \label{eq: inversion2}
    \frac{\tilde{a}^{2d}_{\mathcal O}}{\chi  D^{\alpha}}
=
\frac{1}{\pi}\frac{2}{1+\delta_{J,0}}
\operatorname*{Res}_{\alpha=-\Delta}
\int_{\Lambda}^{\infty}
\text{d}p\,
p^{-\alpha} 
\int_{-1}^{1} \text{d}\eta\;
\frac{\eta^2 (1-\eta^2)^{-\frac12}}{\eta+p (1-\eta^2)} \; T_{J}(\eta) \ .
\end{equation}
The integrand has two real poles $\eta^{(\pm)}$ located at the values:
\begin{equation} \nonumber
    \eta^{(-)}=P \ , \quad \eta^{(+)}=-\frac{1}{P} \ , \quad P=\frac{1-\sqrt{1+4 p^2}}{2p} \in (-1,0) \ ,
\end{equation}
and so it can be reshaped in the convenient form, where we also exchange the variable $p$ for $P$:
\begin{equation} \nonumber
    \frac{\tilde{a}^{2d}_{\mathcal O}}{\chi  D^{\alpha}}
=
-\frac{1}{\pi}\frac{2}{1+\delta_{J,0}}
\operatorname*{Res}_{\alpha=-\Delta}
\int_{-1}^{P_{\Lambda}}
\frac{\text{d}P}{P^2-1}
\left(P-\frac{1}{P} \right)^{\alpha}
\int_{-1}^{1} \text{d}\eta\;
\frac{\eta^2 \, T_{J}(\eta)}{\sqrt{1-\eta^2}}  \; \left( \frac{1}{\eta-P} - \frac{1}{\eta+\frac{1}{P}}\right) \ .
\end{equation}
The analytic structure of the integrand and the contour of integration are shown in figure \ref{fig:2Dinversion}. Using the Chebyshev recurrence relation $2 x T_{J}(x)=T_{J+1}(x)+T_{J-1}(x)$, it can be shown that: 
\begin{equation} \label{eq: cheb id}
    \eta^2 T_{J}(\eta)=\frac{1}{4}\left(T_{J+2}+2T_{J}+T_{J-2} \right) \ .
\end{equation}
\paragraph{$\eta$ integral.}
\newcommand{\EtaAnalyticStructureShifted}{%
\begin{tikzpicture}[
    scale=1,
    baseline={([yshift=-.5ex]current bounding box.center)}
]
    \draw[scalar,-Latex] (0,0) -- (6.2,0);
    \draw[scalar,-Latex] (3,-2) -- (3,2);

    \node at (5.7,1.7) {$\eta$};
    \draw[scalar] (5.4,1.9) -- (5.4,1.45) -- (6,1.45);

    \draw[
        decorate,
        decoration={zigzag,segment length=6,amplitude=3},
        line width=1,
        SCRed
    ] (0,0) -- (1,0);

    \draw[
        decorate,
        decoration={zigzag,segment length=6,amplitude=3},
        line width=1,
        SCRed
    ] (1,0) -- (2,0);

    \draw[
        decorate,
        decoration={zigzag,segment length=6,amplitude=3},
        line width=1,
        SCRed
    ] (4,0) -- (6,0);

    \draw[
        blue,
        very thick,
        postaction={decorate},
        decoration={
            markings,
            mark=at position .75 with {\arrow{Latex}}
        }
    ] (2,0) -- (3.25,0);

    \draw[
        blue,
        very thick,
        postaction={decorate},
    ] (3.75,0) -- (4,0);

    \draw[fill=WCOrange,WCOrange] (2,0) circle (.075);
    \node[WCOrange] at (2,-.5) {$-1$};

    \draw[fill=WCOrange,WCOrange] (4,0) circle (.075);
    \node[WCOrange] at (4,-.5) {$1$};

    \draw[fill=MediumBlueGrey,MediumBlueGrey]
        (1,0) circle (.075);
    \node[MediumBlueGrey] at (1,-.5) {$\eta^{(+)}$};

\draw[blue,very thick]
    (3.75,0)
    arc[start angle=0,end angle=-180,radius=.25];

    \draw[fill=MediumBlueGrey,MediumBlueGrey]
        (3.5,0) circle (.075);

    \node[MediumBlueGrey,anchor=west]
        at (3.25,.65) {$\eta^{(-)}+i0$};
\end{tikzpicture}%
}
\newcommand{\EtaAnalyticStructure}{%
\begin{tikzpicture}[
    scale=1,
    baseline={([yshift=-.5ex]current bounding box.center)}
]
    \draw[scalar,-Latex] (0,0) -- (6.2,0);
    \draw[scalar,-Latex] (3,-2) -- (3,2);

    \node at (5.7,1.7) {$\eta$};
    \draw[scalar] (5.4,1.9) -- (5.4,1.45) -- (6,1.45);

    \draw[
        decorate,
        decoration={zigzag,segment length=6,amplitude=3},
        line width=1,
        SCRed
    ] (4,0) -- (6,0);

    \draw[fill=WCOrange,WCOrange] (4,0) circle (.075);
    \node[WCOrange] at (4,-.5) {$1$};

    \draw[fill=WCOrange,WCOrange] (2,0) circle (.075);
    \node[WCOrange] at (2,-.5) {$-1$};

    \draw[
        decorate,
        decoration={zigzag,segment length=6,amplitude=3},
        line width=1,
        SCRed
    ] (0,0) -- (2,0);

    \draw[fill=MediumBlueGrey,MediumBlueGrey]
        (3.5,0) circle (.075);
    \node[MediumBlueGrey] at (3.5,-.5) {$\eta^{(-)}$};

    \draw[fill=MediumBlueGrey,MediumBlueGrey]
        (1,0) circle (.075);
    \node[MediumBlueGrey] at (1,-.5) {$\eta^{(+)}$};
\end{tikzpicture}%
}

\begin{figure}
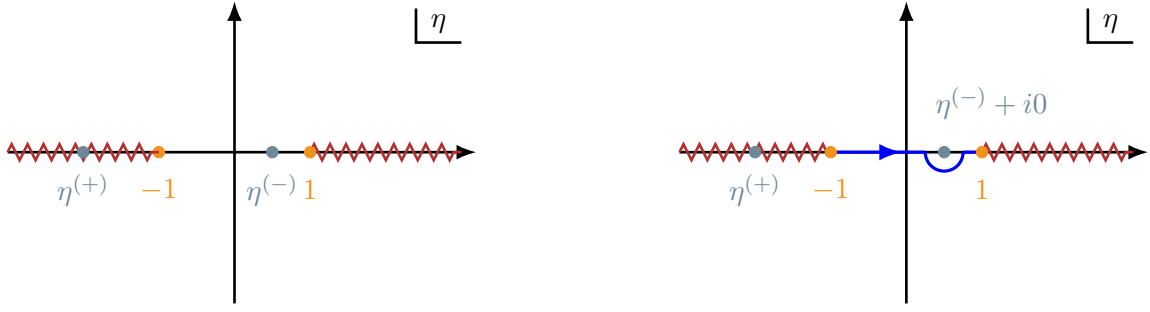

    \centering

    \EtaAnalyticStructure
    \hfill
    \EtaAnalyticStructureShifted

    \caption{The analytic structure in the $\eta$ plane of the integrand in the $2d$ inversion formula \eqref{eq: inversion2}. The pole $\eta^{(-)}$ always lie on the integration path: to perform the computation it is necessary to use
    the $+i0$ prescription. This generates an additional residue which provides the imaginary part for the extracted OPE coefficients (see \eqref{eq: coeffim}).}
    \label{fig:2Dinversion}
\end{figure}
We focus on the $\eta$ integral, which splits into two fundamental ones:
\begin{equation} \nonumber
    I_{\ell}(P)=I_{\ell}^{(1)}(P)-I_{\ell}^{(2)}(P)=\int_{-1}^{1} \text{d}\eta\;
\frac{T_{\ell}(\eta)}{\sqrt{1-\eta^2}}  \; \frac{1}{\eta-P} -\int_{-1}^{1} \text{d}\eta\;
\frac{T_{\ell}(\eta)}{\sqrt{1-\eta^2}}  \;  \frac{1}{\eta+\frac{1}{P}} \ ,
\end{equation}
where $\ell=J+2, J, J-2$, following \eqref{eq: cheb id}. 

The integral $I_{\ell}^{(1)}(P)$ requires $\pm i 0$ prescription in order to be evaluated. It splits into a principal value and a residue in $\eta=P$:
\begin{equation}
    I_{\ell}^{(1)}(P)=\text{PV}\, I_{\ell}^{(1)}(P)\mp \pi i \frac{T_{\ell}(P)}{\sqrt{1-P^2}} \ .
\end{equation}
The principal value can be computed by recursion. For this purpose, we switch to angular coordinates $\eta= \cos \vartheta$:
\begin{equation}
    I_{\ell}^{(1)}(P)=\int_{0}^{\pi} \text{d}\vartheta  \; \frac{\cos(\ell \vartheta)}{\cos \vartheta-P}
\end{equation}
and we compute the intial conditions:
\begin{equation}
    \text{PV} \, I_{0}^{(1)}(P)=0 \ , \quad \text{PV}\, I_{1}^{(1)}(P)= \pi + P \; \text{PV}\, I_{0}^{(1)}(P)=\pi  \ ,
\end{equation} 
and consider the recursion relation:
\begin{equation}
    \cos((\ell+1)\theta)=2 \cos(\theta) \cos(\ell \theta)-\cos((\ell-1)\theta) \ ,
\end{equation}
which leads to a recursion relation for the principal values:
\begin{equation}
    \text{PV} \, I_{\ell+1}^{(1)}(P)=2P \; \text{PV} \, I_{\ell}^{(1)}(P)-\text{PV} \, I_{\ell-1}^{(1)}(P) 
\end{equation}
which coincides with the definition of the Chebyschev polynomials of the second kind $ \pi U_{\ell-1}(P)$.  With the $+i0$ prescription, the first integral evaluates to
\begin{equation}
    I_{\ell}^{(1)}(P)=\pi U_{|\ell|-1}(P)- \pi i \frac{T_{\ell}(P)}{\sqrt{1-P^2}} \ , \quad U_{-1}(P)\equiv 0 \ .
\end{equation}

By contrast, the integral $I_{\ell}^{(2)}(P)$ converges on the integration path. It is again useful to switch to angular coordinates: 
\begin{equation} \label{eq: compaeri}
    I_{\ell}^{(2)}(P)=\int_{-1}^{1} \text{d}\eta\;
\frac{T_{\ell}(\eta)}{\sqrt{1-\eta^2}}  \;  \frac{1}{\eta+\frac{1}{P}}=P \int_{0}^{\pi} \text{d}\vartheta\;
 \frac{\cos(\ell \vartheta)}{1+P \cos \vartheta} \ .
\end{equation}
Defining the variables
\begin{equation} \label{eq: new variables}
    s=\sqrt{1-P^2} \ , \quad t=\frac{P}{1+s} \ ,
\end{equation}
it can be shown that the denominator can be put in the form:
\begin{equation} \nonumber
    \frac{1}{1+P \cos \vartheta}=\frac{1}{s}\frac{1-t^2}{1+2t \cos \vartheta+t^2} \ ,
\end{equation}
where we recognize the Poisson kernel, which is useful since it admits the decomposition:
\begin{equation} \nonumber
    \frac{1-t^2}{1+2t \cos \vartheta+q^2}=1+2 \sum_{m=1}^{\infty}(-t)^m \cos(m \vartheta) \ .
\end{equation}
Substituting the Poisson kernel into the integral reduces the latter to an orthogonality relation for the cosine:
\begin{equation} \nonumber
    I_{\ell}^{(2)}(P)=2 \frac{P}{\sqrt{1-P^2}} \sum_{m=1}^{\infty} (-t)^m \int_{0}^{\pi} \text{d}\vartheta \, \cos(\ell \vartheta)\cos(m \vartheta)=(-1)^{|\ell|} \frac{\pi P}{\sqrt{1-P^2}} \left(\frac{P}{1+\sqrt{1-P^2}}\right)^{|\ell|} \ ,
\end{equation}
where we introduced $|\ell|$ since $T_{\ell}(\eta)=T_{-\ell}(\eta)$.
Using the original expression \eqref{eq: compaeri}, it can be shown that the same final formula also applies to $\ell=0$, leading to the final result:
\begin{equation} \nonumber
    I_{\ell}^{(2)}(P)=(-1)^{|\ell|} \frac{\pi P}{\sqrt{1-P^2}} \left(\frac{P}{1+\sqrt{1-P^2}}\right)^{|\ell|}. 
\end{equation}

In conclusion, the full integral is a rather complicated algebraic function of $P$:
\begin{equation} \nonumber
    I_{\ell}(P)=\pi U_{|\ell|-1}(P)-(-1)^{|\ell|} \frac{\pi P}{\sqrt{1-P^2}} \left(\frac{P}{1+\sqrt{1-P^2}}\right)^{|\ell|}- \pi i \frac{T_{\ell}(P)}{\sqrt{1-P^2}} \ , \quad U_{-1}(P)=0 \ .
\end{equation}
\paragraph{$p$ integral.} We now need to evaluate the second main integral entering the inversion formula.
In terms of the variable $P$, it reads
\begin{equation} \nonumber
    H_{\ell}(\alpha)=\int_{-1}^{P_{\Lambda}}
\frac{\text{d}P}{P^2-1}
\left(P-\frac{1}{P} \right)^{\alpha} \, I_{\ell}(P) \ .
\end{equation}
It is useful to switch to the intermediate $t$ coordinate previously defined in~\eqref{eq: new variables}.
The integral reads
\begin{equation} \nonumber
    H_{\ell}(\alpha)=\int_{-1}^{t_{\Lambda}} \frac{\text{d}t}{1-t^2}\left[-\frac{(1-t^2)^2}{2t (1+t^2)} \right]^{\alpha}I_{\ell}\left(t\right) \ ,
\end{equation}
where the $\eta$ integral takes the form
\begin{equation} \label{eq: kernel}
    I_{\ell}(t)=\pi U_{|\ell|-1}\left(\frac{2t}{1+t^2} \right)-\frac{2 \pi (-1)^{|\ell|} t^{|\ell|+1}}{1-t^2}- i \pi \frac{1+t^2}{1-t^2} T_{\ell}\left(\frac{2t}{1+t^2} \right) \ .
\end{equation}
The final change of coordinates moves the lower endpoint from $-1$ to $0$, allowing the reconstruction of the poles via a Laurent expansion. If we impose $t=-e^{-u}$, then the integral reads:
\begin{equation} \nonumber
    H_{\ell}(\alpha)=\int_{0}^{u_{\Lambda}} \text{d}u \frac{(\sinh u)^{2\alpha-1}}{(\cosh u)^\alpha}I_{\ell}\left(u\right) \ ,
\end{equation}
with the $\eta$ integral now reading:
\begin{equation} \nonumber
    I_{\ell}\left(u\right)=\pi U_{|\ell|-1}(-\text{sech} u)+\frac{\pi e^{-|\ell|u}}{\sinh u}-i\pi  \coth u \, T_{\ell}(-\text{sech} u) \ .
\end{equation}
To evaluate the $u$ integral, the original large-$q$ asymptotic has been replaced by a Laurent series in $u$. We can use it to compute the first terms of $H_{\ell}(\alpha)$. The truncated Laurent expansion of the kernel in \eqref{eq: kernel} is:
\begin{equation} \nonumber
    \frac{(\sinh u)^{2\alpha-1}}{(\cosh u)^\alpha}=u^{-1+2\alpha}-\frac{1}{6} (\alpha +1) u^{1+2\alpha}+\frac{1}{360} (\alpha +7) (5 \alpha +1) u^{3+2\alpha}+ \dots 
\end{equation}
The truncated Laurent expansion of the $I_{\ell}(u)$ integral depends on the spin $J$.
For simplicity, we focus on operators with spin $J=0$, but the same method applies to arbitrary spin.
According to~\eqref{eq: cheb id}, we need $\ell=0,2$.
The Laurent expansions are the following:
\begin{align*}
    I_{0}(u)&=\frac{(1-i) \pi }{u}-\left(\frac{1}{6}+\frac{i}{3}\right) \pi  u+\left(\frac{7}{360}+\frac{i}{45}\right) \pi  u^3+\dots \ , \\
    I_{2}(u)&=-4 \pi+\frac{(1-i) \pi }{u}+\left(\frac{11}{6}+\frac{5 i}{3}\right) \pi  u+\left(\frac{127}{360}-\frac{29 i}{45}\right) \pi  u^3+\dots \ 
\end{align*}
In conclusion, we can iteratively compute the $H_{\ell}(\alpha)$ integrals and put them together according to \eqref{eq: cheb id}. The final formula for the coefficients is:
\begin{equation} \label{eq: master form}
     \frac{\tilde{a}^{2d}_{\mathcal O}}{\chi  D^{\alpha}}
=
-\frac{1}{2\pi}\frac{1}{1+\delta_{J,0}}\operatorname*{Res}_{\alpha=-\Delta}  \left(H_{J+2}(\alpha)+2 H_{J}(\alpha)+H_{J-2}(\alpha) \right) \ .
\end{equation}
For $J=0$, the formula \eqref{eq: master form} reduces to:\footnote{In the formula, every residue has already been evaluated to its corresponding pole for clarity purposes.}
\begin{equation} \nonumber
    \tilde{a}^{2d}_{\Delta,0}
=\frac{\chi}{D^{\Delta}}
\frac{1}{\pi}\operatorname*{Res}_{\alpha=-\Delta} \left[-\frac{\pi}{2}\,
\frac{1-i}{\alpha-\frac12}
+\frac{\pi}{\alpha}
-\frac{3\pi}{8}\,
\frac{1+i}{\alpha+\frac12}-
\frac{5\pi}{64}\,
\frac{1-i}{\alpha+\frac32}
+\dots \right]   \ .
\end{equation}

\bibliography{./main.bib}
\bibliographystyle{./auxi/JHEP}

\end{document}